\documentclass[12pt,a4paper]{article}
\pdfoutput=1
\usepackage{ifthen}
\newboolean{pdflatex}
\setboolean{pdflatex}{true}

\newboolean{articletitles}
\setboolean{articletitles}{true}

\newboolean{uprightparticles}
\setboolean{uprightparticles}{false}

\newboolean{inbibliography}
\setboolean{inbibliography}{false}

\usepackage[top=1in, bottom=1.25in, left=1in, right=1in]{geometry}

\usepackage{microtype}
\usepackage{lineno}  
\usepackage{xspace} 
\usepackage{caption} 

\usepackage{graphicx}  
\usepackage{color}
\usepackage{colortbl}
\graphicspath{{./figs/}} 

\usepackage{amsmath} 
\usepackage{amssymb}
\usepackage{amsfonts}
\usepackage{upgreek} 

\newcommand*\patchAmsMathEnvironmentForLineno[1]{%
\expandafter\let\csname old#1\expandafter\endcsname\csname #1\endcsname
\expandafter\let\csname oldend#1\expandafter\endcsname\csname
end#1\endcsname
 \renewenvironment{#1}%
   {\linenomath\csname old#1\endcsname}%
   {\csname oldend#1\endcsname\endlinenomath}%
}
\newcommand*\patchBothAmsMathEnvironmentsForLineno[1]{%
  \patchAmsMathEnvironmentForLineno{#1}%
  \patchAmsMathEnvironmentForLineno{#1*}%
}
\AtBeginDocument{%
\patchBothAmsMathEnvironmentsForLineno{equation}%
\patchBothAmsMathEnvironmentsForLineno{align}%
\patchBothAmsMathEnvironmentsForLineno{flalign}%
\patchBothAmsMathEnvironmentsForLineno{alignat}%
\patchBothAmsMathEnvironmentsForLineno{gather}%
\patchBothAmsMathEnvironmentsForLineno{multline}%
\patchBothAmsMathEnvironmentsForLineno{eqnarray}%
}

\usepackage{hyperref}    
\usepackage[all]{hypcap} 

\usepackage{cite} 
\usepackage{mciteplus}
\usepackage{soul}

\usepackage{longtable}

\usepackage[separate-uncertainty=true]{siunitx}
\usepackage{lhcb-definitions}
\usepackage{charmproduction-definitions}
\usepackage{measurement-definitions}
\usepackage{subcaption}
\usepackage{booktabs}
\usepackage{lscape}
\usepackage{adjustbox}
\usepackage{hvfloat}

\begin{document}

\renewcommand{\thefootnote}{\fnsymbol{footnote}}
\setcounter{footnote}{1}


\newcommand{\DzXsec}{\SI[parse-numbers=false]{1374 \pm \phantom{1}3 \pm \phantom{1}74}{\micro\barn}}
\newcommand{\DpXsec}{\SI[parse-numbers=false]{551 \pm \phantom{1}5 \pm \phantom{1}48}{\micro\barn}}
\newcommand{\DzOnePtXsec}{\SI[parse-numbers=false]{1004 \pm 3 \pm 54}{\micro\barn}}
\newcommand{\DpOnePtXsec}{\SI[parse-numbers=false]{402 \pm 2 \pm 30}{\micro\barn}}
\newcommand{\DspOnePtXsec}{\SI[parse-numbers=false]{170 \pm 4 \pm 16}{\micro\barn}}
\newcommand{\DstpOnePtXsec}{\SI[parse-numbers=false]{421 \pm 5 \pm 36}{\micro\barn}}
\newcommand{\ccbarXsec}{\SI[parse-numbers=false]{1193 \pm 3 \pm 67\pm 58}{\micro\barn}}
\newcommand{\ccbarOnePtXsec}{\SI[parse-numbers=false]{874 \pm 2 \pm 49\pm 40}{\micro\barn}}
\newcommand{\DpDzint}{$\phantom{1}0.943 ^{+\phantom{1}0.005}_{-\phantom{1}0.005}$$^{+\phantom{1}0.029}_{-\phantom{1}0.029}$}
\newcommand{\DspDzint}{$0.0983 ^{+0.0021}_{-0.0021}$$^{+0.0035}_{-0.0035}$}
\newcommand{\DstpDzint}{$\phantom{1}0.283 ^{+\phantom{1}0.004}_{-\phantom{1}0.004}$$^{+\phantom{1}0.016}_{-\phantom{1}0.016}$}
\newcommand{\DspDpint}{$0.1042 ^{+0.0022}_{-0.0022}$$^{+0.0031}_{-0.0031}$}
\newcommand{\DstpDpint}{$\phantom{1}0.300 ^{+\phantom{1}0.004}_{-\phantom{1}0.004}$$^{+\phantom{1}0.015}_{-\phantom{1}0.015}$}
\newcommand{\DspDstpint}{$\phantom{1}0.348 ^{+\phantom{1}0.008}_{-\phantom{1}0.009}$$^{+\phantom{1}0.018}_{-\phantom{1}0.018}$}

\newcommand{\suDzstat}{0--10}
\newcommand{\suDzmcstat}{0--10}
\newcommand{\suDztracking}{3--5}
\newcommand{\suDzpidstat}{0--1}
\newcommand{\suDzpidbin}{0--30}
\newcommand{\suDzmcagreement}{0.3}
\newcommand{\suDzfit}{0--3}
\newcommand{\suDztruthmatching}{0.04}
\newcommand{\suDzsignalwindow}{0--1}

\newcommand{\suDpstat}{0--10}
\newcommand{\suDpmcstat}{0--10}
\newcommand{\suDptracking}{5--7}
\newcommand{\suDppidstat}{0--1}
\newcommand{\suDppidbin}{0--10}
\newcommand{\suDpmcagreement}{0.7}
\newcommand{\suDpfit}{0--3}
\newcommand{\suDptruthmatching}{0.08}
\newcommand{\suDpsignalwindow}{0--1}

\newcommand{\suDstpstat}{0--20}
\newcommand{\suDstpmcstat}{1--10}
\newcommand{\suDstptracking}{5--7}
\newcommand{\suDstppidstat}{0--2}
\newcommand{\suDstppidbin}{0--20}
\newcommand{\suDstpmcagreement}{2}
\newcommand{\suDstpfit}{0.0--1.0}
\newcommand{\suDstptruthmatching}{0.5}
\newcommand{\suDstpsignalwindow}{0--2}

\newcommand{\suDspstat}{0--20}
\newcommand{\suDspmcstat}{2--9}
\newcommand{\suDsptracking}{4--7}
\newcommand{\suDsppidstat}{0--2}
\newcommand{\suDsppidbin}{0--20}
\newcommand{\suDspmcagreement}{0.6}
\newcommand{\suDspfit}{0--3}
\newcommand{\suDsptruthmatching}{0.2}
\newcommand{\suDspsignalwindow}{0--3}

\begin{titlepage}
\pagenumbering{roman}

\vspace*{-1.5cm}
\centerline{\large EUROPEAN ORGANIZATION FOR NUCLEAR RESEARCH (CERN)}
\vspace*{1.5cm}
\noindent
\begin{tabular*}{\linewidth}{lc@{\extracolsep{\fill}}r@{\extracolsep{0pt}}}
\ifthenelse{\boolean{pdflatex}}
{\vspace*{-2.7cm}\mbox{\!\!\!\includegraphics[width=.14\textwidth]{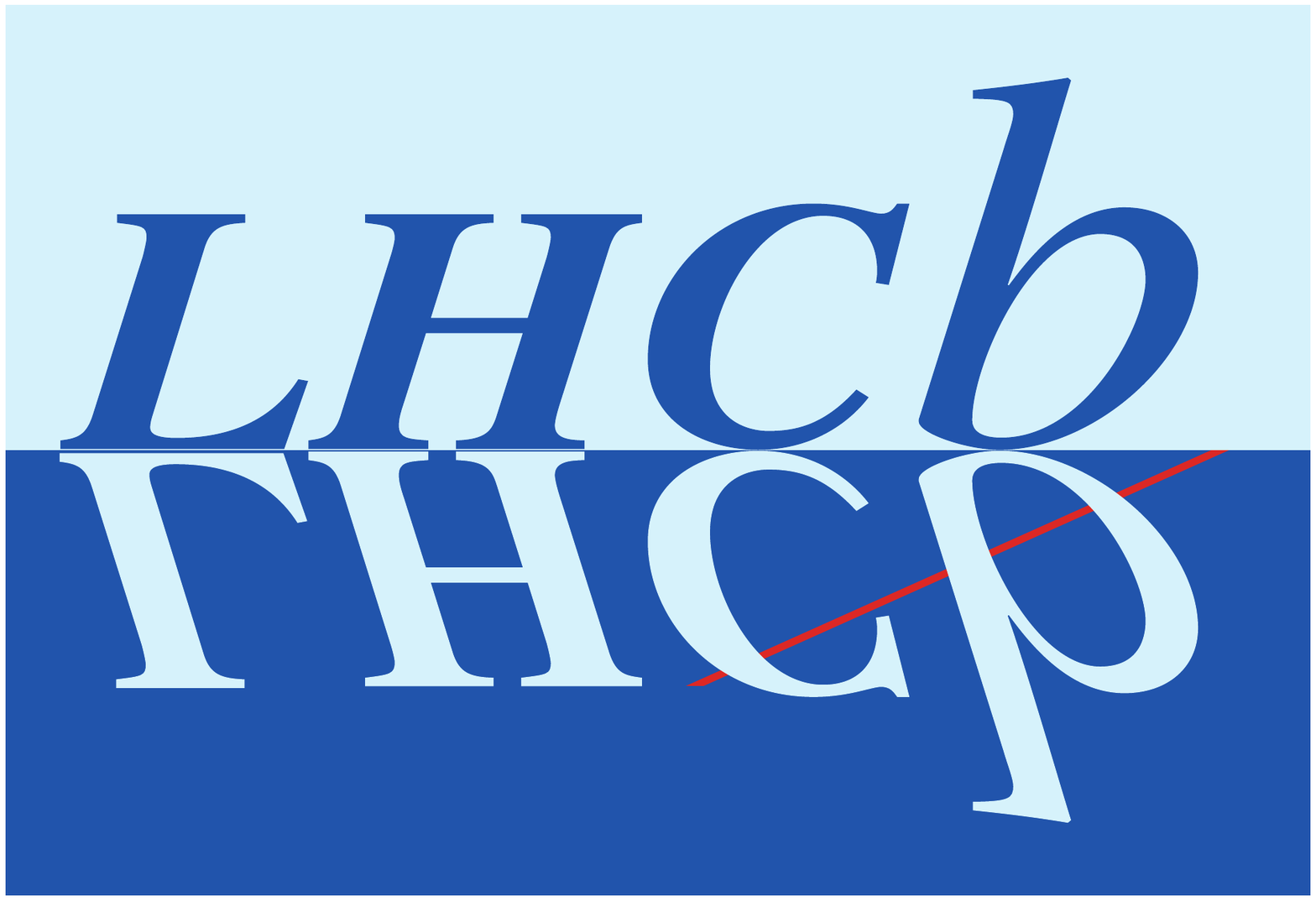}} & &}%
{\vspace*{-1.2cm}\mbox{\!\!\!\includegraphics[width=.12\textwidth]{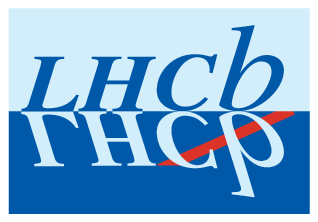}} & &}%
\\
 & & CERN-EP-2016-244 \\  
 & & LHCb-PAPER-2016-042 \\  
 & & \today \\ 
\end{tabular*}

\vspace*{1.5cm}

{\bf\boldmath\huge
\begin{center}
  Measurements of prompt charm production cross-sections in \pp collisions at $\sqrt{s} = 5$\,TeV
\end{center}
}

\vspace*{0.5cm}

\begin{center}
The LHCb collaboration\footnote{Authors are listed at the end of this paper.}
\end{center}

\vspace{0.5cm}

\begin{abstract}
  \noindent
  Production cross-sections of prompt charm mesons are measured using data from \pp collisions at the LHC at a centre-of-mass energy of \SI{5}{\TeV}.
  The data sample corresponds to an integrated luminosity of
  \totlumi collected by the \lhcb experiment.
  The production cross-sections of \Dz, \Dp, \Dsp, and \Dstp mesons are measured in bins
  of charm meson transverse momentum, \pT, and rapidity, \rapidity. They cover the rapidity range \yrange and transverse momentum ranges $0 < \pT < \SI{10}{\gevc}$
  for \Dz and \Dp and $1 < \pT < \SI{10}{\gevc}$ for \Dsp and \Dstp mesons.
  The inclusive cross-sections for the four mesons, including charge-conjugate states, within the range of $1 < \pT < \SI{8}{\gevc}$ are determined to be
  \begin{align*}
    \begin{array}{lcr}
      \sigma(\decay{\pp}{\Dz X})  &=& \DzOnePtXsec, \\
      \sigma(\decay{\pp}{\Dp X})  &=& \DpOnePtXsec, \\
      \sigma(\decay{\pp}{\Dsp X}) &=& \DspOnePtXsec, \\
      \sigma(\decay{\pp}{\Dstp X})&=& \DstpOnePtXsec, \\
    \end{array}
  \end{align*}
where the uncertainties are statistical and systematic, respectively.

\end{abstract}

\vspace*{\fill}

\begin{center}
  Submitted to JHEP
\end{center}

\vspace{\fill}

{\footnotesize 
\centerline{\copyright~CERN on behalf of the \lhcb collaboration, licence \href{http://creativecommons.org/licenses/by/4.0/}{CC-BY-4.0}.}}
\vspace*{2mm}

\end{titlepage}


\newpage
\setcounter{page}{2}
\mbox{~}
%

\cleardoublepage

\renewcommand{\thefootnote}{\arabic{footnote}}
\setcounter{footnote}{0}

\pagestyle{plain}
\setcounter{page}{1}
\pagenumbering{arabic}


\section{Introduction}
\label{sec:Introduction}
Measurements of charm production cross-sections in proton-proton collisions are important tests of perturbative quantum chromodynamics~\cite{Gauld:2015yia,Cacciari:2015fta,Kniehl:2012ti,Maciula:2013wg}.
Predictions of charm meson cross-sections have been made 
at next-to-leading order using the generalised mass variable flavour number scheme~(GMVFNS)~\cite{Kniehl:2004fy,Kniehl:2005gz,Kniehl:2005ej,Kneesch:2007ey,Kniehl:2009ar,Kniehl:2012ti} and 
at fixed order with next-to-leading-log resummation~(FONLL)~\cite{Gauld:2015yia,Cacciari:1998it,Cacciari:2003zu,Cacciari:2005uk,Cacciari:2012ny, Cacciari:2015fta}.
These are based on a factorisation approach, where the cross-sections are calculated as a convolution of three terms: the parton distribution functions 
of the incoming protons; the partonic hard scattering rate, estimated as a perturbative series in the coupling constant of the strong interaction; and a fragmentation function 
that parametrises the hadronisation of the charm quark into a given type of charm hadron. Predictions are also obtained from Monte Carlo programs implementing NLO perturbative QCD, where the partonic final state is matched to parton shower simulations~\cite{Gauld:2015yia}.
The range of rapidity, \rapidity, and of the momentum component transverse to the beam axis, \pT, accessible to the \lhcb experiment enables quantum chromodynamics calculations to be tested in a region where the momentum fraction, $x$, of the initial state partons 
can reach values below $10^{-4}$. In this region the uncertainties on the gluon density function are large, exceeding 30$\%$~\cite{Gauld:2015yia,Zenaiev:2015rfa}, and 
\lhcb measurements can be used to constrain it.
This paper presents measurements of charm hadron production cross-sections at a proton-proton centre-of-mass energy of
\comenergy.
Ratios of cross-sections at different \sqrts benefit from cancellations in experimental uncertainties on the measured cross-sections as well as theoretical uncertainties on the predictions \cite{Cacciari:2015fta}, allowing for precise comparisons.
Measurements of charm production cross-sections in $pp$ collisions at {\comenergy} also define the reference for the determination of nuclear modification factors in heavy ion collisions at that nucleon-nucleon centre-of-mass energy, recorded in 2015 at the Large Hadron Collider (LHC).

Measurements of charm production cross-sections in hadronic collisions have been performed in different kinematic regions and for various centre-of-mass energies in the TeV range.
Measurements by the CDF experiment cover the central rapidity region $|y| < 1$ and transverse momenta, \pT, 
between \SI{5.5}{\gevc} and \SI{20}{\gevc} at  $\sqrts = \SI{1.96}{\TeV} $ in \ppbar collisions~\cite{Acosta:2003ax}.
At the LHC, charm cross-sections in \pp collisions have been measured in 
the $|y| < 0.5$ region for $\pT>\SI{1}{\gevc}$ at $\sqrts =\SI{2.76}{\TeV}$ 
and for $\pT>\SI{0}{\gevc}$ at $\sqrts = \SI{7}{\TeV}$  by the ALICE experiment~\cite{Abelev:2012sx,ALICE:2012ik,ALICE:2011aa,Adam:2016ich},
and for pseudorapidity $|\Eta| < 2.1$ in the \pT region $3.5 < \pT < \SI{100}{\gevc}$ at $\sqrts = \SI{7}{\TeV}$ by the ATLAS experiment~\cite{Aad:2015zix}.
The \lhcb experiment has recorded the world's largest dataset of charm hadrons to date and this has led to numerous high-precision measurements of their production and decay properties. 
\lhcb measured the cross-sections in the forward region $2.0 < y < 4.5$ for  $0<\pT<\SI{8}{\gevc}$ at $\sqrts = \SI{7}{\TeV}$~\cite{LHCb-PAPER-2012-041} and for  $0<\pT<\SI{15}{\gevc}$ at $\sqrts = \SI{13}{\TeV}$~\cite{LHCb-PAPER-2015-041}.

Charm mesons produced at the \pp collision point, either directly or as decay 
products of excited charm resonances, are referred to as promptly produced.
No attempt is made to distinguish between these two sources.
This paper presents measurements of the cross-sections for the prompt production of \Dz, \Dp, \Dsp, and $\PD^*(2010)^+$ (henceforth denoted as \Dstarp) mesons, 
based on data corresponding to an integrated luminosity of \totlumi.
Charm mesons produced through the decays of \bquark hadrons are referred to as secondary charm and are considered here as a background process.

The analysis techniques described in this paper are nearly identical to those used in the measurements made at $\sqrts = \SI{13}{\TeV}$~\cite{LHCb-PAPER-2015-041}, allowing for very precise determination of the ratios between the two results.

Section~\ref{sec:Detector} describes the detector, data acquisition conditions, and the simulation; this is followed by a summary of the data analysis in Sec.~\ref{sec:analysis}.
The differential cross-section results are given in Sec.~\ref{sec:measurements} while Sec.~\ref{sec:ratios} presents the measurements of integrated cross-sections and of the ratios of the cross-sections measured at $\sqrts = \SI{5}{\TeV}$ to those at \SI{13}{\TeV}.
The theoretical predictions and their comparison with the results of this paper are discussed in Sec.~\ref{sec:theory}.
Section~\ref{sec:summary} provides a summary.

\section{Detector and simulation}
\label{sec:Detector}
The \lhcb detector~\cite{Alves:2008zz,LHCb-DP-2014-002} is a single-arm forward
spectrometer covering the \mbox{pseudorapidity} range $2<\eta <5$,
designed for the study of particles containing \bquark or \cquark
quarks. The detector includes a high-precision tracking system
consisting of a silicon-strip vertex detector surrounding the $pp$
interaction region, a large-area silicon-strip detector located
upstream of a dipole magnet with a bending power of about
$4{\rm\,Tm}$, and three stations of silicon-strip detectors and straw
drift tubes placed downstream of the magnet.
The tracking system provides a measurement of momentum of charged particles with
a relative uncertainty that varies from 0.5\% at low momentum to 1.0\% at \SI{200}{\gevc}.
The minimum distance of a track to a primary vertex~(PV), the impact parameter~(IP), is measured with a resolution of $(15+29/\pt)\,\si{\micro\meter}$,
where \pt is the component of the momentum transverse to the beam, in\,\si{\gevc}.
Different types of charged hadrons are distinguished by information
from two ring-imaging Cherenkov detectors. 
Photons, electrons and hadrons are identified by a calorimeter system consisting of
scintillating-pad and preshower detectors, an electromagnetic
calorimeter and a hadronic calorimeter. Muons are identified by a
system composed of alternating layers of iron and multiwire
proportional chambers.

The online event selection is performed by a trigger. This consists of a hardware stage, which, for this analysis, randomly selects 
a pre-defined fraction of all beam-beam crossings at a rate of \SI{300}{\kilo\hertz}, followed by a software 
stage.
In between the hardware and software stages, an alignment and calibration of 
the detector is performed in near real-time~\cite{Dujany:2015lxd} and updated 
constants are made available for the trigger.
The same alignment and calibration information is propagated to the offline reconstruction, ensuring consistent and high-quality
particle identification~(\pid) information between the trigger and offline software.
The identical performance of the online and offline 
reconstruction offers the opportunity to perform physics analyses directly using candidates
reconstructed in the trigger~\cite{LHCb-DP-2012-004,LHCb-DP-2016-001}, which the present analysis exploits.
The storage of only the triggered candidates enables a reduction
in the event size by an order of magnitude.

In the simulation, $pp$ collisions are generated with
\pythia~8.1~\cite{Sjostrand:2006za, Sjostrand:2007gs} using a specific \lhcb
configuration similar to the one described in Ref.~\cite{LHCb-PROC-2010-056}.  Decays of hadronic particles
are described by \evtgen~\cite{Lange:2001uf} in which final-state
radiation is generated with \photos~\cite{Golonka:2005pn}. The
implementation of the interaction of the generated particles with the detector, and its response,
uses the \geant
toolkit~\cite{Allison:2006ve, *Agostinelli:2002hh} as described in
Ref.~\cite{LHCb-PROC-2011-006}.

\section{Analysis strategy}
\label{sec:analysis}

The analysis methodology is very similar to that used to measure the prompt production
cross-sections at $\sqrts = \SI{13}{\TeV}$~\cite{LHCb-PAPER-2015-041}, 
of which a summary is given below.
The measurement at $\sqrts = \SI{5}{\TeV}$ reconstructs the same final states:
\DzToKpi, \DpToKpipi, \DsTophipi and \DstarpTopipDzToKmpip. Throughout this paper, charge
conjugation is implied and thus the \DzToKpi sample contains the sum of the Cabibbo-favoured decays
\DzToKpi and the doubly Cabibbo-suppressed decays \DzbToKpi.
The \DsTophipi sample comprises all reconstructed \DsToKKpi decays where the invariant mass
of the $\Km\Kp$ pair falls into a $\pm\SI{20}{\mevcc}$ window around the nominal
$\phi(1020)$ mass, taken to be \SI{1020}{\mevcc}.

The cross-sections are measured in two-dimensional bins of \pT and \rapidity of 
the reconstructed mesons, where \pT and \rapidity are measured in the \pp 
centre-of-mass frame with respect to the \pp collision axis.  The bin widths 
are $0.5$ in \rapidity covering a range of \yrange and  \SI{1}{\gevc} in \pT 
for $0 < \pT < \SI{10}{\gevc}$.

\subsection{Selection of candidates}
\label{sec:ana_sel}

Candidates for \Dz, \Dp and \Dsp mesons are formed in events containing at least one reconstructed PV by combining tracks that have been positively identified as kaons or pions by the \lhcb \pid system.
These tracks are required to be above a transverse momentum threshold that depends on the decay mode, and
must be consistent with originating from a common vertex.  Due to the long lifetimes of the studied charm mesons,
this common vertex is required to be significantly displaced from any reconstructed PV and the displacement vector with respect to the closest PV must align with the combined momentum of the tracks. Candidates for \Dstp mesons are formed
by combining a \Dz and a charged pion candidate, requiring that both form a good quality vertex.
Applying the selection, 1\% of all events have more than one selected candidate, all of which are considered in this analysis.

The efficiencies for the reconstruction and selection of charm meson candidates are obtained for each \pTy bin in a near-identical manner to the $\sqrts = \SI{13}{\TeV}$ measurement~\cite{LHCb-PAPER-2015-041}.
All efficiencies are evaluated using the event simulation, except for the 
efficiencies for identifying kaons and pions and the tracking efficiencies.
Kaon and pion identification efficiencies are evaluated using a high 
purity calibration sample of \DstarpTopipDzToKmpip decays which have been 
selected without \pid requirements. Only the correct assignment of the kaon and pion hypotheses for the decay products of the \Dz candidates yields the expected shapes in the mass distributions. The numbers of true kaon and pion tracks in the calibration sample are determined from maximum likelihood fits to the {\Dz} and {\Dstp} candidate invariant mass distributions, as described in Section{~\ref{sec:ana_yie}}.
The identification efficiency for a true kaon or pion to pass a given {\pid} requirement is computed by performing these fits in bins of track multiplicity and track kinematics before and after the selection.
  This technique differs slightly from that used in the analysis of the ${\sqrts} = {\SI{13}{\TeV}}$ data,
where a single maximum likelihood fit was performed on the calibration sample integrated across all bins, and the per-bin kaon and pion yields were computed by summing 
sWeights~\cite{Pivk:2004ty}.
The change leads to a stable solution for sparsely populated bins and is introduced due to the substantially
smaller size of the calibration sample compared to the $\sqrts = \SI{13}{\TeV}$ measurement.
Tracking efficiencies in the simulation are corrected with a factor derived from data as described in Ref.~\cite{LHCb-DP-2013-002} with typical values in the range of 0.97 to 1.02.

\subsection{Determination of signal yields}
\label{sec:ana_yie}

The data contain a mixture of prompt signal decays, secondary charm mesons 
produced in decays of \bquark hadrons, and combinatorial background. While combinatorial
background can be distinguished from signal decays in the invariant mass distribution of charm meson candidates, both prompt signal decays and secondary
charm mesons have the same mass shape.
However, secondary charm mesons will, on average, have a greater \IP with respect to the closest PV than prompt signal,
and this is exploited by using the spectrum of \lnipchisq of the charm meson candidates, where \ipchisq is defined as the difference in \chisq of the PV fit, performed with and without the particle under consideration.

The prompt signal yield in each \pTy bin is obtained from a fit to the \lnipchisq distribution of the charm meson within a {\SI{20}{\mevcc}} window
around the known mass~\cite{PDG2014} of the \Dz, \Dp, or \Dsp meson, corresponding to approximately 2.5 times the mass resolution. For the \Dstp measurements, an additional signal window around the
nominal $\deltam = m(\Dstp) - m(\Dz)$ value of \SI{145.43}{\mevcc}~\cite{PDG2014} is used, and the fit is made to the \lnipchisq distribution of the \Dz meson. Candidates outside the signal region
are used to construct templates for the combinatorial background shape in the \lnipchisq distributions.
For the \Dz, \Dp, and \Dsp measurements, a fit to the invariant mass distributions is used to constrain the number of
combinatorial background candidates in the \lnipchisq fit. A fit to the \deltam distribution is used for the \Dstp measurements. The fits to the invariant mass, \deltam, and \lnipchisq distributions are performed as extended binned maximum likelihood fits, performed simultaneously across all \pTy bins.
The fit model definitions, and the choice of model parameters that are shared across \pTy bins, are identical to those used in the $\sqrts = \SI{13}{\TeV}$ measurement~\cite{LHCb-PAPER-2015-041}.

\begin{figure}
  \begin{subfigure}[b]{0.5\textwidth}
    \includegraphics[width=\textwidth]{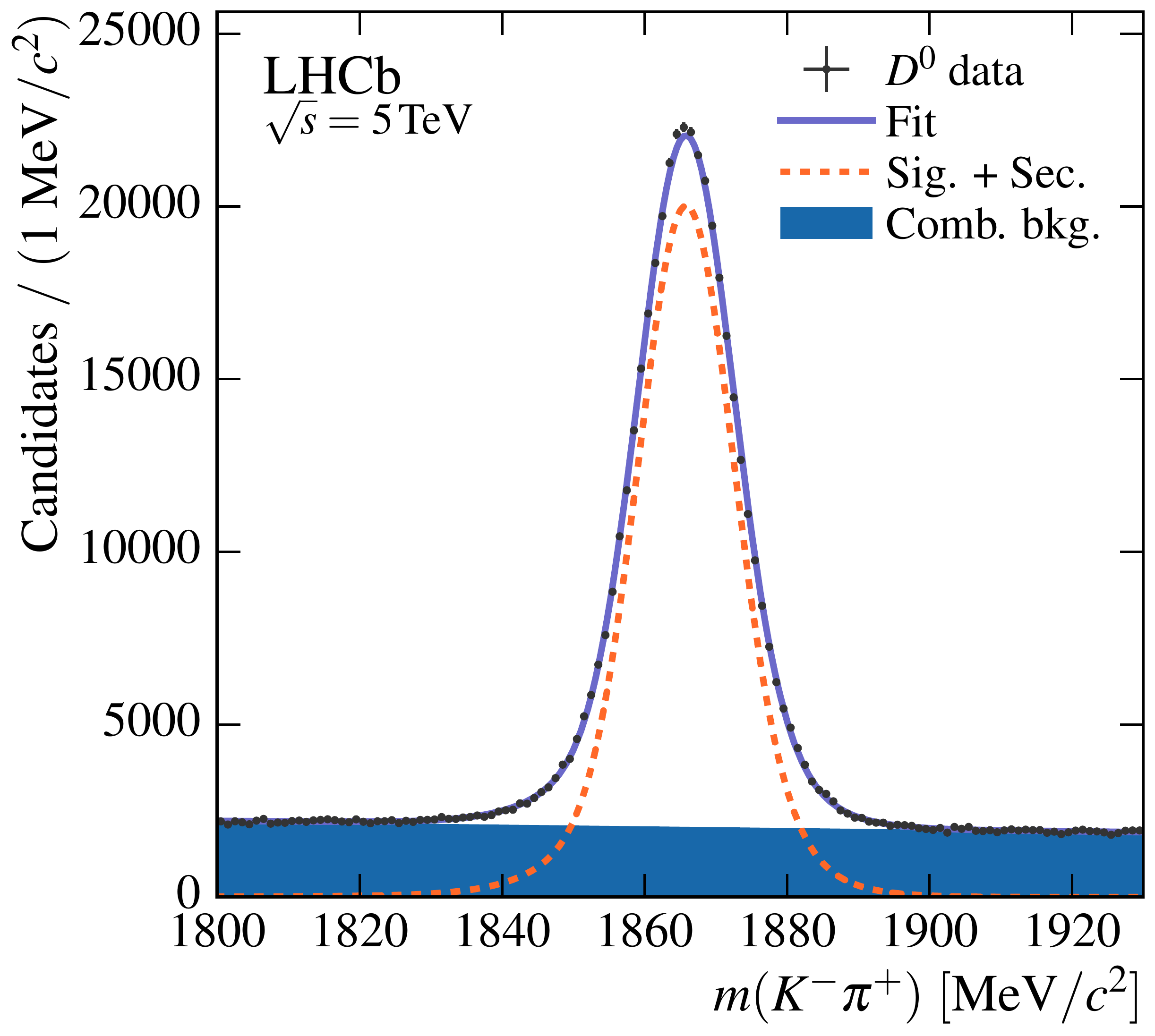}
  \end{subfigure}
  \begin{subfigure}[b]{0.5\textwidth}
    \includegraphics[width=\textwidth]{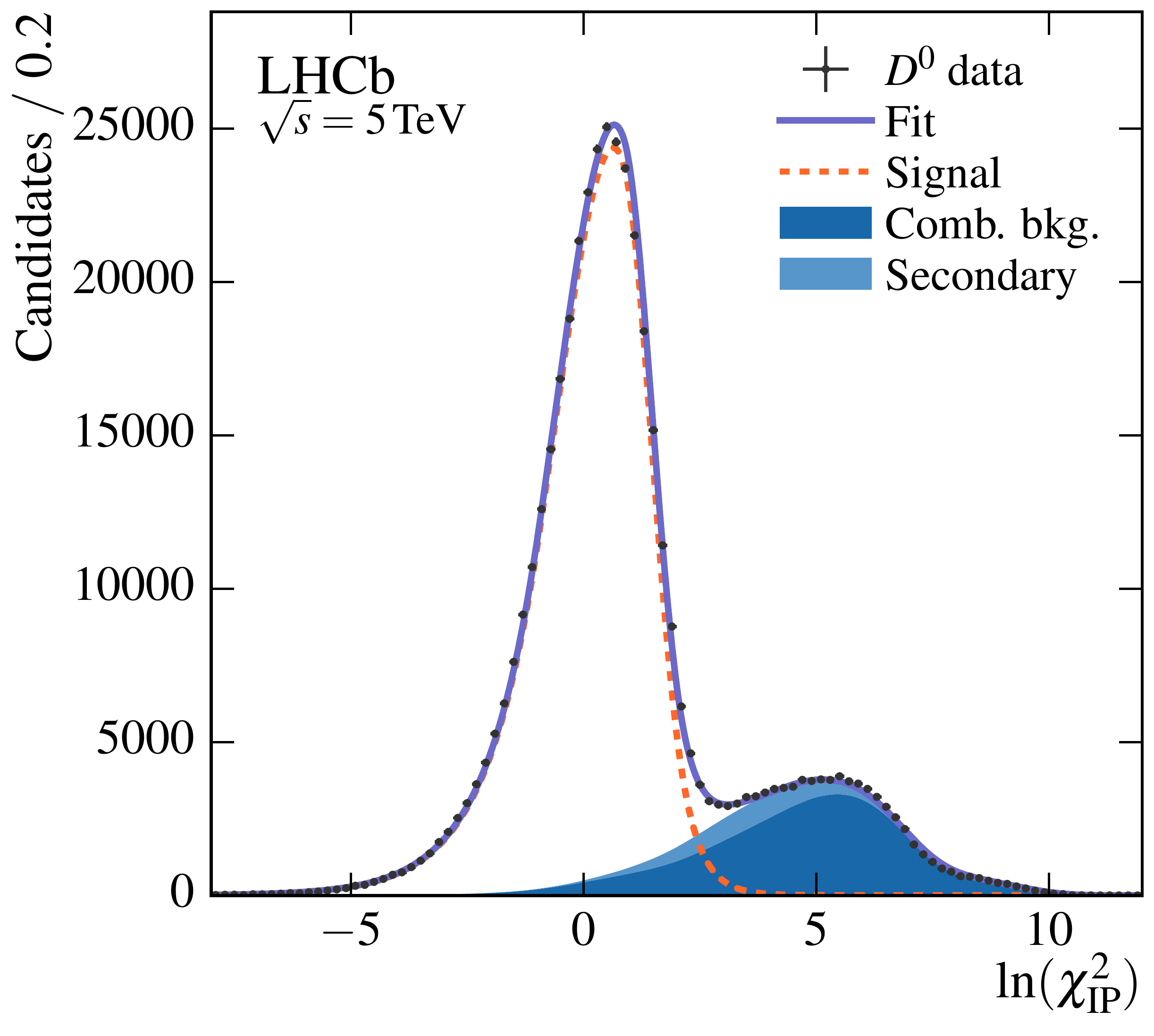}
  \end{subfigure}
  \caption{%
    Distributions for selected \DzToKpi candidates: (left) $\Km\pip$ invariant mass and (right) \lnipchisq for a mass window of $\pm\SI{20}{\mevcc}$ around the nominal \Dz mass.
    The sum of the simultaneous likelihood fits in each 
    \pTy bin is shown, with components as indicated in the legends.
  }
  \label{fig:analysis:fits:D0ToKpi}
\end{figure}

\begin{figure}
  \begin{subfigure}[b]{0.5\textwidth}
    \includegraphics[width=\textwidth]{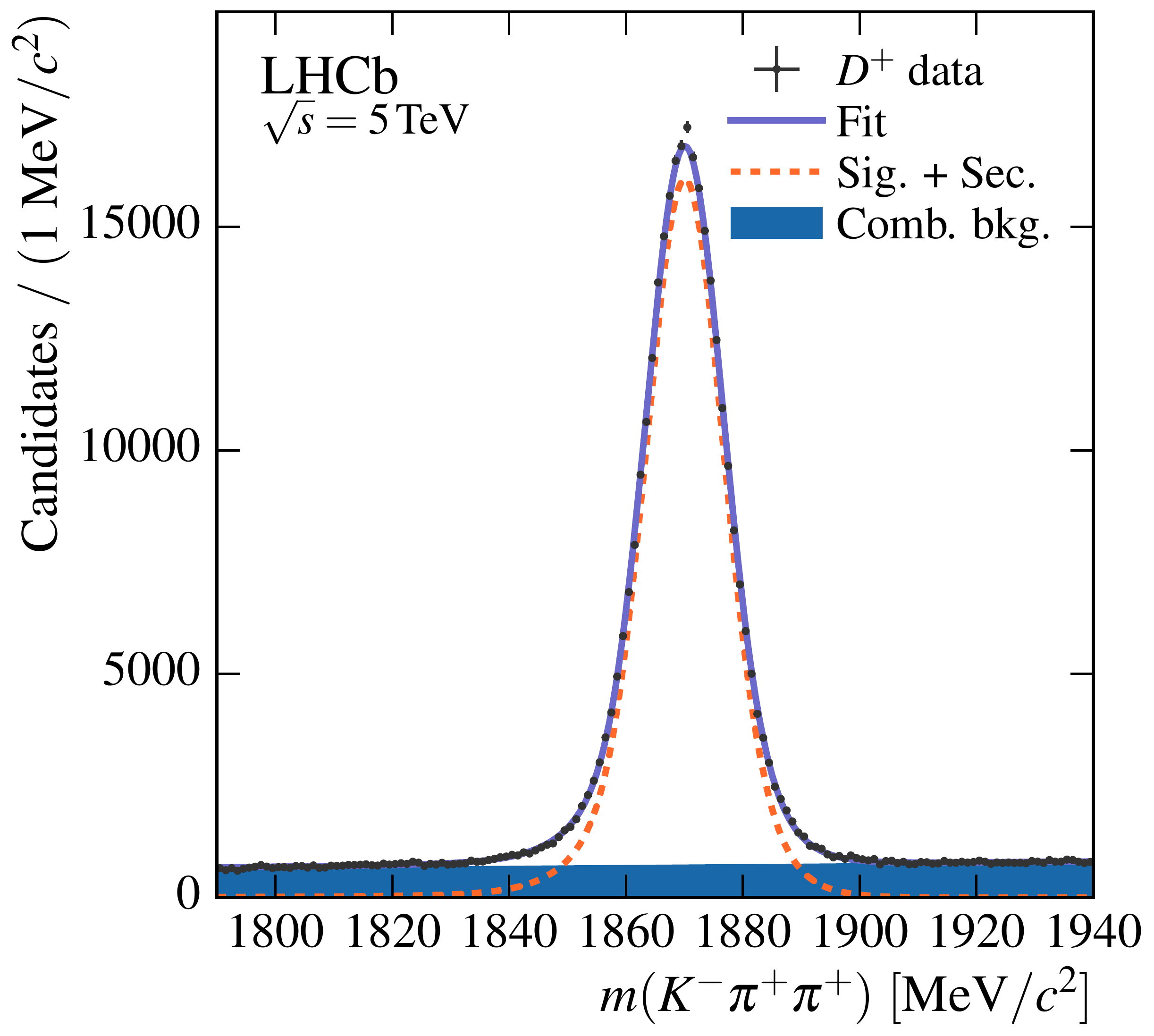}
  \end{subfigure}
  \begin{subfigure}[b]{0.5\textwidth}
    \includegraphics[width=\textwidth]{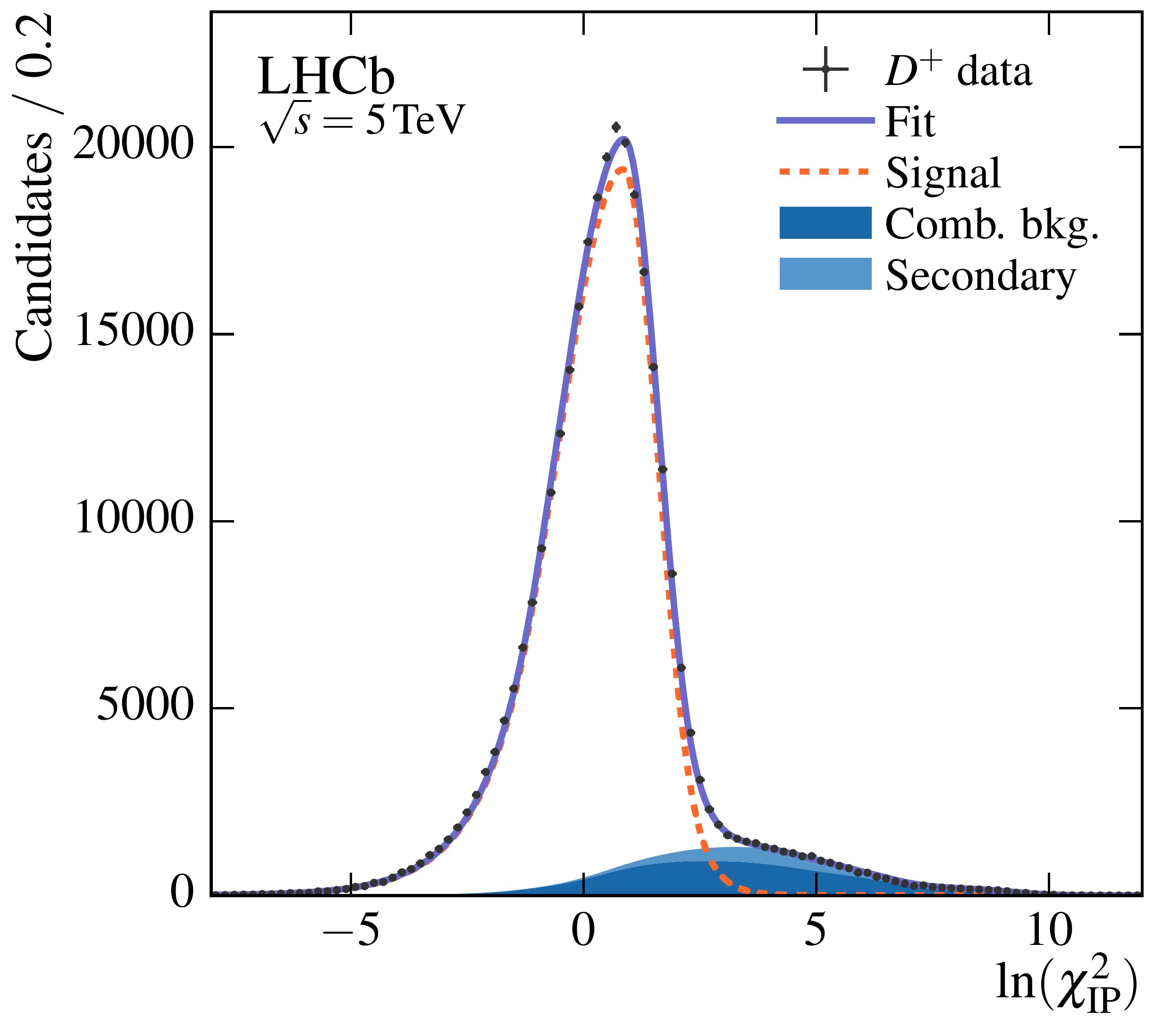}
  \end{subfigure}
  \caption{%
    Distributions for selected \DpToKpipi candidates: (left) $\Km\pip\pip$ invariant mass and (right) \lnipchisq for a mass window of $\pm\SI{20}{\mevcc}$ around the nominal \Dp mass.
    The sum of the simultaneous likelihood fits in each 
    \pTy bin is shown, with components as indicated in the legends.
  }
  \label{fig:analysis:fits:DpToKpipi}
\end{figure}

\begin{figure}
  \begin{subfigure}[b]{0.5\textwidth}
    \includegraphics[width=\textwidth]{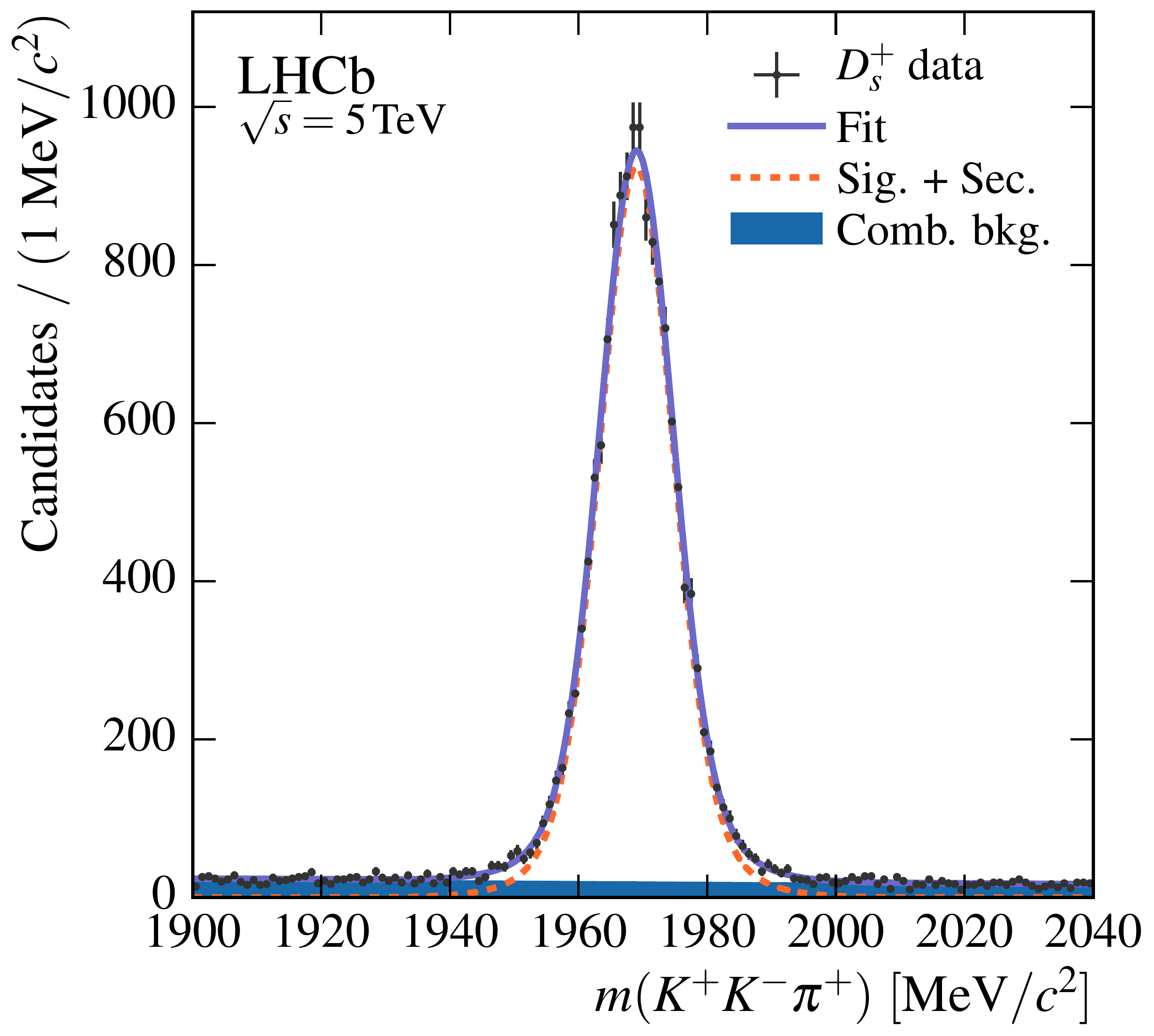}
  \end{subfigure}
  \begin{subfigure}[b]{0.5\textwidth}
    \includegraphics[width=\textwidth]{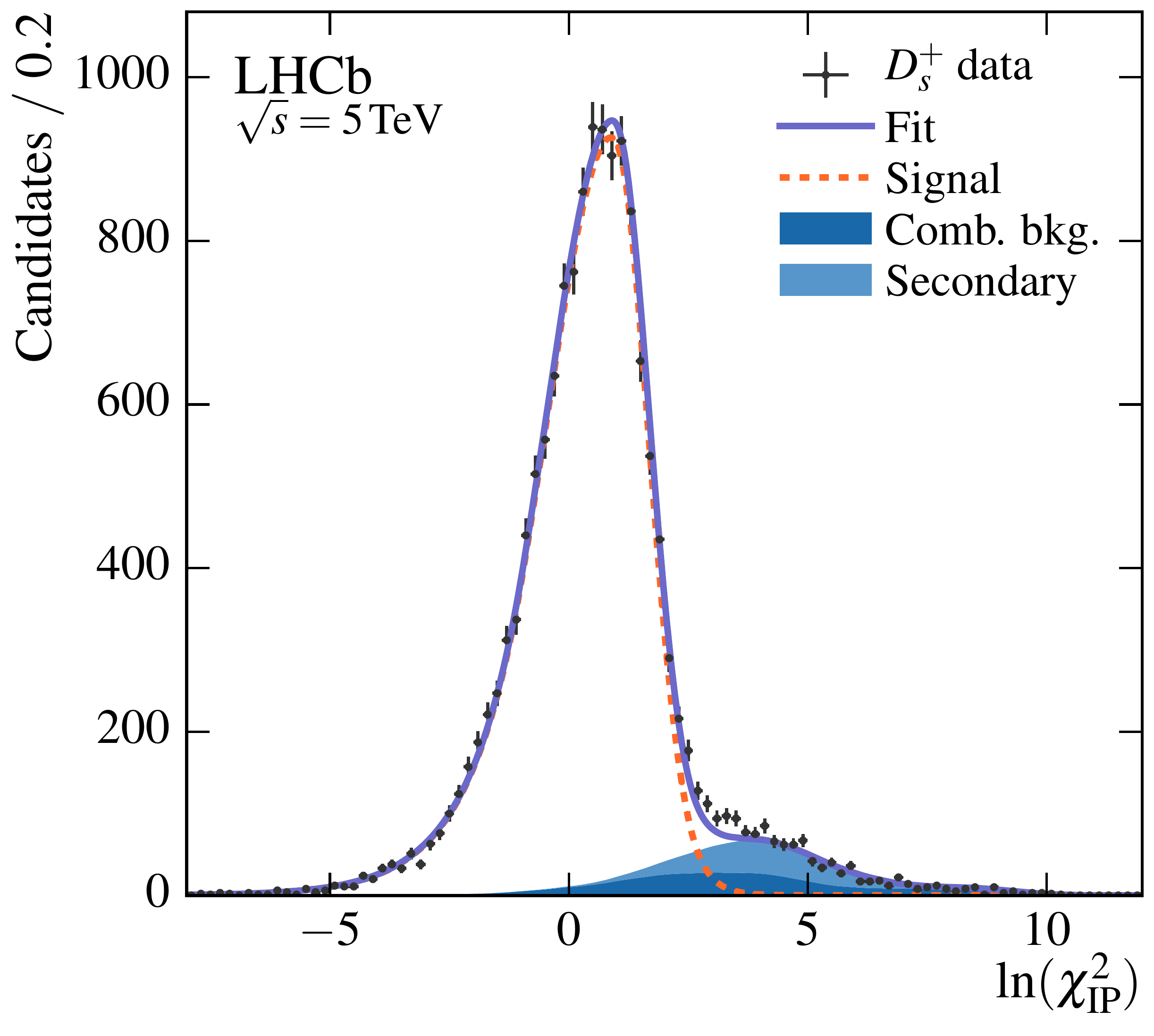}
  \end{subfigure}
  \caption{%
    Distributions for selected \DsTophipi candidates: (left) $\Kp\Km\pip$ invariant mass and (right) \lnipchisq for a mass window of $\pm\SI{20}{\mevcc}$ around the nominal \Dsp mass.
    The sum of the simultaneous likelihood fits in each 
    \pTy bin is shown, with components as indicated in the legends.
  }
  \label{fig:analysis:fits:DsTophipi}
\end{figure}

\begin{figure}
  \begin{subfigure}[b]{0.5\textwidth}
    \includegraphics[width=\textwidth]{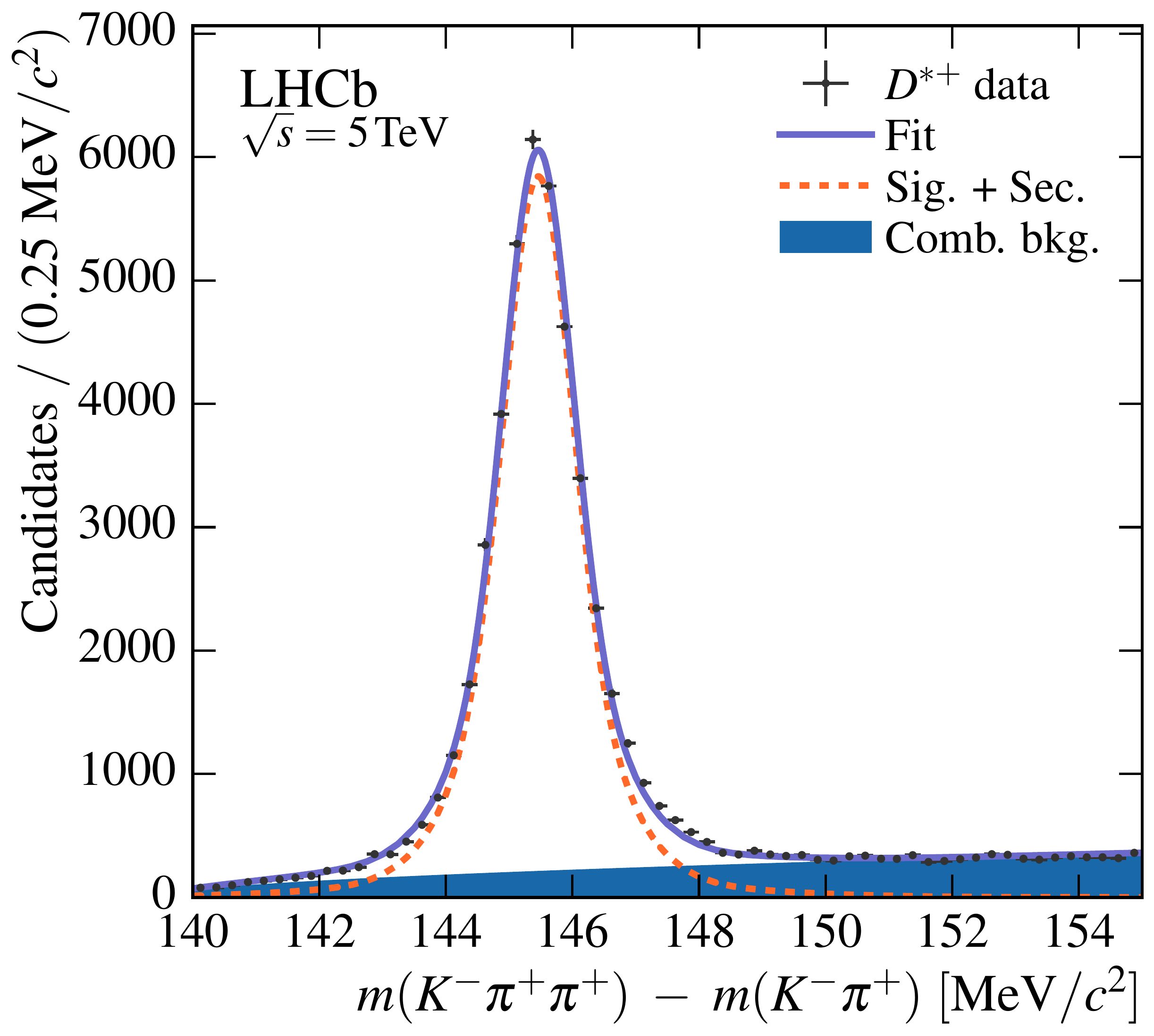}
  \end{subfigure}
  \begin{subfigure}[b]{0.5\textwidth}
    \includegraphics[width=\textwidth]{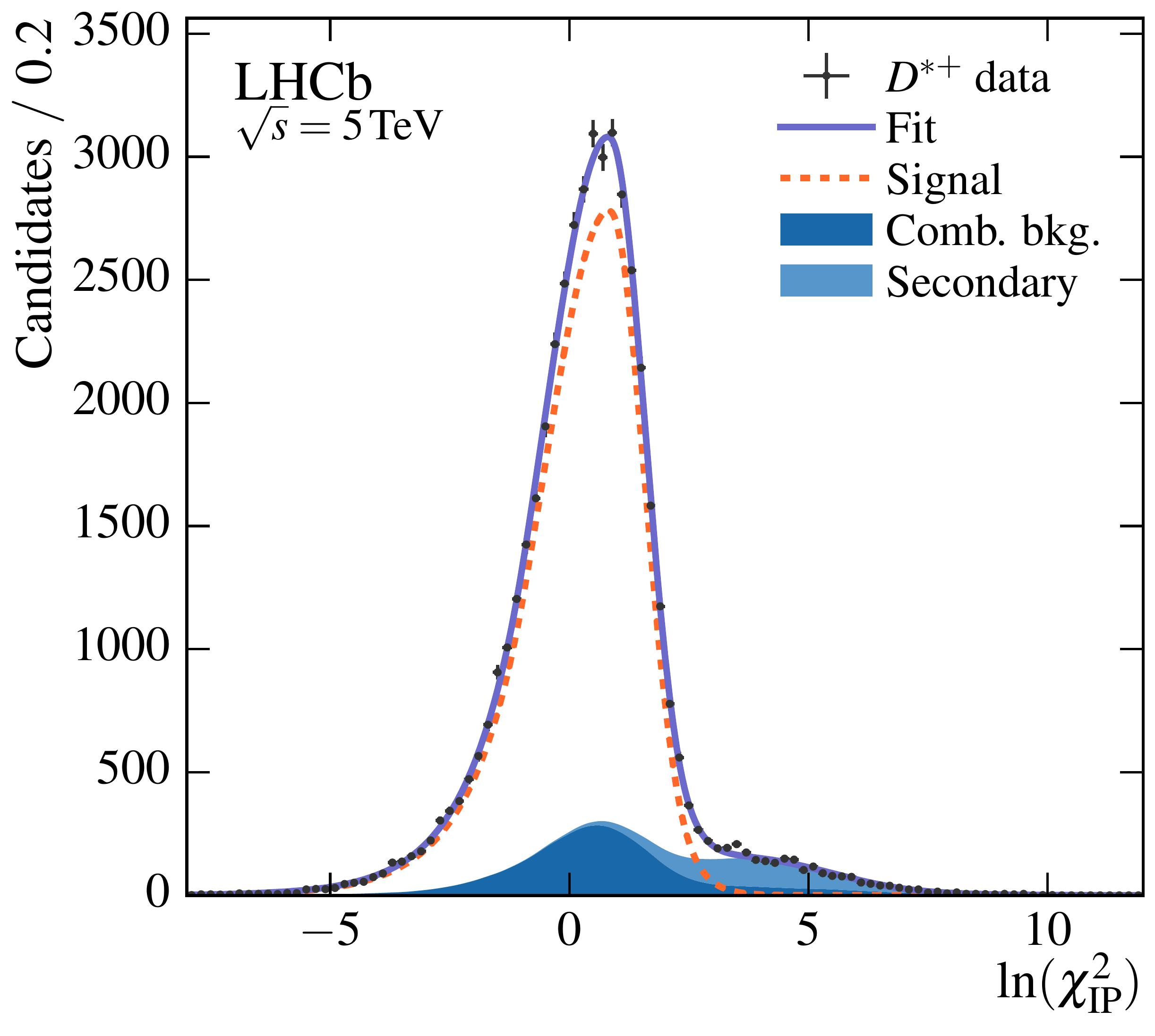}
  \end{subfigure}
  \caption{%
    Distributions for selected \DstToDzpi candidates, with \DzToKpi: (left) 
    \mbox{$\deltam = m(\Dstp) - m(\Dz)$} for a mass window of $\pm\SI{20}{\mevcc}$ 
    around the nominal \Dz mass and (right) \lnipchisq of the \Dz with an additional mass 
    window of $\pm\SI{3}{\mevcc}$ around the nominal \Dstp{\kern 0.2em}-\,\Dz mass difference.
    The sum of the simultaneous likelihood fits in each \pTy bin is shown, with 
    components as indicated in the legends.
  }
  \label{fig:analysis:fits:DstToD0pi_D0ToKpi}
\end{figure}

The sums of the fits over all \pTy bins are given in 
Figs.~\ref{fig:analysis:fits:D0ToKpi}--\ref{fig:analysis:fits:DstToD0pi_D0ToKpi}.
The fits generally describe the data well. Inaccuracies in the description of the data by the fit model are found to have only a small effect on the estimated prompt signal yield and are taken into account as systematic uncertainties.
The sums of the prompt signal yields, as determined by the fits, are given in 
Table~\ref{table:analysis:yields}.

\begin{table}
  \caption{%
    Prompt signal yields in the fully selected dataset, summed over all \pTy 
    bins in which a measurement is made. Only statistical uncertainties are given.
  }
  \label{table:analysis:yields}
  \centering{%
    \begin{tabular}{lr}
      Hadron & Prompt signal yield            \\
      \midrule
      \Dz    & $(34.4 \pm 0.7) \times 10^{4}$ \\
      \Dp    & $(27.6 \pm 0.6) \times 10^{4}$ \\
      \Dsp   & $(13.2 \pm 0.1) \times 10^{3}$ \\
      \Dstp  & $(39.0 \pm 0.2) \times 10^{3}$ \\
    \end{tabular}
  }
\end{table}

\clearpage

\section{Cross-section measurements}
\label{sec:measurements}

In each \pTy bin $i$, the bin averaged differential cross-section for producing the charm meson \Hc is calculated from the relation
\begin{equation}
  \frac{\text{d}^2\sigma_i(\Hc)}{\text{d}\pT\,\text{d}y} =
    \frac{1}{\Delta\pT\Delta\rapidity} \cdot
    \frac{%
      N_i(\decay{\Hc}{f} + \text{c.c.})
    }{%
      \varepsilon_{i,\text{tot}}(\decay{\Hc}{f}) \bfrac(\decay{\Hc}{f}) \,\kappa\,\lum_{\text{int}}
    },
  \label{eq:xsec}
\end{equation}
where $\Delta\pT$ and $\Delta\rapidity$ are the widths in \pT and \rapidity of bin $i$, 
$N_{i}(\decay{\Hc}{f} + \text{c.c.})$ is the measured yield of prompt \Hc decays
to the final state $f$ in bin $i$ from the \lnipchisq fit including the charge-conjugate decay, 
and $\varepsilon_{i,\text{tot}}(\decay{\Hc}{f})$ 
is the total efficiency for observing the signal decay in bin $i$.
The total integrated luminosity collected, $\lum_{\text{int}}$, is \totlumi and $\kappa=1.86\%$ is the efficiency of the
hardware trigger.
The integrated luminosity of the dataset is evaluated from the number of visible $\proton\proton$ collisions and a constant of proportionality that is measured in a dedicated calibration dataset.
The absolute luminosity for the calibration dataset is determined from the beam currents, which are measured by LHC instruments, and the beam profiles and overlap integral, which are measured with a beam-gas imaging method~\cite{LHCb-PAPER-2014-047}. In contrast to Ref.~\cite{LHCb-PAPER-2014-047}, no van der Meer scan is used. The correlation coefficient between the uncertainties of the luminosity measurements at \comenergy and \comenergyold is 32\%.

The values for the branching fractions, taken from Ref.~\cite{PDG2014nu}, are identical to those used in the $\sqrts = \SI{13}{\TeV}$ measurement~\cite{LHCb-PAPER-2015-041}, thus ensuring a complete cancellation in ratios between cross-sections measured at 13 and \SI{5}{\TeV}.
The values \mbox{$\bfrac(\DpToKmpippip)$}, \mbox{$\bfrac(\DstarpTopipDzToKmpip)$},~and \mbox{$\bfrac(\decay{\Dz}{\Kmp\pipm})$} are \mbox{$(9.13 \pm 0.19)\%$}, \mbox{$(2.63 \pm 0.04)\%$},~and \mbox{$(3.89 \pm 0.05)\%$}, respectively.
For the \Dsp measurement the fraction of \DspToKmKppip decays with a $\Km\Kp$ invariant mass in the range $1000<m_{\Km\Kp}<\SI{1040}{\mevcc}$ is taken as $(2.24 \pm 0.13)\%$~\cite{Alexander:2008aa}.

Several sources of systematic uncertainty are identified and evaluated for each decay mode and \pTy bin as described in Ref.~\cite{LHCb-PAPER-2015-041}.
For all decay modes, the dominant uncertainties in most bins are due to the luminosity and the estimation of the tracking efficiencies. The calibration of the tracking efficiencies is performed independently for datasets taken at different centre-of-mass energies, leading to different relative uncertainties compared to the ${\sqrts} = {\SI{13}{\TeV}}$ analysis. The simulated sample gives rise to systematic uncertainties
due to its finite size and imperfect modelling of the selection variables. Uncertainties are also evaluated for the \pid calibration procedure
to account for the finite size of the calibration sample and residual differences of the kinematic distributions between the calibration sample and the final state tracks.
Additionally, an uncertainty is evaluated to account for the choice of fit models used in the determination of the signal yields.
Table~\ref{tab:sys:summary} lists the fractional systematic uncertainties for the different decay modes and their correlations between different \pTy bins and decay modes.

\begin{figure}[htb]
  \centering
  \vspace{1cm}
  \includegraphics[width=\textwidth]{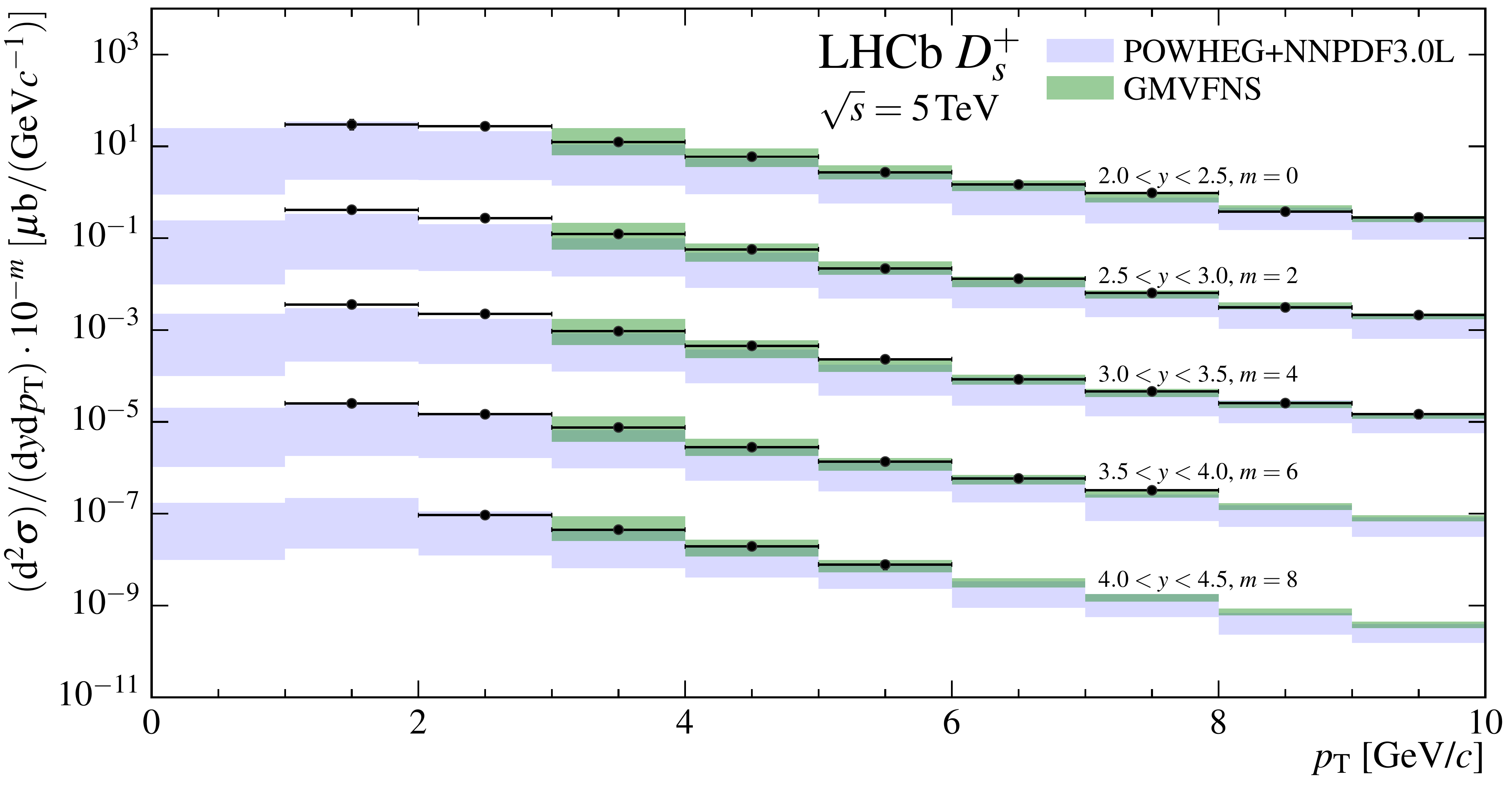}
  \vspace{1cm}
  \includegraphics[width=\textwidth]{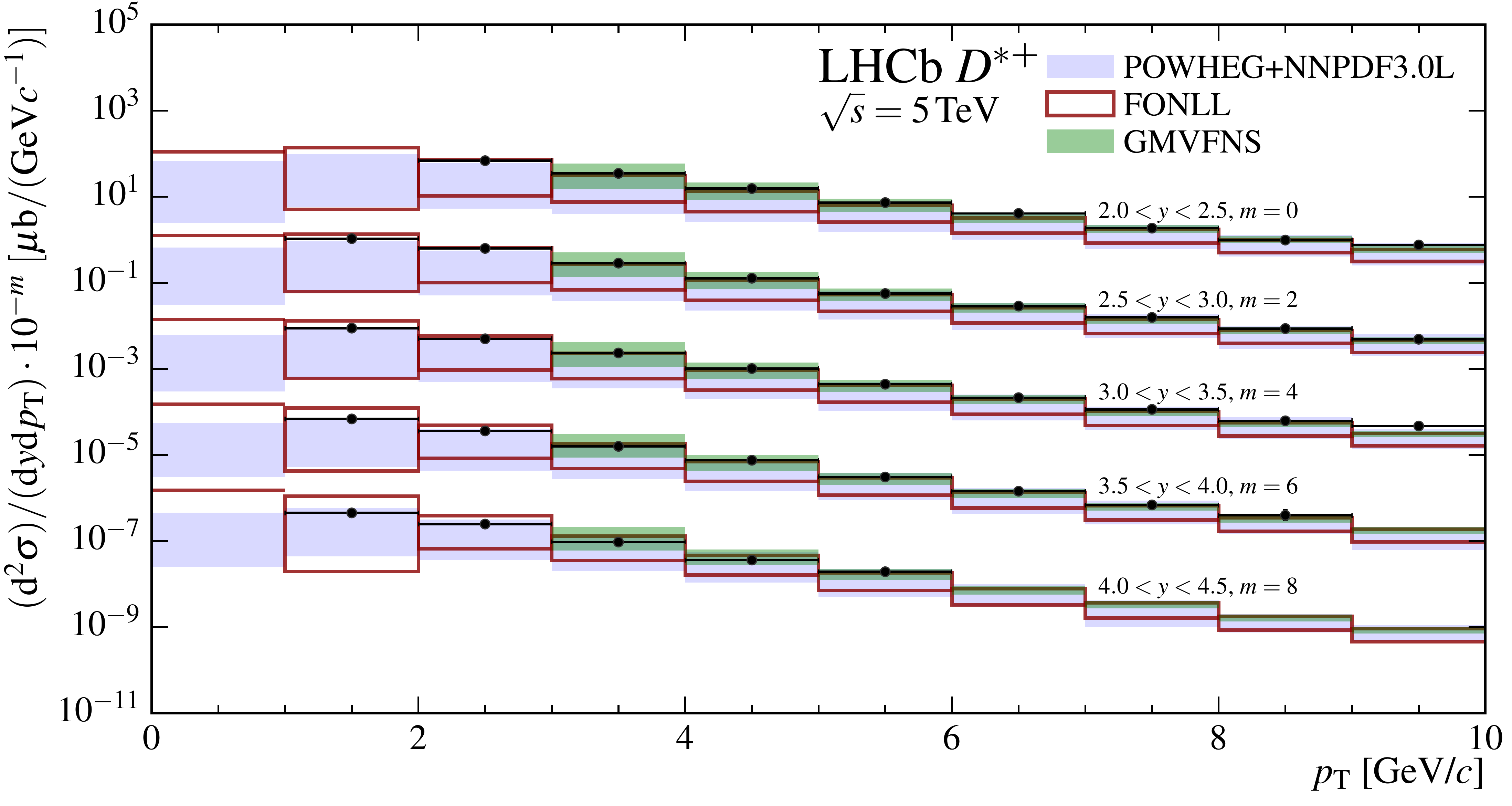}
  \caption{Measurements and predictions for the absolute prompt (top) \Dsp, and (bottom) \Dstarp cross-sections at \comenergy.
           Each set of measurements and predictions in a given rapidity bin is 
           offset by a multiplicative factor $10^{-m}$, where the factor $m$ is 
           shown on the plots.
           The boxes indicate the $\pm1\sigma$ uncertainty band on the theory 
           predictions, where only the upper edge is shown if the uncertainty 
           exceeds two orders of magnitude.
  \label{fig:theory_absolute_Dsp_Dstar}}
\end{figure}

The measured differential cross-sections are tabulated in Appendix~\ref{app:xsec}.
Figures~\ref{fig:theory_absolute_D0_Dp} and~\ref{fig:theory_absolute_Dsp_Dstar} show the \Dz, \Dp, \Dsp, and \Dstarp cross-section 
measurements and predictions~\cite{Gauld:2015yia,Cacciari:2015fta,Kniehl:2012ti}, which are discussed in Sec.~\ref{sec:theory}.
\begin{table}
  \caption[Systematic uncertainties summary]{%
    Fractional systematic uncertainties, in percent.
    Uncertainties that are computed bin-by-bin are 
    expressed as ranges giving the minimum to maximum values.
    Ranges for the correlations between \pT-\rapidity bins and between modes 
    are also given, expressed in percent.
  }
  \label{tab:sys:summary}
  \centering
  \begin{tabular}{lcccccc}
                              & \multicolumn{4}{c}{Uncertainties (\%)} & \multicolumn{2}{c}{Correlations (\%)} \\
                              & \Dz                     & \Dp                 & \Dsp               & \Dstp               & Bins     & Decay modes  \\
    \midrule
    Luminosity                & \multicolumn{4}{c}{3.8}                                                                 & 100      & 100    \\
    Tracking                  & \suDztracking            & \suDptracking      & \suDsptracking     & \suDstptracking     & 90--100   & 90--100 \\
    Branching fractions       & 1.2                      & 2.1                & 5.8                & 1.5                 & 100      & 0--95  \\
    Simulation sample size    & \suDzmcstat              & \suDpmcstat        & \suDspmcstat       & \suDstpmcstat       & 0      & 0    \\
    Simulation modelling      & \suDzmcagreement         & \suDpmcagreement   & \suDspmcagreement  & \suDstpmcagreement  & 0      & 0    \\
    \pid sample size          & \suDzpidstat             & \suDppidstat       & \suDsppidstat      & \suDstppidstat      & 0--100    & 0--100    \\
    \pid binning              & \suDzpidbin              & \suDppidbin        & \suDsppidbin       & \suDstppidbin       & 0      & 0    \\
    Fit model shapes          & \suDzfit                 & \suDpfit           & \suDspfit          & \suDstpfit          & 0      & 0    \\

  \end{tabular}
\end{table}

\clearpage
\begin{figure}[htb]
  \centering
  \vspace{1cm}
  \includegraphics[width=\textwidth]{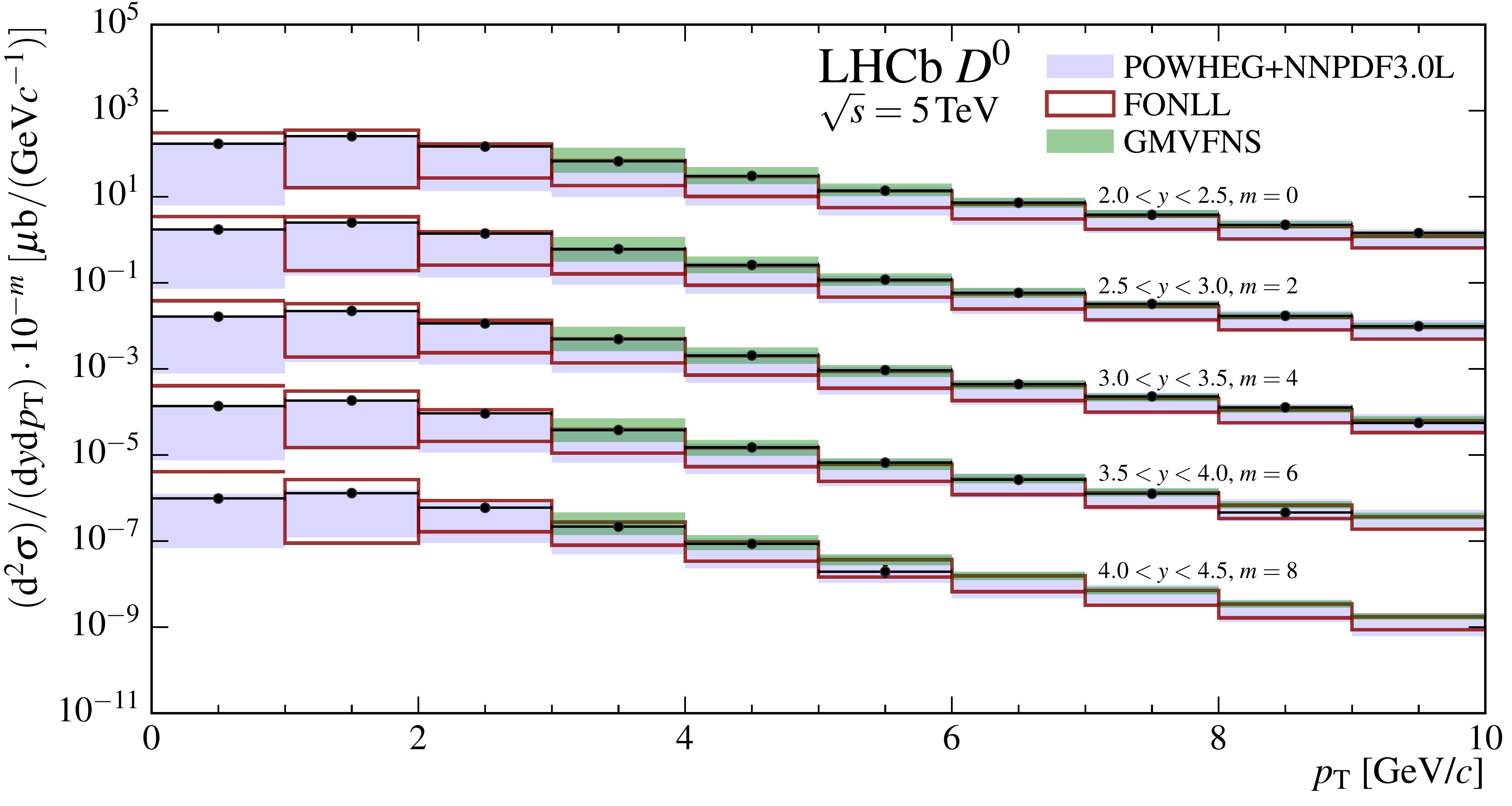}
  \vspace{1cm}
  \includegraphics[width=\textwidth]{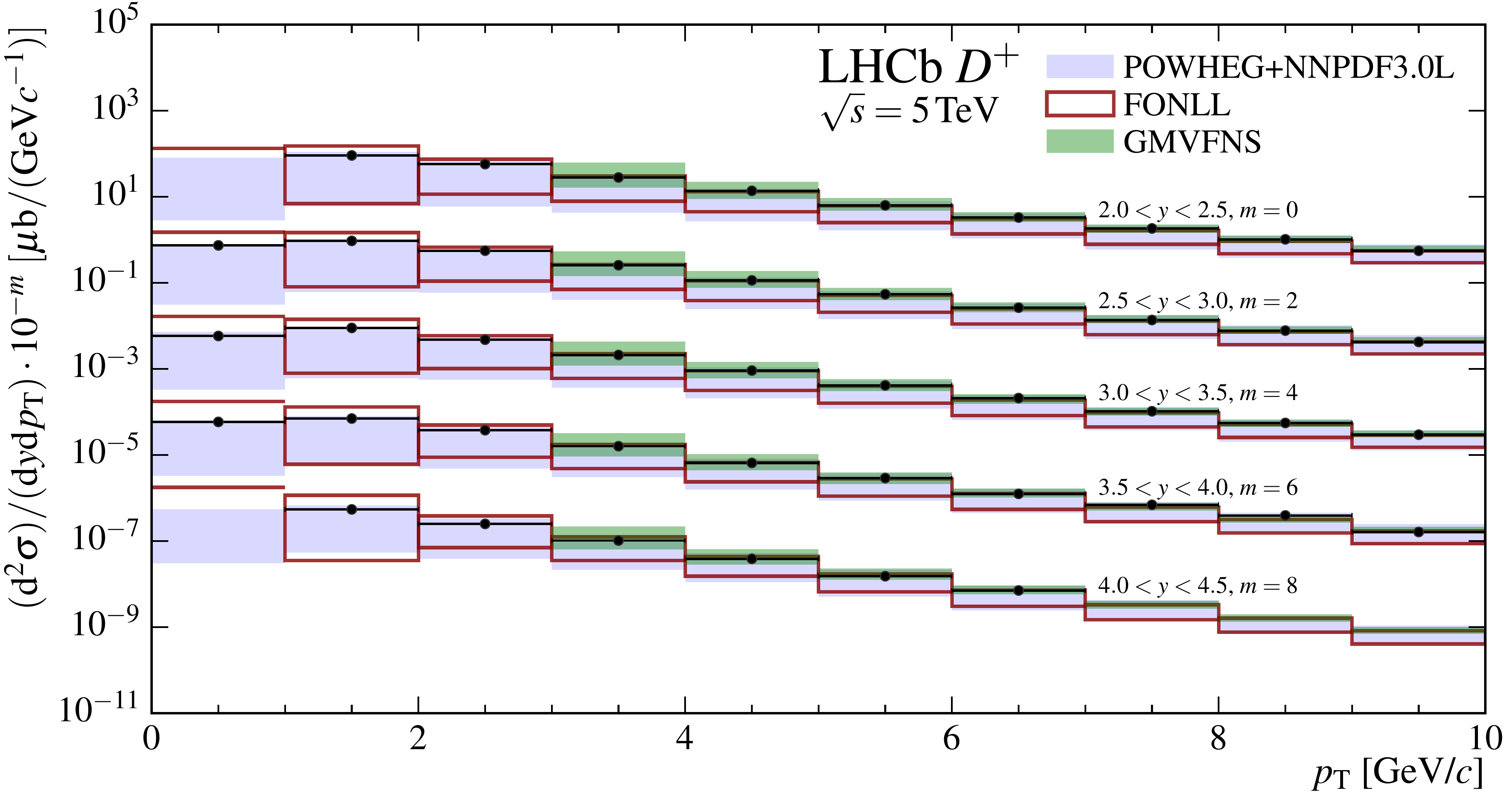}
  \caption{Measurements and predictions for the absolute prompt (top) \Dz, and (bottom) \Dp cross-sections at \comenergy.
           Each set of measurements and predictions in a given rapidity bin is 
           offset by a multiplicative factor $10^{-m}$, where the factor $m$ is 
           shown on the plots.
           The boxes indicate the $\pm1\sigma$ uncertainty band on the theory 
           predictions, where only the upper edge is shown if the uncertainty 
           exceeds two orders of magnitude.
  \label{fig:theory_absolute_D0_Dp}}
\end{figure}

\clearpage

\section{Production ratios and integrated cross-sections}
\label{sec:ratios}
\subsection{Production ratios}

The present analysis uses the same techniques as the \lhcb measurement at \comenergyold. Therefore,
several sources of systematic uncertainty are highly correlated, leading to 
cross-section ratios between $\sqrts=\SI{13}{\TeV}$ and $\sqrts=\SI{5}{\TeV}$, $R_{13/5}$, with greatly reduced relative uncertainties compared to the differential cross-sections.
Furthermore, the predicted ratios of prompt charm production cross-sections between different centre-of-mass energies have cancellations of several theoretical
uncertainties~\cite{Gauld:2015yia,Cacciari:2015fta,Kniehl:2012ti}.
The $\sqrts=\SI{13}{\TeV}$ measurements are rebinned to match the binning used for the present results and the 
production ratios are presented for  $0<\pT<\SI{8}{\gevc}$ and $2.0 < y < 4.5$ in Appendix~\ref{app:ratios}.
Figure~\ref{fig:TheoryPredRatio} shows
the measured ratios for \Dz, \Dp, \Dsp, and \Dstp mesons compared with predictions from theory calculations~\cite{Gauld:2015yia,Cacciari:2015fta,Kniehl:2012ti}, discussed in Sec.~\ref{sec:theory}.

A second set of differential ratios is obtained by dividing the prompt charm production cross-sections of different charm mesons. These can be compared with the ratios of the cross-sections measured at $\ep\en$\/ colliders operating at 
a centre-of-mass energy close to the $\Upsilon(4S)$ resonance~\cite{Artuso:2004pj,Seuster:2005tr,Aubert:2002ue}. Those measurements were performed with the same final states as the analysis presented here. Therefore, a more precise comparison is made by taking ratios of \xsectimesbfrac. Differential ratios are shown in Figs.~\ref{fig:MesonRatio}~and~\ref{fig:MesonRatio2}, and tabulated 
results are presented in Appendix~\ref{app:MesonRatio}.
The measurements in each \pT bin are shown integrated over \rapidity for clearer visualisation.
They exhibit a \pT dependence that is consistent with heavier particles 
having a harder \pT spectrum.

\subsection{Integrated cross-sections}

Integrated production cross-sections, $\sigma(D)$, for each charm meson are computed as the 
sum of the measurements in each bin, where the uncertainty on the sum takes into 
account the correlations between bins.
Integrated cross-sections are computed for all four mesons in the kinematic region $1 < \pT < \SI{8}{\gevc}$ and $2.0 < \rapidity < 4.5$
and additionally down to $\pT=\SI{0}{\gevc}$ for \Dz and \Dp.
The upper limit of $\pT=\SI{8}{\gevc}$ is chosen to match that of the measurements at $\sqrt{s}=\SI{7}{\TeV}$ and \comenergyold.

Contributions from bins within the integration range for which a measurement 
was not possible are estimated using a theory-based correction factor,
computed as the ratio between the predicted integrated cross-section within the 
considered kinematic region and the sum of all cross-section 
predictions for bins for which a measurement exists, as in 
Ref.~\cite{LHCb-PAPER-2015-041}. \Rhorry \cite{Gauld:2015yia} predictions, discussed in
the next section, are used to compute the extrapolation factor. 
From the differences between the central value and the upper and 
lower bounds on the prediction, the larger of the two is assigned
as systematic uncertainty of the extrapolation factor.
Table~\ref{table:ratios:integrated} gives the integrated cross-sections for 
\Dz, \Dp, \Dsp, and \Dstp mesons and Table~\ref{table:ratios:yearintegrated}
gives the corresponding values for the ratios of integrated cross-sections
measured at \comenergyold and \SI{5}{\TeV}.

The integrated \ccbar production cross-section, $\sigma(\ppToccbarX)$, is calculated as 
$\sigma(D)/(2f(\decay{\cquark}{D}))$ for each decay mode.
The term $f(\decay{\cquark}{D})$ is the quark-to-hadron transition 
probability, and the factor 2 accounts for the inclusion of charge-conjugate 
states in the measurement.
Measurements at $\ep\en$ 
colliders operating at a centre-of-mass energy close to the $\Upsilon(4S)$ 
resonance are used to determine the transition probabilities~\cite{Amsler:2008zzb:Frag}
\mbox{$f(\decay{\cquark}{\Dz}) = 0.565 \pm 0.032$},
\mbox{$f(\decay{\cquark}{\Dp}) = 0.246 \pm 0.020$},
\mbox{$f(\decay{\cquark}{\Dsp}) = 0.080 \pm 0.017$}, and
\mbox{$f(\decay{\cquark}{\Dstarp}) = 0.224 \pm 0.028$}.
The fragmentation fraction $f(\decay{\cquark}{\Dz})$ has an overlapping 
contribution
from $f(\decay{\cquark}{\Dstarp})$.

\begin{figure}[tbh]
  \centering
  \includegraphics[width=0.49\textwidth]{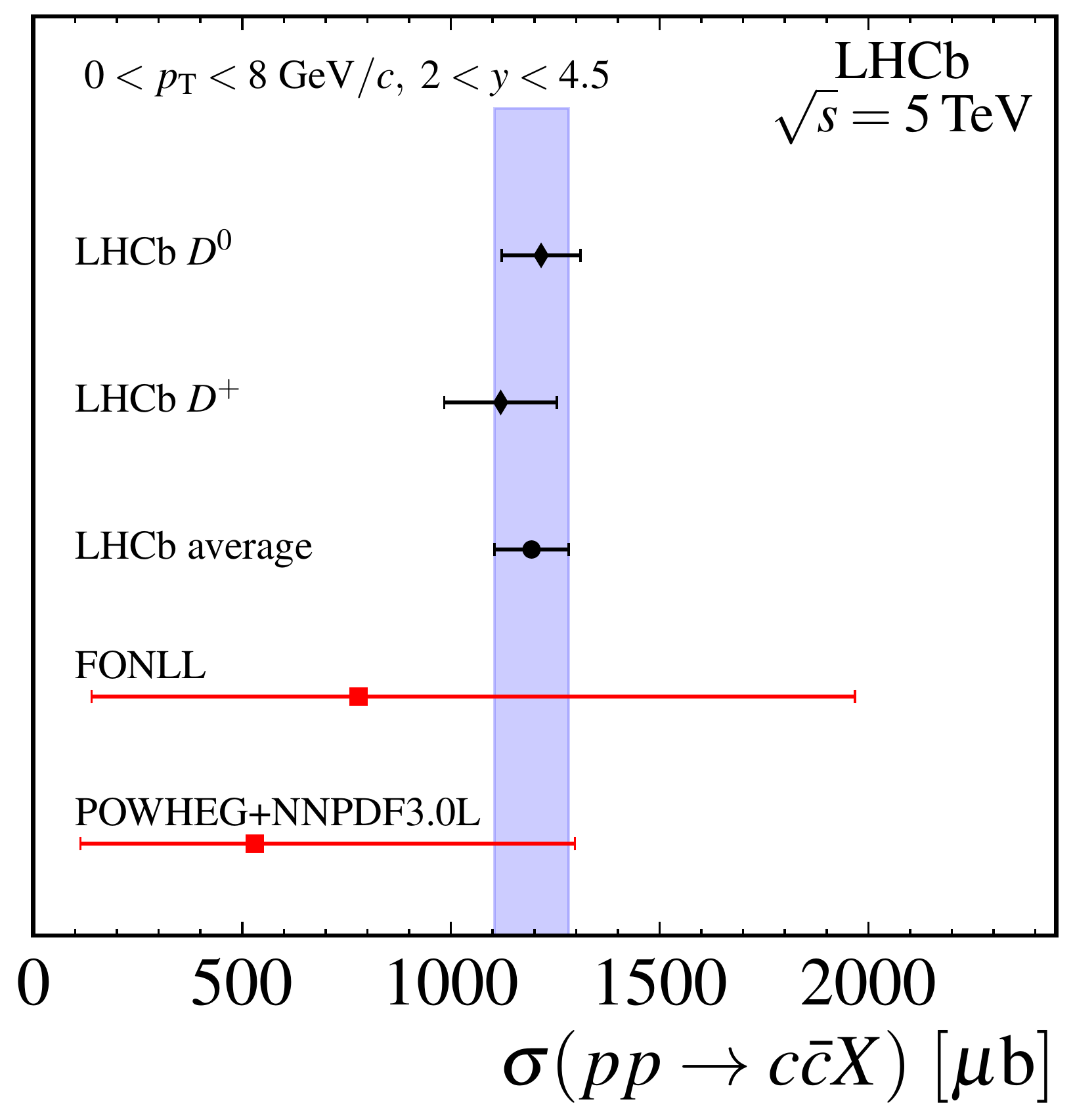}
  \includegraphics[width=0.49\textwidth]{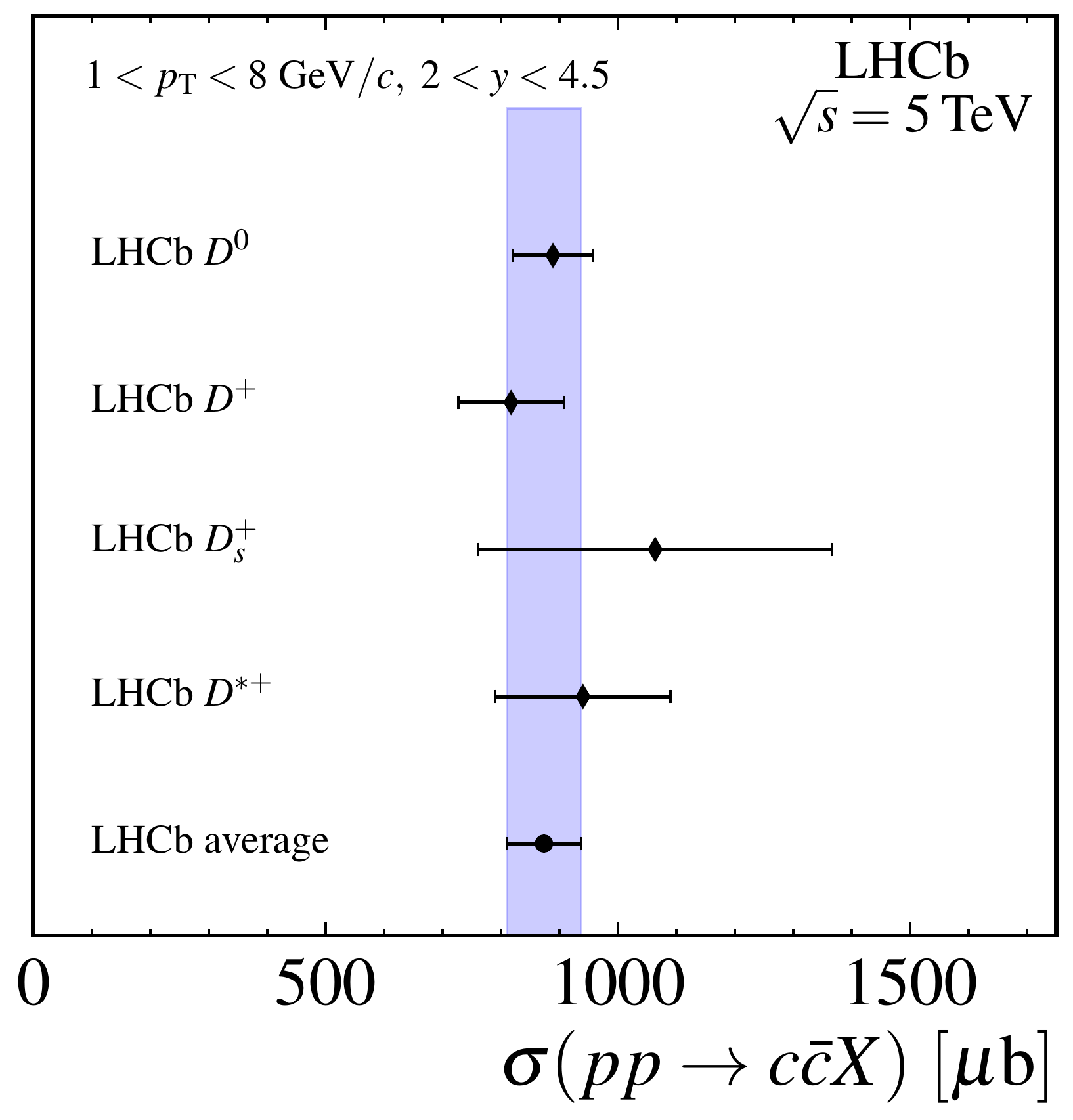}
    \caption{Integrated \ccbar cross-sections (black diamonds), their average (black circle and blue band) and theory predictions (red squares)~\cite{Gauld:2015yia,Cacciari:2015fta} are shown (left) based on the \Dz and \Dp for $0 < \pT < \SI{8}{\gevc}$ and (right) for measurements based on all four mesons for $1 < \pT < \SI{8}{\gevc}$. 
  \label{fig:TotalCrossSec}}
\end{figure}

The combination of the individual \Dz and \Dp measurements of the \ccbar 
cross-section is performed by means of a weighted average that minimises
the variance of the estimate~\cite{Lyons:1988rp}, giving
\begin{equation*}
  \sigma{(\ppToccbarX)}_{\pT\,<\,8\,\si{\gevc},\,2.0\,<\,y\,<\,4.5} = \ccbarXsec,
\end{equation*}
where the uncertainties are due to statistical, systematic and fragmentation 
fraction uncertainties, respectively. The specified kinematic range refers to the produced charm hadron, not the individual
charm quarks.
A comparison with predictions is given in~Fig.~\ref{fig:TotalCrossSec}.
The same Figure also shows a comparison of $\sigma(\ppToccbarX)$ for $1 < \pT < \SI{8}{\gevc}$ based on the measurements of the four meson species.
Ratios of the integrated measurements of cross-section times branching fraction are 
given in Table~\ref{table:ratios:integrated_ratios}.

\begin{table}[tb]
  \caption[Integrated charm hadron cross-sections]{%
    Prompt {\PD}-meson production cross-sections in the kinematic ranges given.
    The computation of the extrapolation factors is described in Ref~\cite{LHCb-PAPER-2015-041}.
    The first uncertainty on the cross-section is statistical, and the second 
    is systematic and includes the contribution from the extrapolation factor.
    A dash indicates that measurements are available in all bins and no extrapolation factor is needed. 
    Integrated numbers in the reduced acceptance $2.5 < \rapidity < 4.0$ are quoted as
    reference for future heavy ion measurements.
  }
  \label{table:ratios:integrated}
  \begin{adjustbox}{center}
\begin{tabular}{ccccr}
&&& Extrapolation factor & Cross-section~(\si{\micro\barn}) \\
\midrule
\Dz   & $0 < \pT < \SI{8}{\GeVc}$ & $2 < \rapidity < 4.5$ & $1.0013 \pm 0.0019$  & $1374 \pm \phantom{1}3  \pm \phantom{1}74$\\
\Dp   & $0 < \pT < \SI{8}{\GeVc}$ & $2 < \rapidity < 4.5$ & $1.108 \pm 0.049$  & $551 \pm \phantom{1}5  \pm \phantom{1}48$\\
\midrule
\Dz   & $1 < \pT < \SI{8}{\GeVc}$ & $2 < \rapidity < 4.5$ & $1.0017 \pm 0.0020$  & $1004 \pm \phantom{1}3  \pm \phantom{1}54$\\
\Dp   & $1 < \pT < \SI{8}{\GeVc}$ & $2 < \rapidity < 4.5$ & $1.00062 \pm 0.00099$  & $402 \pm \phantom{1}2  \pm \phantom{1}30$\\
\Dsp   & $1 < \pT < \SI{8}{\GeVc}$ & $2 < \rapidity < 4.5$ & $1.0734 \pm 0.0080$  & $170 \pm \phantom{1}4  \pm \phantom{1}16$\\
\Dstp   & $1 < \pT < \SI{8}{\GeVc}$ & $2 < \rapidity < 4.5$ & $1.122 \pm 0.046$  & $421 \pm \phantom{1}5  \pm \phantom{1}36$\\
\midrule
\Dz   & $0 < \pT < \SI{8}{\GeVc}$ & $2.5 < \rapidity < 4$ & ---  & $866 \pm \phantom{1}2  \pm \phantom{1}45$\\
\Dp   & $0 < \pT < \SI{8}{\GeVc}$ & $2.5 < \rapidity < 4$ & ---  & $349 \pm \phantom{1}4  \pm \phantom{1}27$\\
\midrule
\Dz   & $1 < \pT < \SI{8}{\GeVc}$ & $2.5 < \rapidity < 4$ & ---  & $630 \pm \phantom{1}2  \pm \phantom{1}33$\\
\Dp   & $1 < \pT < \SI{8}{\GeVc}$ & $2.5 < \rapidity < 4$ & ---  & $253 \pm \phantom{1}1  \pm \phantom{1}18$\\
\Dsp   & $1 < \pT < \SI{8}{\GeVc}$ & $2.5 < \rapidity < 4$ & ---  & $110 \pm \phantom{1}2  \pm \phantom{1}10$\\
\Dstp   & $1 < \pT < \SI{8}{\GeVc}$ & $2.5 < \rapidity < 4$ & ---  & $267 \pm \phantom{1}3  \pm \phantom{1}21$\\
\end{tabular}
\end{adjustbox}

\end{table}

\begin{table}[tb]
  \caption[Integrated charm hadron cross-sections]{%
    Ratios of integrated prompt {\PD}-meson production cross-sections between measurements at \comenergyold and \comenergy.
    The first uncertainty on the ratio is statistical, and the second is systematic. 
  }
  \label{table:ratios:yearintegrated}
  \begin{adjustbox}{center}
\begin{tabular}{ccccr}
&& & $R_{13/5}$ \\
\midrule
\Dz   & $0 < \pT < \SI{8}{\GeVc}$ & $2 < \rapidity < 4.5$   & $1.977 \pm 0.005  \pm 0.120$\\
\Dp   & $0 < \pT < \SI{8}{\GeVc}$ & $2 < \rapidity < 4.5$   & $\phantom{1}2.02 \pm \phantom{1}0.02  \pm  \phantom{1}0.22$\\
\midrule
\Dz   & $1 < \pT < \SI{8}{\GeVc}$ & $2 < \rapidity < 4.5$   & $2.070 \pm 0.006  \pm 0.121$\\
\Dp   & $1 < \pT < \SI{8}{\GeVc}$ & $2 < \rapidity < 4.5$   & $\phantom{1}2.09 \pm  \phantom{1}0.01  \pm  \phantom{1}0.19$\\
\Dsp   & $1 < \pT < \SI{8}{\GeVc}$ & $2 < \rapidity < 4.5$   & $\phantom{1}2.09 \pm  \phantom{1}0.07  \pm  \phantom{1}0.23$\\
\Dstp   & $1 < \pT < \SI{8}{\GeVc}$ & $2 < \rapidity < 4.5$   & $\phantom{1}1.87 \pm  \phantom{1}0.03  \pm  \phantom{1}0.23$\\
\midrule
\end{tabular}
\end{adjustbox}

\end{table}

\begin{table}[tb]
  \renewcommand{\arraystretch}{1.3}
  \caption[Ratios of the measurements of cross-section times branching fraction]{%
    Ratios of the measurements of cross-section times branching fraction in 
    the kinematic range $1 < \pT < \SI{8}{\gevc}$ and $2 < \rapidity < 4.5$. 
    The first uncertainty on the ratio is statistical and the second 
    is systematic. The notation $\sigma(\decay{D}{f})$ is shorthand for \xsectimesbfrac.
  }
  \label{table:ratios:integrated_ratios}
  \begin{adjustbox}{center}
  \begin{tabular}{lc}
    Quantity                                & Measurement \\
    \midrule
    $\sigma(\DpToKpipi)/\sigma(\DzToKpi)$                                                       & \DpDzint    \\
    $\sigma(\Dsp \to [K^{-}K^{+}]_{\phi}\pi^{+})/\sigma(\DzToKpi)$                              & \DspDzint    \\
    $\sigma(\Dstp \to [K^{-}\pi^{+}]_{\Dz}\pi^{+})/\sigma(\DzToKpi)$                            & \DstpDzint   \\
    \midrule                                                                                      
    $\sigma(\Dsp \to [K^{-}K^{+}]_{\phi}\pi^{+})/\sigma(\DpToKpipi)$                            & \DspDpint    \\
    $\sigma(\Dstp \to [K^{-}\pi^{+}]_{\Dz}\pi^{+})/\sigma(\DpToKpipi)$                          & \DstpDpint   \\
    \midrule                                                                                      
    $\sigma(\Dsp \to [K^{-}K^{+}]_{\phi}\pi^{+})/\sigma(\Dstp \to [K^{-}\pi^{+}]_{\Dz}\pi^{+})$ & \DspDstpint   \\
  \end{tabular}
  \end{adjustbox}
\end{table}

\begin{figure}[p]
  \centering
  \vspace{0.5cm}
  \includegraphics[width=0.49\textwidth]{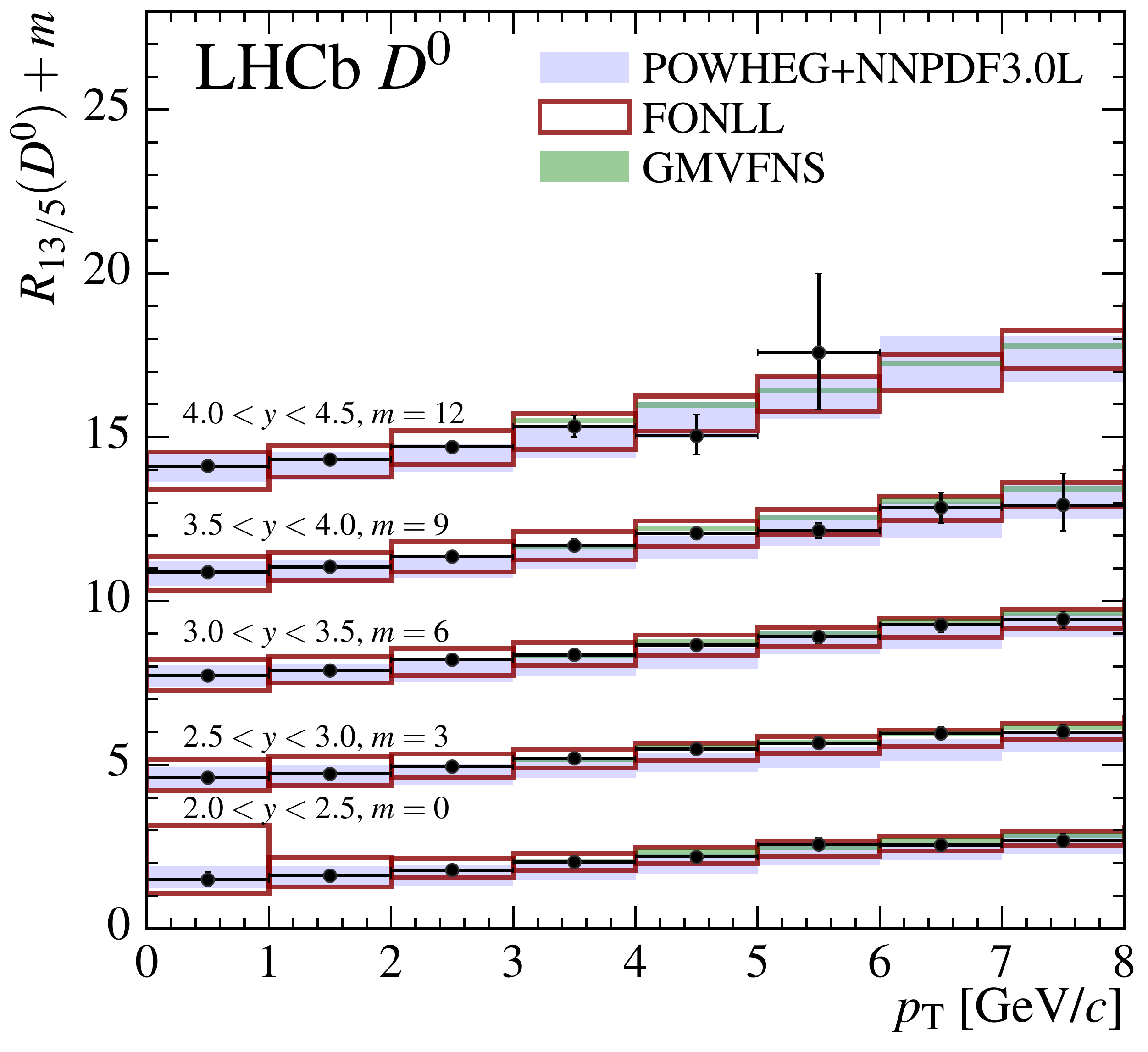}
  \includegraphics[width=0.49\textwidth]{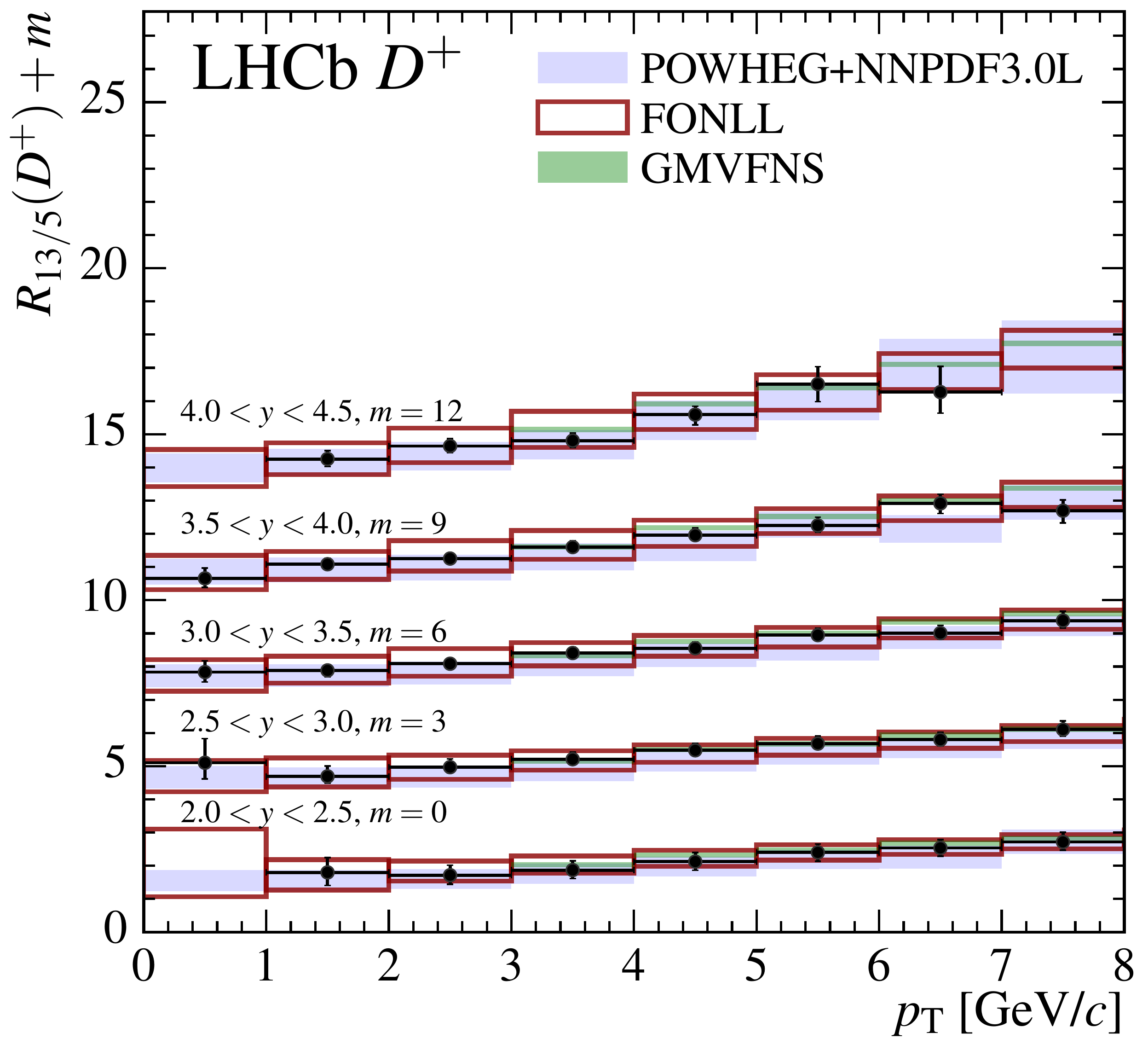}\\
  \vspace{0.5cm}
  \includegraphics[width=0.49\textwidth]{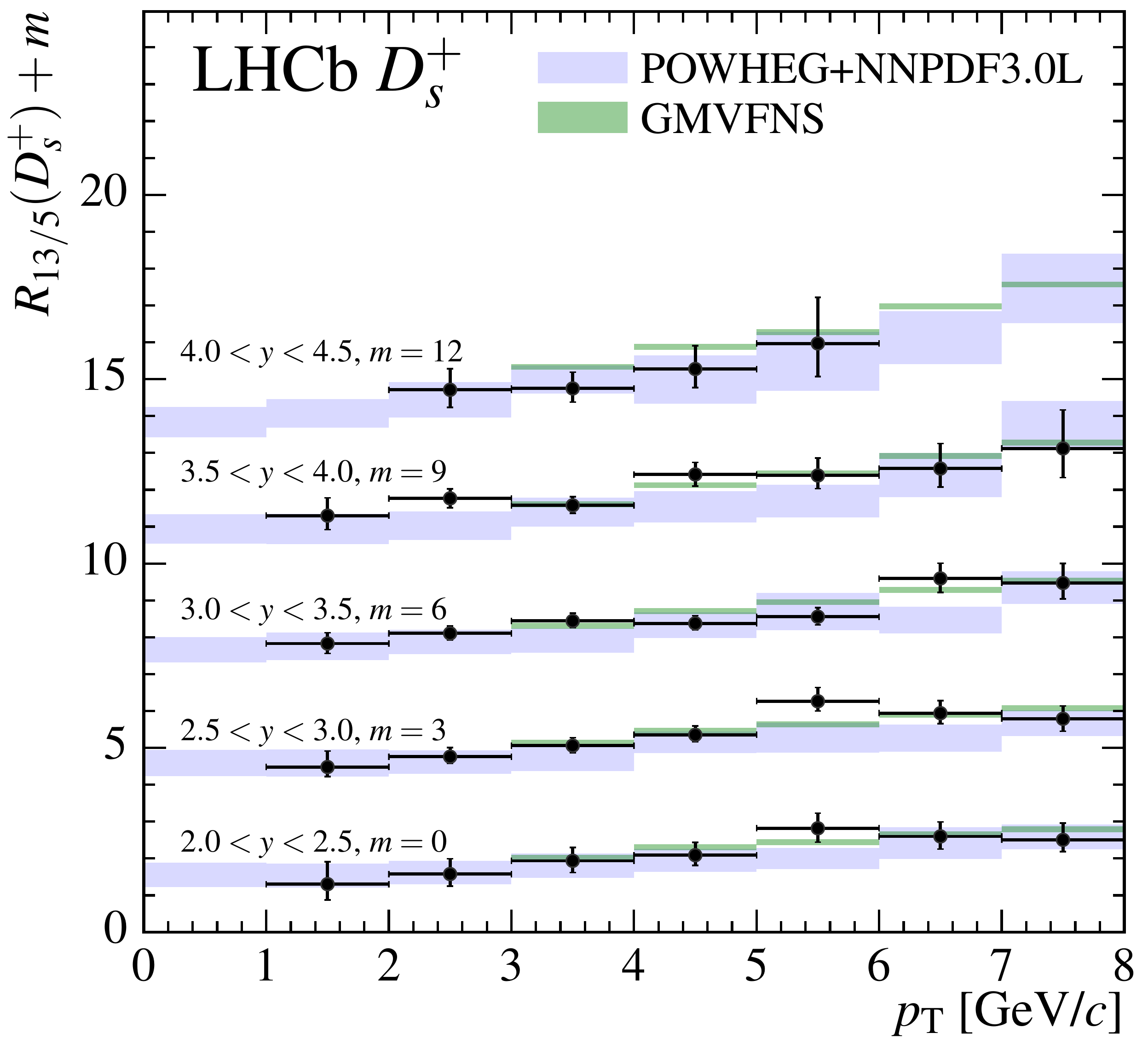}
  \includegraphics[width=0.49\textwidth]{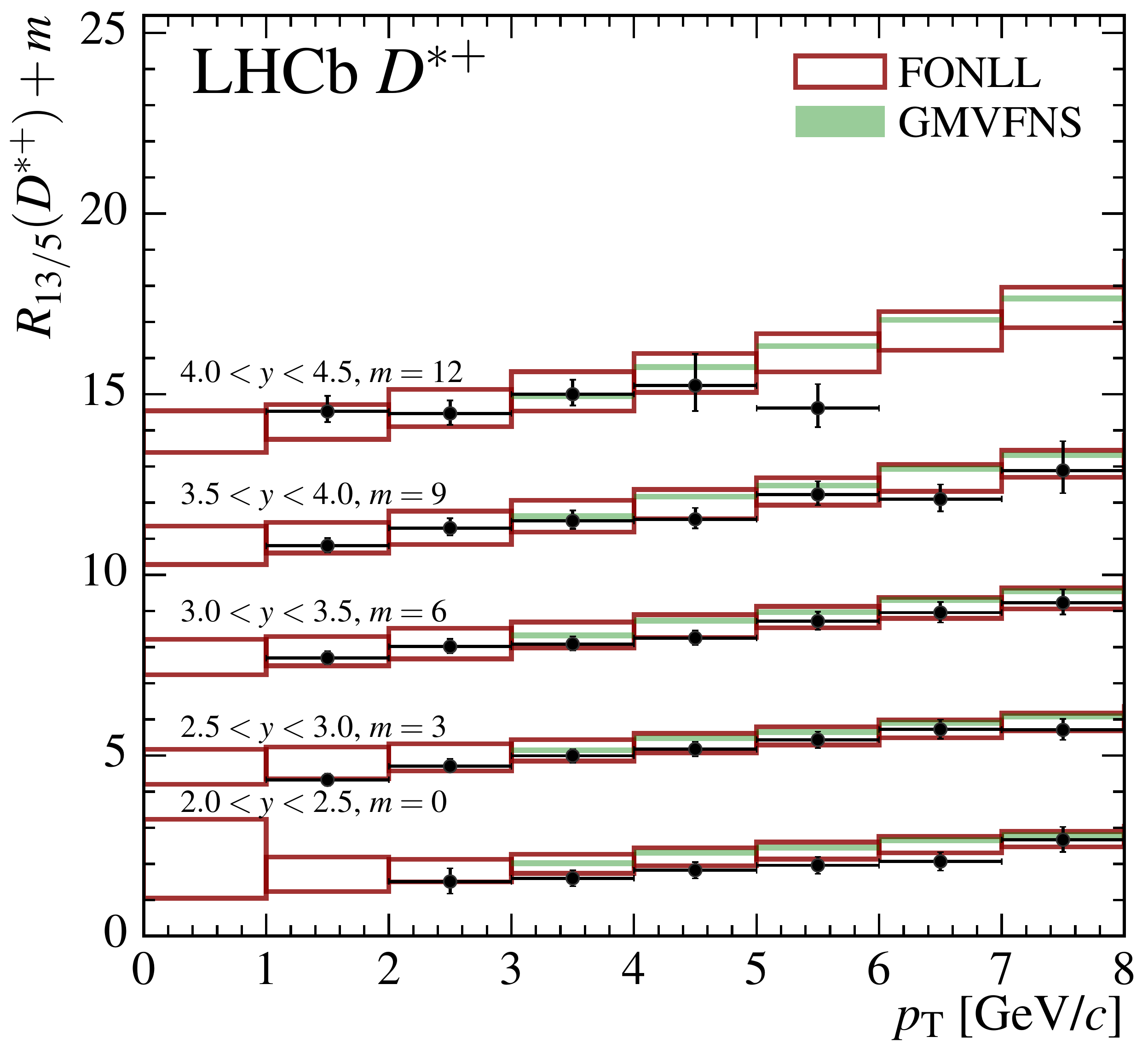}
    \caption{%
      Measurements and predictions of the prompt \Dz, \Dp, \Dsp, and \Dstarp 
      cross-section ratios between $\sqrts = 13$ and \SI{5}{\TeV}.
      Each set of measurements and predictions in a given rapidity bin is 
      offset by an additive constant $m$, which is shown on the plot.
      Only central values are provided for the \Hubert predictions.
      \label{fig:TheoryPredRatio}}
\end{figure}
\begin{figure}[p]
  \centering
  \includegraphics[width=0.70\textwidth]{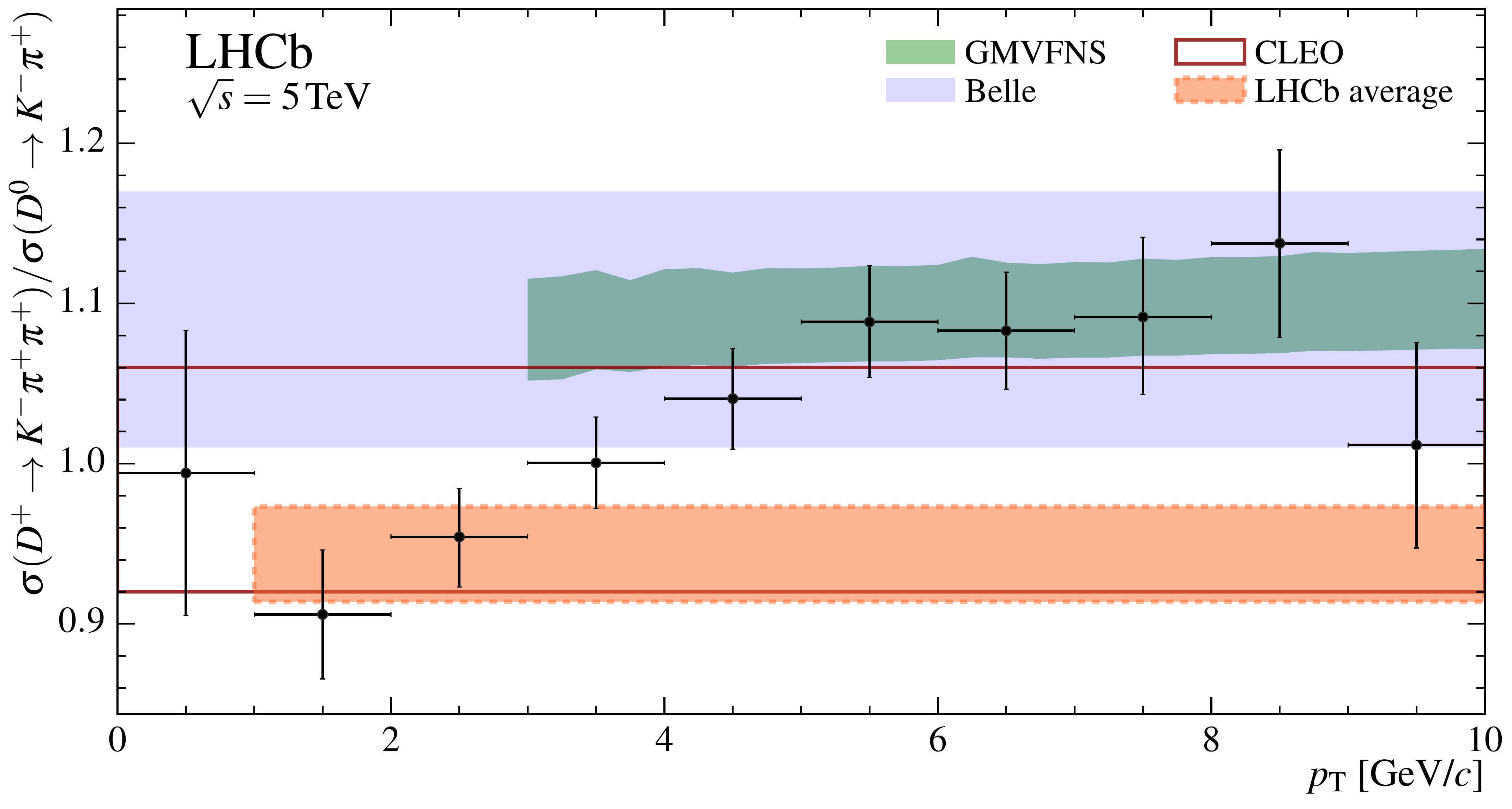}
  \includegraphics[width=0.70\textwidth]{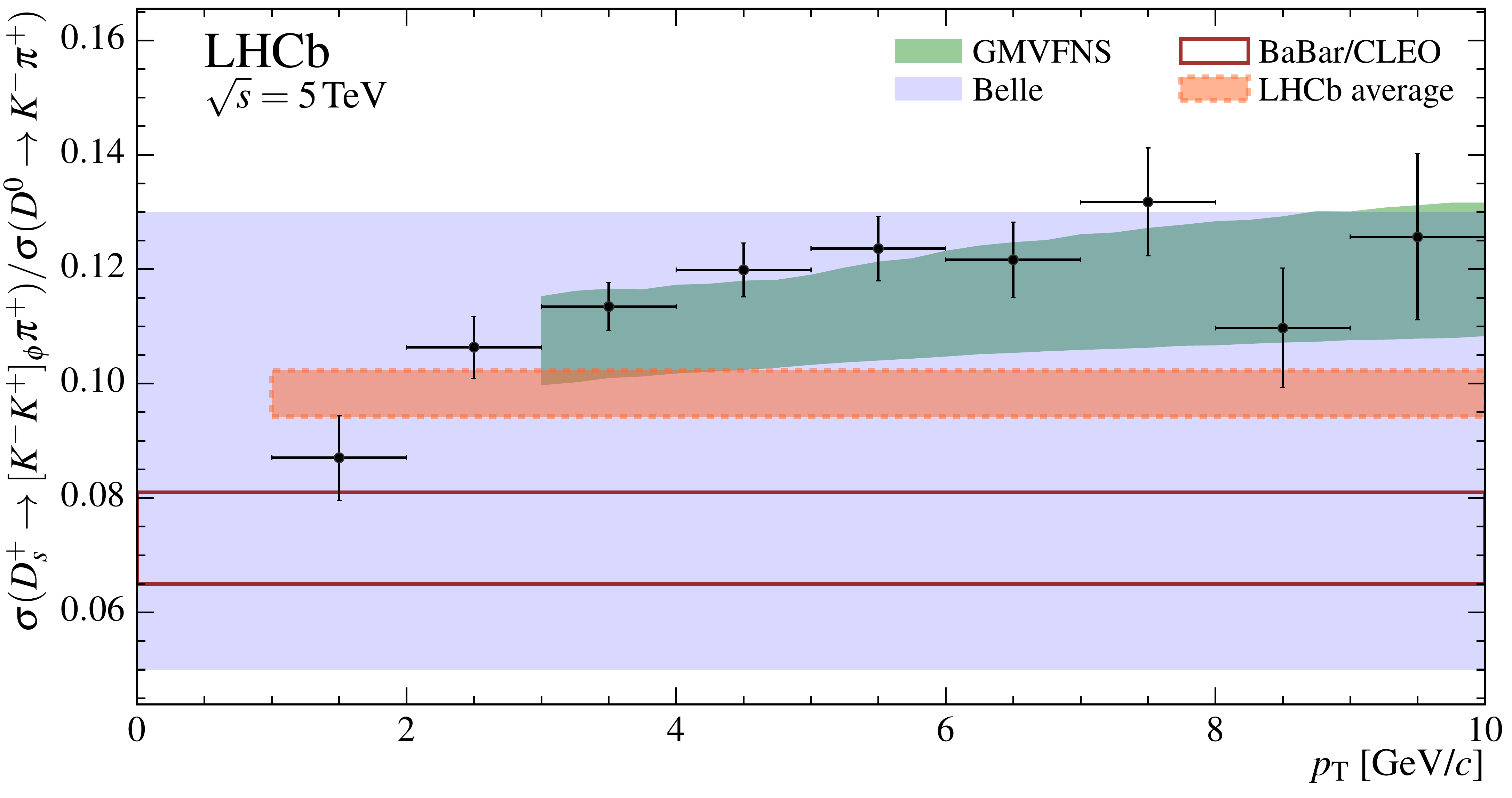}
  \includegraphics[width=0.70\textwidth]{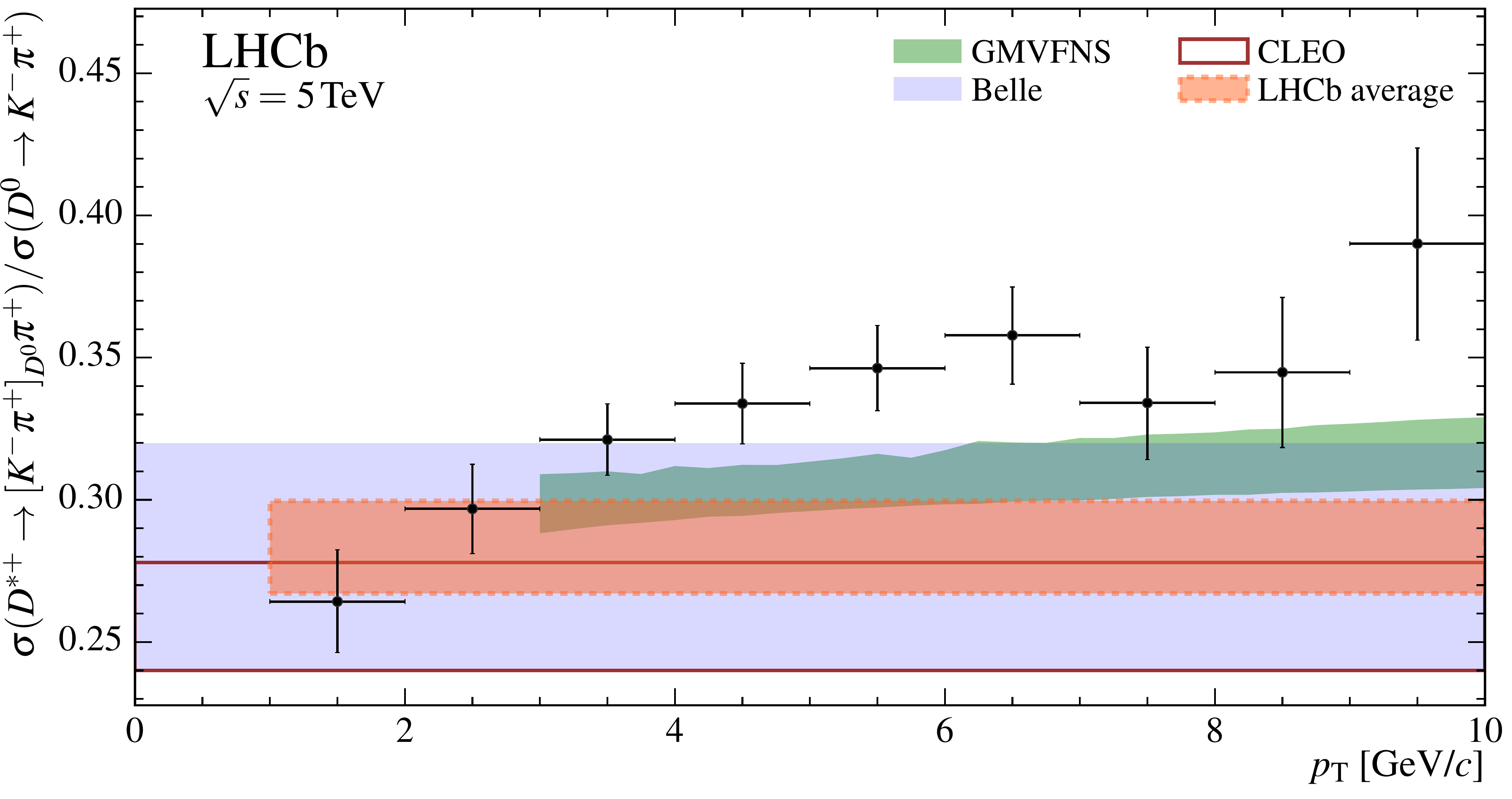}
    \caption{%
      Ratios of the measurements of cross-section times branching fraction of (top)
      \Dp, (middle) \Dsp, and (bottom) \Dstarp mesons with respect to the \Dz 
      measurements.
      The bands indicate the corresponding ratios computed using measurements 
      from $\ep\en$ collider 
      experiments~\cite{Artuso:2004pj,Seuster:2005tr,Aubert:2002ue}.
      The ratios are given as a function of \pT integrated over \rapidity.
      The notation $\sigma(\decay{D}{f})$ is shorthand for \xsectimesbfrac.
      \label{fig:MesonRatio}}
\end{figure}

\begin{figure}[h!]
  \centering
  \includegraphics[width=0.70\textwidth]{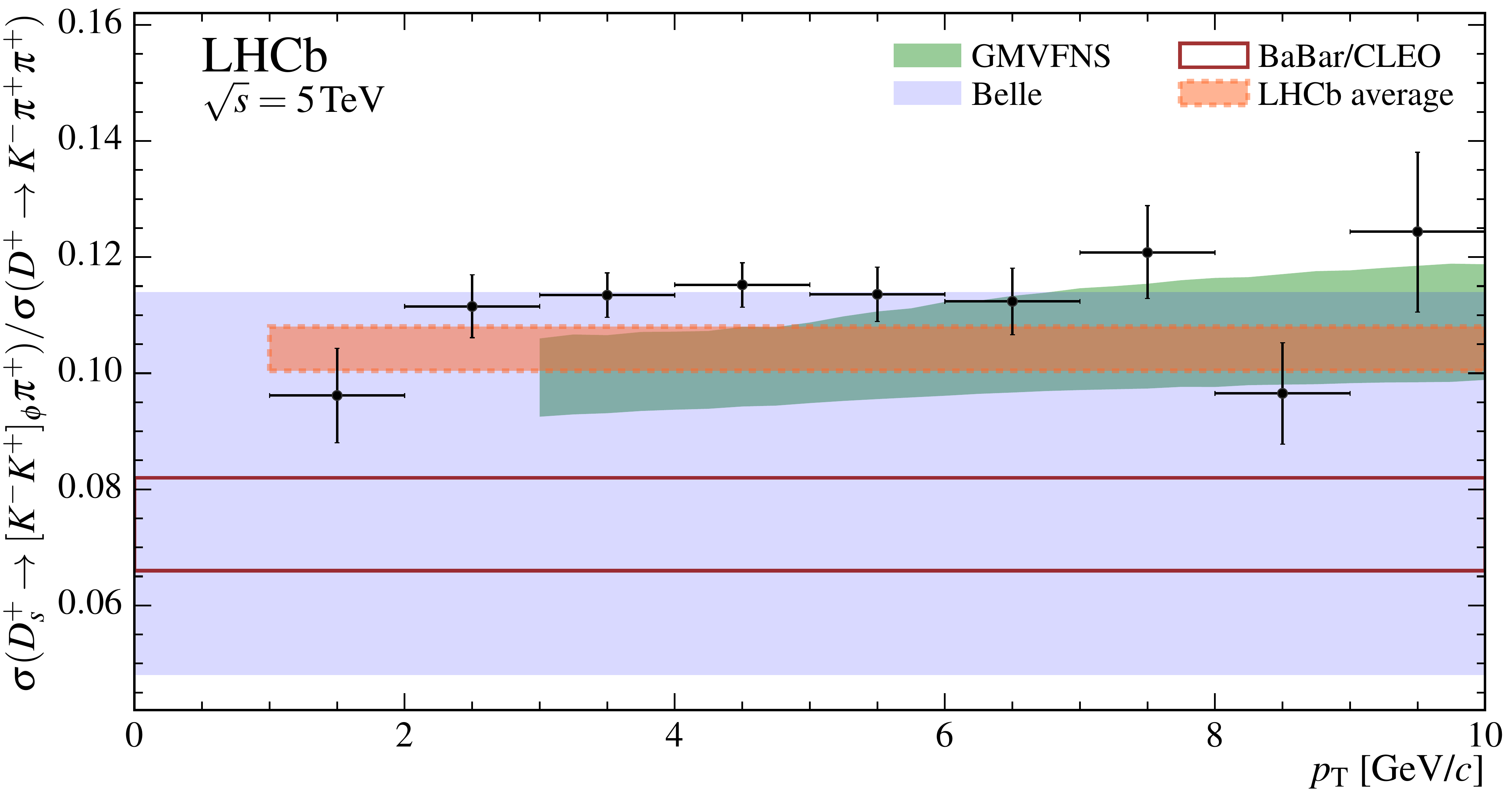}
  \includegraphics[width=0.70\textwidth]{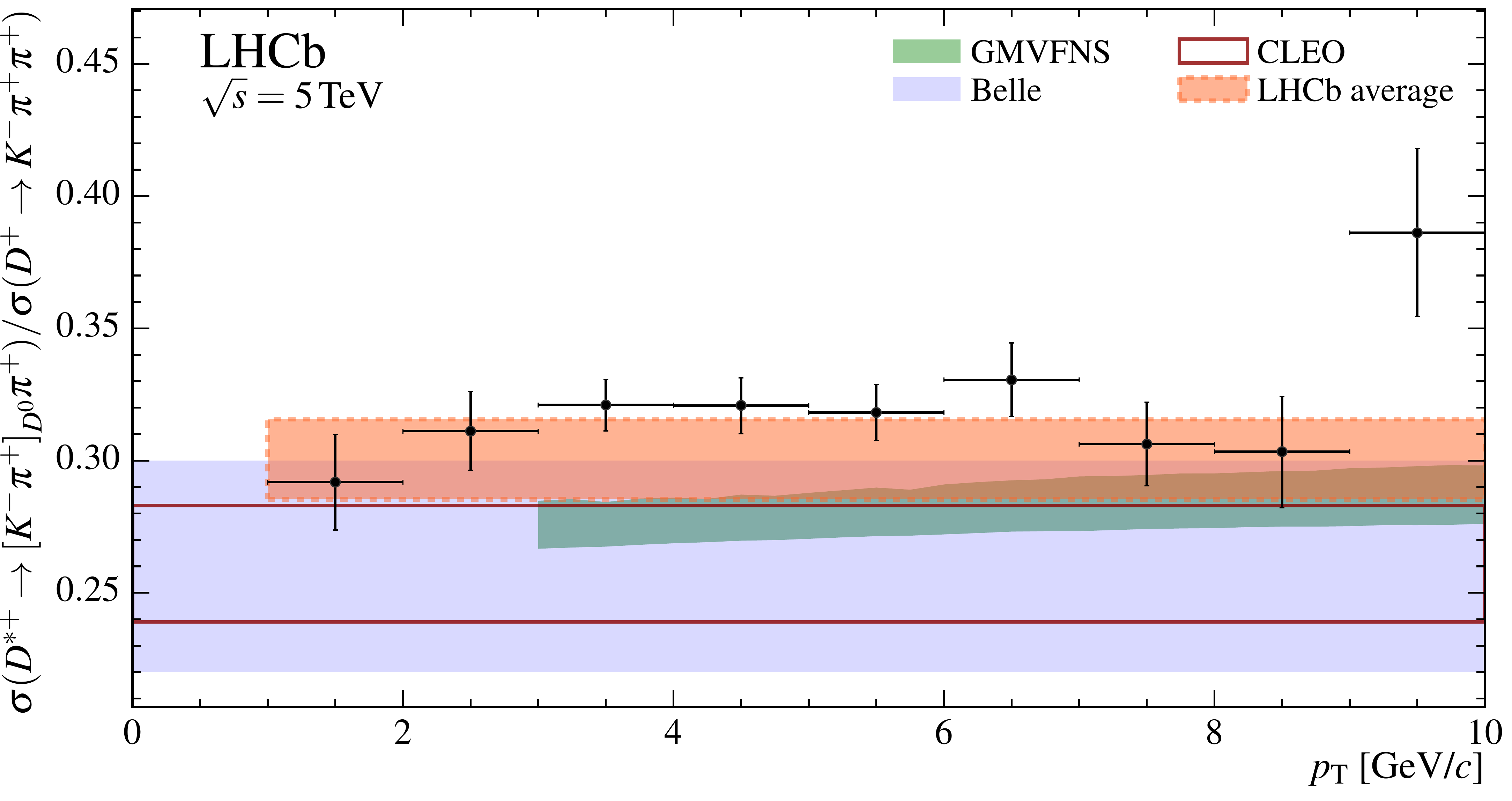}
  \includegraphics[width=0.70\textwidth]{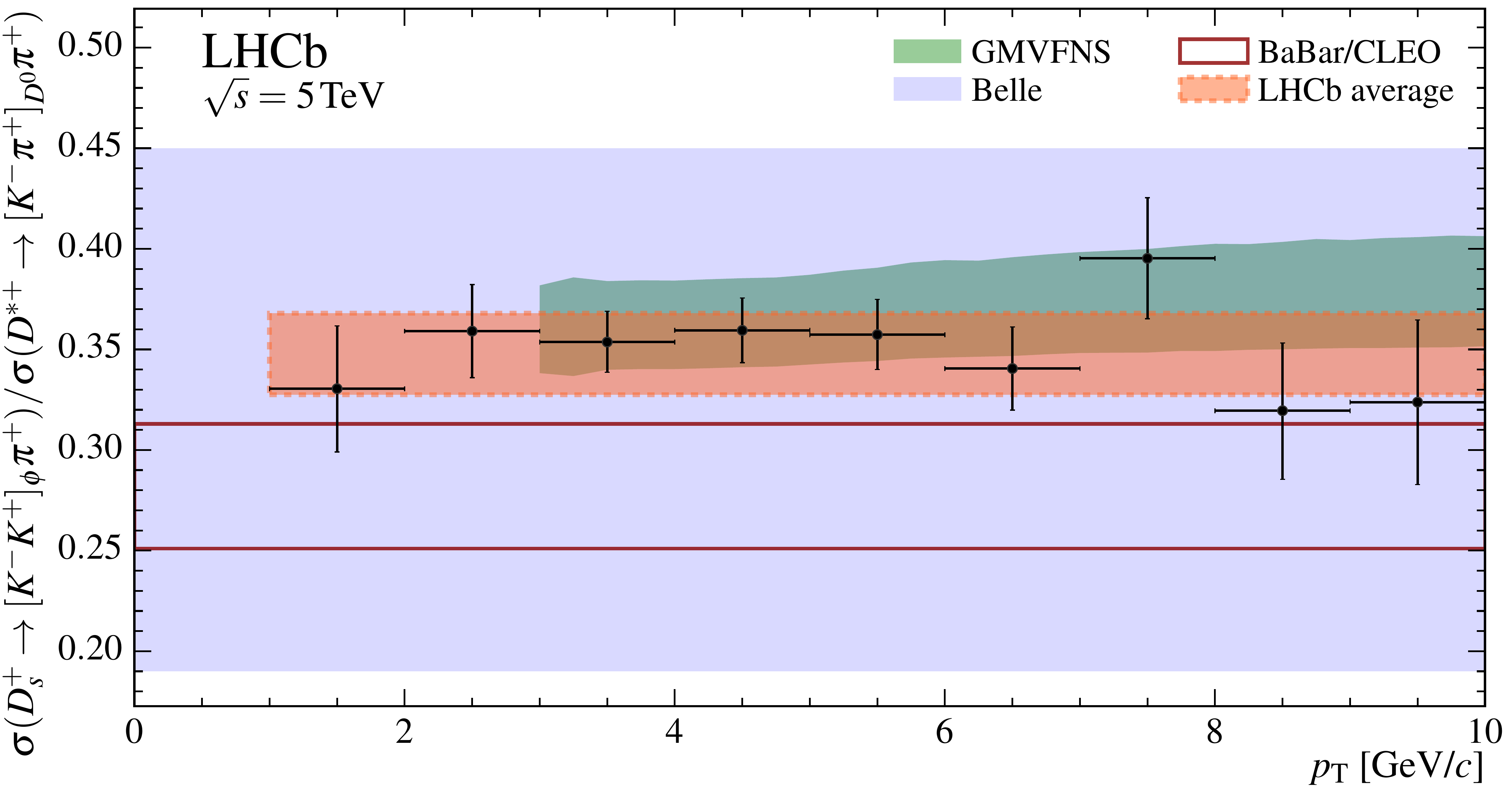}
    \caption{%
      Ratios of the measurements of cross-section times branching fraction of (top) 
      \Dstarp, and (middle) \Dsp mesons with respect to \Dp cross-sections, and 
      (bottom) \Dsp over \Dstarp mesons.
      The bands indicate the corresponding ratios computed using measurements 
      from $\ep\en$ collider 
      experiments~\cite{Artuso:2004pj,Seuster:2005tr,Aubert:2002ue}.
      The ratios are given as a function of \pT integrated over \rapidity.
      The notation $\sigma(\decay{D}{f})$ is shorthand for \xsectimesbfrac.
      \label{fig:MesonRatio2}}
\end{figure}

\section{Comparison to theory}
\label{sec:theory}

Theoretical calculations for charm meson production cross-sections in
\pp collisions at \comenergy, according to the same methods described in 
Refs.~\cite{Gauld:2015yia} (\Rhorry), \cite{Cacciari:2015fta} (\Matteo) and 
\cite{Kniehl:2012ti} (\Hubert), have been provided by the authors.
All sets of calculations are performed at NLO precision, and each
includes estimates of theoretical uncertainties due to the renormalisation
and factorisation scales.
The theoretical uncertainties provided with the \Rhorry predictions
also include contributions due to uncertainties in the effective charm quark mass and the 
parton distribution functions.

 The \Matteo predictions are provided in the form of \Dz, \Dp,
 and \Dstp production cross-sections for \pp collisions at \comenergy for each
 phase space bin in the range \mbox{$\pT < \SI{10}{\gevc}$} and
 \mbox{$2.0 < y < 4.5$}.
 Ratios of these cross-sections to those computed for \pp collisions at
 \SI{13}{\TeV}\ are also supplied.
 The calculations use the NNPDF3.0 NLO~\cite{Ball:2014uwa} parton densities.
 These \Matteo calculations of the meson differential production cross-sections 
 assume $f(\decay{\cquark}{\PD}) = 1$ and are multiplied by the
 transition probabilities measured at $\ep\en$\/ colliders for comparison to the current measurements.
 No dedicated \Matteo cross-section calculation for \Dsp production is available.

\clearpage

The \Rhorry predictions provide \Dz, \Dp, and \Dsp integrated production cross-sections for
\pp collisions at \comenergy in the same \pT and \rapidity binning used for the measurement.
Ratios of \comenergyold to \SI{5}{\TeV} cross-sections are also predicted for \Dz, \Dp and \Dsp.
They are obtained with \powheg~\cite{Alioli:2010xd} matched to
\pythiaeight~\cite{Sjostrand:2014zea} parton showers and an improved
 version of the NNPDF3.0 NLO parton distribution function set 
 designated NNPDF3.0+LHCb~\cite{Gauld:2015yia}. To produce this improved set, 
the authors of Ref.~\cite{Gauld:2015yia}
weight the NNPDF3.0 NLO set in order to match FONLL calculations to the \lhcb
charm cross-section measurements at \SI{7}{\TeV}~\cite{LHCb-PAPER-2012-041}.
This results in a significant reduction in the uncertainties for the gluon
distribution function at small momentum fraction $x$.
Predictions for the integrated cross-section are provided for each meson.

The \Hubert calculations include theoretical predictions for the absolute cross-sections for all mesons
studied in this analysis.
Predictions of central values are also provided for the cross-section ratios between mesons.
Results are provided for \mbox{$3 < \pT < \SI{10}{\gevc}$}.
Here the CT10~\cite{Lai:2010vv} set of parton distributions is used.
The \Hubert theoretical framework includes the convolution with fragmentation
functions describing the transition \decay{\cquark}{\Hc}
that are normalised to the respective total transition
probabilities~\cite{Kneesch:2007ey,Kniehl:2006mw}.
The fragmentation functions are taken from a fit to production
measurements at $\ep\en$\/ colliders, where no attempt is made to
separate direct production and feed-down from higher resonances.

In general, the predicted shapes of the cross-sections at \comenergy agree with the data shown in Figures~\ref{fig:theory_absolute_D0_Dp} and~\ref{fig:theory_absolute_Dsp_Dstar}.
The central values of the measurements generally lie above those of the theory predictions, albeit within the uncertainties provided.
For the \Rhorry and \Matteo predictions, the data tend to lie at the upper edge of the uncertainty band.
The \Hubert predictions provide a better description of the data, although the 
cross-sections decrease with \pT at a higher rate than the data near their low \pT 
limit of \SI{3}{\gevc}.
Similar behaviour was observed for the $\sqrt{s}=\SI{7}{\TeV}$ measurement~\cite{LHCb-PAPER-2012-041}, where only central values were shown for the \Matteo prediction, and for the $\sqrts = \SI{13}{\TeV}$ measurement~\cite{LHCb-PAPER-2015-041}, where in both cases the predictions gave lower cross-sections than the data.

The data generally agree well with the \Matteo, \Rhorry and \Hubert predictions for the ratios of cross-sections at $\sqrts=\SI{13}{\TeV}$ and \SI{5}{\TeV}, shown in Figure~\ref{fig:TheoryPredRatio}.
These ratio measurements are significantly more precise than those 
  previously presented between cross-sections at ${\sqrts}={\SI{13}{\TeV}}$ 
  and {\SI{7}{\TeV}}, and will act as stronger constraints in PDF fits as a 
  result.

\section{Summary}
\label{sec:summary}

A measurement of charm production in \pp collisions at a
centre-of-mass energy of \mbox{\comenergy} has been performed with data 
collected with the \lhcb detector. While the shapes of the differential
cross-sections for \Dz, \Dp, \Dsp, and \Dstarp mesons are found to be in
agreement with NLO perturbative QCD calculations, the measured values tend to lie at the upper 
edge of the predictions.  This is a feature also common to the measurements at 
$\sqrts = 7$ and \SI{13}{\TeV}, which indicates
a general underestimation in the prediction of the absolute value of prompt charm production in the forward region.
The ratios of the production cross-sections for centre-of-mass energies of \SI{13}{TeV} and \SI{5}{TeV} 
have been measured and show consistency with theoretical predictions.
The integrated cross-sections for prompt open charm meson production in \pp 
collisions at \comenergy and
in the range $1 < \pT < \SI{8}{\gevc}$ and $2 < \rapidity < 4.5$ are
\begin{align*}
  \begin{array}{lcr}
    \sigma(\decay{\pp}{\Dz X})  &=& \DzOnePtXsec, \\
    \sigma(\decay{\pp}{\Dp X})  &=& \DpOnePtXsec, \\
    \sigma(\decay{\pp}{\Dsp X}) &=& \DspOnePtXsec, \\
    \sigma(\decay{\pp}{\Dstp X})&=& \DstpOnePtXsec. \\
  \end{array}
\end{align*}

\section*{Acknowledgements}
\noindent The authors would like to thank R.~Gauld and J.~Rojo for the 
provisions of the \Rhorry numbers; M.~Cacciari, M.L.~Mangano, and P.~Nason for 
the \Matteo predictions; and H.~Spiesberger and G.~Kramer for the \Hubert 
calculations.
We express our gratitude to our colleagues in the CERN
accelerator departments for the excellent performance of the LHC. We
thank the technical and administrative staff at the LHCb
institutes. We acknowledge support from CERN and from the national
agencies: CAPES, CNPq, FAPERJ and FINEP (Brazil); NSFC (China);
CNRS/IN2P3 (France); BMBF, DFG and MPG (Germany); INFN (Italy); 
FOM and NWO (The Netherlands); MNiSW and NCN (Poland); MEN/IFA (Romania); 
MinES and FASO (Russia); MinECo (Spain); SNSF and SER (Switzerland); 
NASU (Ukraine); STFC (United Kingdom); NSF (USA).
We acknowledge the computing resources that are provided by CERN, IN2P3 (France), KIT and DESY (Germany), INFN (Italy), SURF (The Netherlands), PIC (Spain), GridPP (United Kingdom), RRCKI and Yandex LLC (Russia), CSCS (Switzerland), IFIN-HH (Romania), CBPF (Brazil), PL-GRID (Poland) and OSC (USA). We are indebted to the communities behind the multiple open 
source software packages on which we depend.
Individual groups or members have received support from AvH Foundation (Germany),
EPLANET, Marie Sk\l{}odowska-Curie Actions and ERC (European Union), 
Conseil G\'{e}n\'{e}ral de Haute-Savoie, Labex ENIGMASS and OCEVU, 
R\'{e}gion Auvergne (France), RFBR and Yandex LLC (Russia), GVA, XuntaGal and GENCAT (Spain), Herchel Smith Fund, The Royal Society, Royal Commission for the Exhibition of 1851 and the Leverhulme Trust (United Kingdom).

\clearpage

{\noindent\bf\Large Appendices}

\appendix

\section{Absolute cross-sections}
\label{app:xsec}

Tables~\ref{table:differential:D0}--\ref{table:differential:Dstp} give the numerical results for the differential cross-sections.

\vspace*{\fill}
\hvFloat[
  nonFloat=true,
  objectAngle=90,
  capPos=l,
  capAngle=90,
  capWidth=h]{table}{
    \centering
    \renewcommand{\arraystretch}{1.3}
\begin{tabular}{c|r@{\hskip+0.2em}c@{\hskip+0.2em}r@{\hskip+0.2em}c@{\hskip+0.2em}rr@{\hskip+0.2em}c@{\hskip+0.2em}r@{\hskip+0.2em}c@{\hskip+0.2em}rr@{\hskip+0.2em}c@{\hskip+0.2em}r@{\hskip+0.2em}c@{\hskip+0.2em}rr@{\hskip+0.2em}c@{\hskip+0.2em}r@{\hskip+0.2em}c@{\hskip+0.2em}rr@{\hskip+0.2em}c@{\hskip+0.2em}r@{\hskip+0.2em}c@{\hskip+0.2em}r}
\toprule&\multicolumn{25}{c}{$\text{$y$}$}\\
$\text{$p_{\text{T}}$ [\gevc]}$ & \multicolumn{5}{c}{$[2.0,2.5]$} & \multicolumn{5}{c}{$[2.5,3.0]$} & \multicolumn{5}{c}{$[3.0,3.5]$} & \multicolumn{5}{c}{$[3.5,4.0]$} & \multicolumn{5}{c}{$[4.0,4.5]$} \\
\midrule$[0,1]$ & $170$ & $^+_-$ & $^{3}_{3}$&$^+_-$ & $^{19}_{17}$ & $173$ & $^+_-$ & $^{1}_{1}$&$^+_-$ & $^{10}_{10}$ & $163.1$ & $^+_-$ & $^{1.3}_{1.3}$&$^+_-$ & $^{8.6}_{9.2}$ & $136.5$ & $^+_-$ & $^{1.5}_{1.5}$&$^+_-$ & $^{8.2}_{7.4}$ & $97.3$ & $^+_-$ & $^{2.4}_{2.4}$&$^+_-$ & $^{8.0}_{7.1}$ \\
$[1,2]$ & $255$ & $^+_-$ & $^{3}_{3}$&$^+_-$ & $^{21}_{19}$ & $252$ & $^+_-$ & $^{1}_{1}$&$^+_-$ & $^{13}_{15}$ & $222$ & $^+_-$ & $^{1}_{1}$&$^+_-$ & $^{11}_{12}$ & $184$ & $^+_-$ & $^{1}_{1}$&$^+_-$ & $^{10}_{10}$ & $130.1$ & $^+_-$ & $^{2.1}_{2.1}$&$^+_-$ & $^{8.2}_{7.9}$ \\
$[2,3]$ & $147.0$ & $^+_-$ & $^{1.5}_{1.5}$&$^+_-$ & $^{9.4}_{9.1}$ & $139.7$ & $^+_-$ & $^{0.8}_{0.8}$&$^+_-$ & $^{6.8}_{8.1}$ & $113.7$ & $^+_-$ & $^{0.7}_{0.7}$&$^+_-$ & $^{6.1}_{5.9}$ & $92.2$ & $^+_-$ & $^{0.8}_{0.8}$&$^+_-$ & $^{5.2}_{5.1}$ & $59.2$ & $^+_-$ & $^{1.1}_{1.1}$&$^+_-$ & $^{3.5}_{3.4}$ \\
$[3,4]$ & $66.7$ & $^+_-$ & $^{0.7}_{0.7}$&$^+_-$ & $^{3.9}_{3.7}$ & $60.8$ & $^+_-$ & $^{0.4}_{0.4}$&$^+_-$ & $^{3.1}_{3.3}$ & $49.0$ & $^+_-$ & $^{0.4}_{0.4}$&$^+_-$ & $^{2.6}_{2.5}$ & $37.9$ & $^+_-$ & $^{0.4}_{0.4}$&$^+_-$ & $^{2.2}_{2.1}$ & $21.6$ & $^+_-$ & $^{0.6}_{0.6}$&$^+_-$ & $^{1.8}_{1.6}$ \\
$[4,5]$ & $30.4$ & $^+_-$ & $^{0.4}_{0.4}$&$^+_-$ & $^{1.6}_{1.8}$ & $26.0$ & $^+_-$ & $^{0.2}_{0.2}$&$^+_-$ & $^{1.4}_{1.4}$ & $20.5$ & $^+_-$ & $^{0.2}_{0.2}$&$^+_-$ & $^{1.2}_{1.0}$ & $14.96$ & $^+_-$ & $^{0.23}_{0.23}$&$^+_-$ & $^{0.85}_{0.77}$ & $8.6$ & $^+_-$ & $^{0.4}_{0.4}$&$^+_-$ & $^{1.5}_{1.1}$ \\
$[5,6]$ & $13.83$ & $^+_-$ & $^{0.23}_{0.23}$&$^+_-$ & $^{0.74}_{0.79}$ & $11.84$ & $^+_-$ & $^{0.15}_{0.15}$&$^+_-$ & $^{0.61}_{0.67}$ & $9.34$ & $^+_-$ & $^{0.14}_{0.14}$&$^+_-$ & $^{0.58}_{0.47}$ & $6.62$ & $^+_-$ & $^{0.16}_{0.16}$&$^+_-$ & $^{0.46}_{0.39}$ & $1.95$ & $^+_-$ & $^{0.28}_{0.28}$&$^+_-$ & $^{0.67}_{0.41}$ \\
$[6,7]$ & $7.25$ & $^+_-$ & $^{0.15}_{0.15}$&$^+_-$ & $^{0.40}_{0.42}$ & $5.83$ & $^+_-$ & $^{0.11}_{0.11}$&$^+_-$ & $^{0.32}_{0.34}$ & $4.41$ & $^+_-$ & $^{0.09}_{0.09}$&$^+_-$ & $^{0.27}_{0.22}$ & $2.65$ & $^+_-$ & $^{0.10}_{0.10}$&$^+_-$ & $^{0.28}_{0.22}$ & \multicolumn{5}{c}{ } \\
$[7,8]$ & $3.80$ & $^+_-$ & $^{0.10}_{0.10}$&$^+_-$ & $^{0.22}_{0.23}$ & $3.23$ & $^+_-$ & $^{0.08}_{0.08}$&$^+_-$ & $^{0.21}_{0.18}$ & $2.29$ & $^+_-$ & $^{0.07}_{0.07}$&$^+_-$ & $^{0.16}_{0.13}$ & $1.26$ & $^+_-$ & $^{0.09}_{0.09}$&$^+_-$ & $^{0.25}_{0.18}$ & \multicolumn{5}{c}{ } \\
$[8,9]$ & $2.23$ & $^+_-$ & $^{0.08}_{0.08}$&$^+_-$ & $^{0.15}_{0.15}$ & $1.71$ & $^+_-$ & $^{0.06}_{0.06}$&$^+_-$ & $^{0.12}_{0.11}$ & $1.27$ & $^+_-$ & $^{0.05}_{0.05}$&$^+_-$ & $^{0.12}_{0.10}$ & $0.46$ & $^+_-$ & $^{0.06}_{0.06}$&$^+_-$ & $^{0.11}_{0.08}$ & \multicolumn{5}{c}{ } \\
$[9,10]$ & $1.43$ & $^+_-$ & $^{0.06}_{0.06}$&$^+_-$ & $^{0.10}_{0.11}$ & $0.978$ & $^+_-$ & $^{0.041}_{0.041}$&$^+_-$ & $^{0.076}_{0.066}$ & $0.556$ & $^+_-$ & $^{0.035}_{0.035}$&$^+_-$ & $^{0.077}_{0.060}$ & \multicolumn{5}{c}{ } & \multicolumn{5}{c}{ } \\
\bottomrule\end{tabular}

}{Differential production cross-sections, $\text{d}^2\sigma/(\text{d}\pT\,\text{d}y)$, in 
$\si{\micro\barn}/(\si{\gevc})$ for prompt $\Dz + \Dzbar$ mesons in bins 
of \pTy.
The first uncertainty is statistical, and the second is the total 
systematic.}{table:differential:D0}

\vspace*{\fill}

\newpage
\vspace*{\fill}
\hvFloat[
  nonFloat=true,
  objectAngle=90,
  capPos=l,
  capAngle=90,
  capWidth=h]{table}{
    \centering
    \renewcommand{\arraystretch}{1.3}
\begin{tabular}{c|r@{\hskip+0.2em}c@{\hskip+0.2em}r@{\hskip+0.2em}c@{\hskip+0.2em}rr@{\hskip+0.2em}c@{\hskip+0.2em}r@{\hskip+0.2em}c@{\hskip+0.2em}rr@{\hskip+0.2em}c@{\hskip+0.2em}r@{\hskip+0.2em}c@{\hskip+0.2em}rr@{\hskip+0.2em}c@{\hskip+0.2em}r@{\hskip+0.2em}c@{\hskip+0.2em}rr@{\hskip+0.2em}c@{\hskip+0.2em}r@{\hskip+0.2em}c@{\hskip+0.2em}r}
\toprule&\multicolumn{25}{c}{$\text{$y$}$}\\
$\text{$p_{\text{T}}$ [\gevc]}$ & \multicolumn{5}{c}{$[2.0,2.5]$} & \multicolumn{5}{c}{$[2.5,3.0]$} & \multicolumn{5}{c}{$[3.0,3.5]$} & \multicolumn{5}{c}{$[3.5,4.0]$} & \multicolumn{5}{c}{$[4.0,4.5]$} \\
\midrule$[0,1]$ & \multicolumn{5}{c}{ } & $74$ & $^+_-$ & $^{7}_{7}$&$^+_-$ & $^{10}_{11}$ & $58.2$ & $^+_-$ & $^{3.5}_{3.5}$&$^+_-$ & $^{7.0}_{6.0}$ & $58.6$ & $^+_-$ & $^{4.1}_{4.1}$&$^+_-$ & $^{6.9}_{6.1}$ & \multicolumn{5}{c}{ } \\
$[1,2]$ & $91$ & $^+_-$ & $^{3}_{3}$&$^+_-$ & $^{14}_{12}$ & $94.3$ & $^+_-$ & $^{1.0}_{1.0}$&$^+_-$ & $^{7.4}_{8.4}$ & $89.6$ & $^+_-$ & $^{0.8}_{0.8}$&$^+_-$ & $^{8.0}_{6.5}$ & $70.1$ & $^+_-$ & $^{0.8}_{0.8}$&$^+_-$ & $^{5.6}_{4.7}$ & $54.2$ & $^+_-$ & $^{1.3}_{1.3}$&$^+_-$ & $^{4.9}_{4.5}$ \\
$[2,3]$ & $57.4$ & $^+_-$ & $^{0.8}_{0.8}$&$^+_-$ & $^{6.2}_{5.7}$ & $55.8$ & $^+_-$ & $^{0.4}_{0.4}$&$^+_-$ & $^{3.7}_{4.7}$ & $47.7$ & $^+_-$ & $^{0.3}_{0.3}$&$^+_-$ & $^{3.7}_{3.1}$ & $37.6$ & $^+_-$ & $^{0.3}_{0.3}$&$^+_-$ & $^{2.9}_{2.4}$ & $25.0$ & $^+_-$ & $^{0.4}_{0.4}$&$^+_-$ & $^{1.9}_{1.9}$ \\
$[3,4]$ & $28.0$ & $^+_-$ & $^{0.4}_{0.4}$&$^+_-$ & $^{2.4}_{2.5}$ & $25.6$ & $^+_-$ & $^{0.2}_{0.2}$&$^+_-$ & $^{1.5}_{2.1}$ & $21.0$ & $^+_-$ & $^{0.1}_{0.1}$&$^+_-$ & $^{1.4}_{1.4}$ & $16.0$ & $^+_-$ & $^{0.1}_{0.1}$&$^+_-$ & $^{1.1}_{1.1}$ & $10.26$ & $^+_-$ & $^{0.19}_{0.19}$&$^+_-$ & $^{0.82}_{0.77}$ \\
$[4,5]$ & $13.7$ & $^+_-$ & $^{0.2}_{0.2}$&$^+_-$ & $^{1.0}_{1.2}$ & $11.39$ & $^+_-$ & $^{0.10}_{0.10}$&$^+_-$ & $^{0.71}_{0.84}$ & $9.15$ & $^+_-$ & $^{0.09}_{0.09}$&$^+_-$ & $^{0.62}_{0.61}$ & $6.52$ & $^+_-$ & $^{0.08}_{0.08}$&$^+_-$ & $^{0.44}_{0.43}$ & $3.86$ & $^+_-$ & $^{0.11}_{0.11}$&$^+_-$ & $^{0.30}_{0.26}$ \\
$[5,6]$ & $6.30$ & $^+_-$ & $^{0.11}_{0.11}$&$^+_-$ & $^{0.45}_{0.49}$ & $5.41$ & $^+_-$ & $^{0.06}_{0.06}$&$^+_-$ & $^{0.34}_{0.39}$ & $4.13$ & $^+_-$ & $^{0.05}_{0.05}$&$^+_-$ & $^{0.28}_{0.27}$ & $2.90$ & $^+_-$ & $^{0.05}_{0.05}$&$^+_-$ & $^{0.20}_{0.20}$ & $1.52$ & $^+_-$ & $^{0.08}_{0.08}$&$^+_-$ & $^{0.15}_{0.13}$ \\
$[6,7]$ & $3.29$ & $^+_-$ & $^{0.07}_{0.07}$&$^+_-$ & $^{0.23}_{0.25}$ & $2.64$ & $^+_-$ & $^{0.04}_{0.04}$&$^+_-$ & $^{0.17}_{0.19}$ & $2.10$ & $^+_-$ & $^{0.04}_{0.04}$&$^+_-$ & $^{0.15}_{0.14}$ & $1.249$ & $^+_-$ & $^{0.034}_{0.034}$&$^+_-$ & $^{0.095}_{0.086}$ & $0.709$ & $^+_-$ & $^{0.068}_{0.068}$&$^+_-$ & $^{0.082}_{0.073}$ \\
$[7,8]$ & $1.83$ & $^+_-$ & $^{0.05}_{0.05}$&$^+_-$ & $^{0.13}_{0.15}$ & $1.366$ & $^+_-$ & $^{0.028}_{0.028}$&$^+_-$ & $^{0.089}_{0.100}$ & $1.037$ & $^+_-$ & $^{0.025}_{0.025}$&$^+_-$ & $^{0.073}_{0.072}$ & $0.693$ & $^+_-$ & $^{0.027}_{0.027}$&$^+_-$ & $^{0.063}_{0.052}$ & \multicolumn{5}{c}{ } \\
$[8,9]$ & $1.025$ & $^+_-$ & $^{0.033}_{0.033}$&$^+_-$ & $^{0.072}_{0.081}$ & $0.775$ & $^+_-$ & $^{0.021}_{0.021}$&$^+_-$ & $^{0.052}_{0.059}$ & $0.554$ & $^+_-$ & $^{0.019}_{0.019}$&$^+_-$ & $^{0.044}_{0.039}$ & $0.393$ & $^+_-$ & $^{0.024}_{0.024}$&$^+_-$ & $^{0.050}_{0.039}$ & \multicolumn{5}{c}{ } \\
$[9,10]$ & $0.555$ & $^+_-$ & $^{0.023}_{0.023}$&$^+_-$ & $^{0.043}_{0.046}$ & $0.423$ & $^+_-$ & $^{0.015}_{0.015}$&$^+_-$ & $^{0.030}_{0.033}$ & $0.295$ & $^+_-$ & $^{0.013}_{0.013}$&$^+_-$ & $^{0.029}_{0.023}$ & $0.162$ & $^+_-$ & $^{0.016}_{0.016}$&$^+_-$ & $^{0.028}_{0.021}$ & \multicolumn{5}{c}{ } \\
\bottomrule\end{tabular}

}{Differential production cross-sections, $\text{d}^2\sigma/(\text{d}\pT\,\text{d}y)$, in 
$\si{\micro\barn}/(\si{\gevc})$ for prompt $\Dp + \Dm$ mesons in bins of 
\pTy. 
The first uncertainty is statistical, and the second is the total 
systematic.}{table:differential:Dp}

\vspace*{\fill}

\newpage
\vspace*{\fill}
\hvFloat[
  nonFloat=true,
  objectAngle=90,
  capPos=l,
  capAngle=90,
  capWidth=h]{table}{
  \centering
  \renewcommand{\arraystretch}{1.3}
\begin{tabular}{c|r@{\hskip+0.2em}c@{\hskip+0.2em}r@{\hskip+0.2em}c@{\hskip+0.2em}rr@{\hskip+0.2em}c@{\hskip+0.2em}r@{\hskip+0.2em}c@{\hskip+0.2em}rr@{\hskip+0.2em}c@{\hskip+0.2em}r@{\hskip+0.2em}c@{\hskip+0.2em}rr@{\hskip+0.2em}c@{\hskip+0.2em}r@{\hskip+0.2em}c@{\hskip+0.2em}rr@{\hskip+0.2em}c@{\hskip+0.2em}r@{\hskip+0.2em}c@{\hskip+0.2em}r}
\toprule&\multicolumn{25}{c}{$\text{$y$}$}\\
$\text{$p_{\text{T}}$ [\gevc]}$ & \multicolumn{5}{c}{$[2.0,2.5]$} & \multicolumn{5}{c}{$[2.5,3.0]$} & \multicolumn{5}{c}{$[3.0,3.5]$} & \multicolumn{5}{c}{$[3.5,4.0]$} & \multicolumn{5}{c}{$[4.0,4.5]$} \\
\midrule$[1,2]$ & $29.8$ & $^+_-$ & $^{4.6}_{4.6}$&$^+_-$ & $^{7.5}_{5.3}$ & $41.6$ & $^+_-$ & $^{2.5}_{2.5}$&$^+_-$ & $^{4.4}_{4.6}$ & $36.0$ & $^+_-$ & $^{2.3}_{2.2}$&$^+_-$ & $^{4.2}_{3.6}$ & $25.3$ & $^+_-$ & $^{2.3}_{2.3}$&$^+_-$ & $^{3.0}_{2.5}$ & \multicolumn{5}{c}{ } \\
$[2,3]$ & $27.3$ & $^+_-$ & $^{1.5}_{1.5}$&$^+_-$ & $^{5.1}_{4.0}$ & $27.4$ & $^+_-$ & $^{0.8}_{0.8}$&$^+_-$ & $^{2.4}_{2.7}$ & $22.4$ & $^+_-$ & $^{0.7}_{0.7}$&$^+_-$ & $^{2.2}_{1.9}$ & $14.6$ & $^+_-$ & $^{0.6}_{0.6}$&$^+_-$ & $^{1.5}_{1.2}$ & $9.3$ & $^+_-$ & $^{1.0}_{1.0}$&$^+_-$ & $^{1.5}_{1.2}$ \\
$[3,4]$ & $12.5$ & $^+_-$ & $^{0.6}_{0.6}$&$^+_-$ & $^{1.7}_{1.4}$ & $12.4$ & $^+_-$ & $^{0.4}_{0.4}$&$^+_-$ & $^{1.0}_{1.1}$ & $9.42$ & $^+_-$ & $^{0.30}_{0.30}$&$^+_-$ & $^{0.91}_{0.75}$ & $7.57$ & $^+_-$ & $^{0.32}_{0.32}$&$^+_-$ & $^{0.72}_{0.61}$ & $4.49$ & $^+_-$ & $^{0.42}_{0.42}$&$^+_-$ & $^{0.52}_{0.42}$ \\
$[4,5]$ & $5.96$ & $^+_-$ & $^{0.30}_{0.30}$&$^+_-$ & $^{0.59}_{0.63}$ & $5.71$ & $^+_-$ & $^{0.20}_{0.20}$&$^+_-$ & $^{0.48}_{0.51}$ & $4.52$ & $^+_-$ & $^{0.18}_{0.18}$&$^+_-$ & $^{0.43}_{0.37}$ & $2.81$ & $^+_-$ & $^{0.16}_{0.16}$&$^+_-$ & $^{0.28}_{0.23}$ & $1.94$ & $^+_-$ & $^{0.24}_{0.24}$&$^+_-$ & $^{0.25}_{0.18}$ \\
$[5,6]$ & $2.73$ & $^+_-$ & $^{0.18}_{0.17}$&$^+_-$ & $^{0.27}_{0.27}$ & $2.18$ & $^+_-$ & $^{0.11}_{0.11}$&$^+_-$ & $^{0.19}_{0.19}$ & $2.30$ & $^+_-$ & $^{0.11}_{0.11}$&$^+_-$ & $^{0.23}_{0.19}$ & $1.36$ & $^+_-$ & $^{0.10}_{0.10}$&$^+_-$ & $^{0.14}_{0.11}$ & $0.78$ & $^+_-$ & $^{0.15}_{0.15}$&$^+_-$ & $^{0.12}_{0.08}$ \\
$[6,7]$ & $1.47$ & $^+_-$ & $^{0.11}_{0.11}$&$^+_-$ & $^{0.15}_{0.14}$ & $1.31$ & $^+_-$ & $^{0.08}_{0.08}$&$^+_-$ & $^{0.13}_{0.11}$ & $0.843$ & $^+_-$ & $^{0.062}_{0.062}$&$^+_-$ & $^{0.089}_{0.070}$ & $0.586$ & $^+_-$ & $^{0.065}_{0.065}$&$^+_-$ & $^{0.074}_{0.057}$ & \multicolumn{5}{c}{ } \\
$[7,8]$ & $0.97$ & $^+_-$ & $^{0.09}_{0.09}$&$^+_-$ & $^{0.10}_{0.10}$ & $0.642$ & $^+_-$ & $^{0.052}_{0.052}$&$^+_-$ & $^{0.070}_{0.053}$ & $0.461$ & $^+_-$ & $^{0.044}_{0.044}$&$^+_-$ & $^{0.052}_{0.039}$ & $0.323$ & $^+_-$ & $^{0.046}_{0.046}$&$^+_-$ & $^{0.057}_{0.039}$ & \multicolumn{5}{c}{ } \\
$[8,9]$ & $0.379$ & $^+_-$ & $^{0.052}_{0.052}$&$^+_-$ & $^{0.041}_{0.039}$ & $0.311$ & $^+_-$ & $^{0.035}_{0.035}$&$^+_-$ & $^{0.038}_{0.027}$ & $0.256$ & $^+_-$ & $^{0.034}_{0.034}$&$^+_-$ & $^{0.034}_{0.025}$ & \multicolumn{5}{c}{ } & \multicolumn{5}{c}{ } \\
$[9,10]$ & $0.284$ & $^+_-$ & $^{0.042}_{0.042}$&$^+_-$ & $^{0.035}_{0.034}$ & $0.212$ & $^+_-$ & $^{0.028}_{0.028}$&$^+_-$ & $^{0.028}_{0.020}$ & $0.147$ & $^+_-$ & $^{0.025}_{0.025}$&$^+_-$ & $^{0.022}_{0.015}$ & \multicolumn{5}{c}{ } & \multicolumn{5}{c}{ } \\
\bottomrule\end{tabular}

}{Differential production cross-sections, $\text{d}^2\sigma/(\text{d}\pT\,\text{d}y)$, in 
$\si{\micro\barn}/(\si{\gevc})$ for prompt $\Dsp + \Dsm$ mesons in bins of 
\pTy.
The first uncertainty is statistical, and the second is the total 
systematic.}{table:differential:Dsp}

\vspace*{\fill}

\newpage
\vspace*{\fill}
\hvFloat[
  nonFloat=true,
  objectAngle=90,
  capPos=l,
  capAngle=90,
  capWidth=h]{table}{
    \centering
    \renewcommand{\arraystretch}{1.3}
\begin{tabular}{c|r@{\hskip+0.2em}c@{\hskip+0.2em}r@{\hskip+0.2em}c@{\hskip+0.2em}rr@{\hskip+0.2em}c@{\hskip+0.2em}r@{\hskip+0.2em}c@{\hskip+0.2em}rr@{\hskip+0.2em}c@{\hskip+0.2em}r@{\hskip+0.2em}c@{\hskip+0.2em}rr@{\hskip+0.2em}c@{\hskip+0.2em}r@{\hskip+0.2em}c@{\hskip+0.2em}rr@{\hskip+0.2em}c@{\hskip+0.2em}r@{\hskip+0.2em}c@{\hskip+0.2em}r}
\toprule&\multicolumn{25}{c}{$\text{$y$}$}\\
$\text{$p_{\text{T}}$ [\gevc]}$ & \multicolumn{5}{c}{$[2.0,2.5]$} & \multicolumn{5}{c}{$[2.5,3.0]$} & \multicolumn{5}{c}{$[3.0,3.5]$} & \multicolumn{5}{c}{$[3.5,4.0]$} & \multicolumn{5}{c}{$[4.0,4.5]$} \\
\midrule$[1,2]$ & \multicolumn{5}{c}{ } & $106$ & $^+_-$ & $^{5}_{5}$&$^+_-$ & $^{11}_{\phantom{0}9}$ & $88.4$ & $^+_-$ & $^{2.0}_{2.0}$&$^+_-$ & $^{7.1}_{8.3}$ & $68.7$ & $^+_-$ & $^{1.9}_{1.9}$&$^+_-$ & $^{5.9}_{6.3}$ & $45.0$ & $^+_-$ & $^{2.7}_{2.7}$&$^+_-$ & $^{3.9}_{4.4}$ \\
$[2,3]$ & $68.8$ & $^+_-$ & $^{6.8}_{6.8}$&$^+_-$ & $^{8.1}_{8.5}$ & $63.0$ & $^+_-$ & $^{1.5}_{1.5}$&$^+_-$ & $^{4.2}_{5.1}$ & $50.0$ & $^+_-$ & $^{0.8}_{0.8}$&$^+_-$ & $^{3.9}_{4.4}$ & $36.1$ & $^+_-$ & $^{0.8}_{0.8}$&$^+_-$ & $^{3.0}_{3.3}$ & $24.8$ & $^+_-$ & $^{1.2}_{1.2}$&$^+_-$ & $^{2.9}_{2.7}$ \\
$[3,4]$ & $34.9$ & $^+_-$ & $^{1.7}_{1.7}$&$^+_-$ & $^{2.8}_{3.0}$ & $28.6$ & $^+_-$ & $^{0.6}_{0.6}$&$^+_-$ & $^{2.1}_{2.1}$ & $23.4$ & $^+_-$ & $^{0.4}_{0.4}$&$^+_-$ & $^{2.0}_{1.8}$ & $15.8$ & $^+_-$ & $^{0.4}_{0.4}$&$^+_-$ & $^{1.3}_{1.4}$ & $9.46$ & $^+_-$ & $^{0.59}_{0.59}$&$^+_-$ & $^{0.72}_{0.79}$ \\
$[4,5]$ & $15.5$ & $^+_-$ & $^{0.7}_{0.7}$&$^+_-$ & $^{1.2}_{1.3}$ & $12.76$ & $^+_-$ & $^{0.29}_{0.29}$&$^+_-$ & $^{0.86}_{0.98}$ & $10.17$ & $^+_-$ & $^{0.23}_{0.23}$&$^+_-$ & $^{0.92}_{0.71}$ & $7.53$ & $^+_-$ & $^{0.24}_{0.24}$&$^+_-$ & $^{0.71}_{0.70}$ & $3.61$ & $^+_-$ & $^{0.40}_{0.40}$&$^+_-$ & $^{0.82}_{0.55}$ \\
$[5,6]$ & $7.30$ & $^+_-$ & $^{0.35}_{0.35}$&$^+_-$ & $^{0.55}_{0.60}$ & $5.63$ & $^+_-$ & $^{0.16}_{0.16}$&$^+_-$ & $^{0.38}_{0.43}$ & $4.40$ & $^+_-$ & $^{0.14}_{0.14}$&$^+_-$ & $^{0.38}_{0.32}$ & $3.07$ & $^+_-$ & $^{0.15}_{0.15}$&$^+_-$ & $^{0.28}_{0.22}$ & $1.93$ & $^+_-$ & $^{0.06}_{0.06}$&$^+_-$ & $^{0.31}_{0.25}$ \\
$[6,7]$ & $4.09$ & $^+_-$ & $^{0.22}_{0.22}$&$^+_-$ & $^{0.30}_{0.34}$ & $2.87$ & $^+_-$ & $^{0.11}_{0.11}$&$^+_-$ & $^{0.20}_{0.22}$ & $2.14$ & $^+_-$ & $^{0.09}_{0.09}$&$^+_-$ & $^{0.18}_{0.16}$ & $1.43$ & $^+_-$ & $^{0.10}_{0.10}$&$^+_-$ & $^{0.12}_{0.12}$ & \multicolumn{5}{c}{ } \\
$[7,8]$ & $1.86$ & $^+_-$ & $^{0.13}_{0.13}$&$^+_-$ & $^{0.14}_{0.16}$ & $1.57$ & $^+_-$ & $^{0.08}_{0.08}$&$^+_-$ & $^{0.11}_{0.12}$ & $1.15$ & $^+_-$ & $^{0.06}_{0.06}$&$^+_-$ & $^{0.10}_{0.09}$ & $0.690$ & $^+_-$ & $^{0.081}_{0.081}$&$^+_-$ & $^{0.081}_{0.071}$ & \multicolumn{5}{c}{ } \\
$[8,9]$ & $0.984$ & $^+_-$ & $^{0.085}_{0.085}$&$^+_-$ & $^{0.079}_{0.090}$ & $0.866$ & $^+_-$ & $^{0.054}_{0.054}$&$^+_-$ & $^{0.063}_{0.069}$ & $0.622$ & $^+_-$ & $^{0.047}_{0.047}$&$^+_-$ & $^{0.060}_{0.046}$ & $0.40$ & $^+_-$ & $^{0.07}_{0.07}$&$^+_-$ & $^{0.11}_{0.07}$ & \multicolumn{5}{c}{ } \\
$[9,10]$ & $0.758$ & $^+_-$ & $^{0.073}_{0.073}$&$^+_-$ & $^{0.067}_{0.075}$ & $0.488$ & $^+_-$ & $^{0.040}_{0.040}$&$^+_-$ & $^{0.039}_{0.042}$ & $0.470$ & $^+_-$ & $^{0.044}_{0.044}$&$^+_-$ & $^{0.064}_{0.050}$ & \multicolumn{5}{c}{ } & \multicolumn{5}{c}{ } \\
\bottomrule\end{tabular}

}{Differential production cross-sections, $\text{d}^2\sigma/(\text{d}\pT\,\text{d}y)$, in 
$\si{\micro\barn}/(\si{\gevc})$ for prompt $\Dstp + \Dstm$ mesons in bins 
of \pTy.
The first uncertainty is statistical, and the second is the total 
systematic.}{table:differential:Dstp}

\vspace*{\fill}

\clearpage
\section{Cross-section ratios at different energies}
\label{app:ratios}

Tables~\ref{table:differential_ratio:D0}--\ref{table:differential_ratio:Dstp} give the numerical results of the cross-section ratios between $\sqrt{s}=13$ and \SI{5}{\TeV}.

\vspace*{\fill}
\hvFloat[
  nonFloat=true,
  objectAngle=90,
  capPos=l,
  capAngle=90,
  capWidth=h]{table}{
    \centering
    \renewcommand{\arraystretch}{1.3}
\begin{tabular}{c|r@{\hskip+0.2em}c@{\hskip+0.2em}r@{\hskip+0.2em}c@{\hskip+0.2em}rr@{\hskip+0.2em}c@{\hskip+0.2em}r@{\hskip+0.2em}c@{\hskip+0.2em}rr@{\hskip+0.2em}c@{\hskip+0.2em}r@{\hskip+0.2em}c@{\hskip+0.2em}rr@{\hskip+0.2em}c@{\hskip+0.2em}r@{\hskip+0.2em}c@{\hskip+0.2em}rr@{\hskip+0.2em}c@{\hskip+0.2em}r@{\hskip+0.2em}c@{\hskip+0.2em}r}
\toprule&\multicolumn{25}{c}{$\text{$y$}$}\\
$\text{$p_{\text{T}}$ [\gevc]}$ & \multicolumn{5}{c}{$[2.0,2.5]$} & \multicolumn{5}{c}{$[2.5,3.0]$} & \multicolumn{5}{c}{$[3.0,3.5]$} & \multicolumn{5}{c}{$[3.5,4.0]$} & \multicolumn{5}{c}{$[4.0,4.5]$} \\
\midrule$[0,1]$ & $1.49$ & $^+_-$ & $^{0.03}_{0.03}$&$^+_-$ & $^{0.23}_{0.20}$ & $1.61$ & $^+_-$ & $^{0.01}_{0.01}$&$^+_-$ & $^{0.13}_{0.15}$ & $1.72$ & $^+_-$ & $^{0.02}_{0.02}$&$^+_-$ & $^{0.13}_{0.10}$ & $1.88$ & $^+_-$ & $^{0.02}_{0.02}$&$^+_-$ & $^{0.12}_{0.11}$ & $2.12$ & $^+_-$ & $^{0.06}_{0.05}$&$^+_-$ & $^{0.20}_{0.19}$ \\
$[1,2]$ & $1.61$ & $^+_-$ & $^{0.02}_{0.02}$&$^+_-$ & $^{0.20}_{0.18}$ & $1.72$ & $^+_-$ & $^{0.01}_{0.01}$&$^+_-$ & $^{0.13}_{0.12}$ & $1.88$ & $^+_-$ & $^{0.01}_{0.01}$&$^+_-$ & $^{0.12}_{0.10}$ & $2.04$ & $^+_-$ & $^{0.02}_{0.02}$&$^+_-$ & $^{0.12}_{0.11}$ & $2.31$ & $^+_-$ & $^{0.04}_{0.04}$&$^+_-$ & $^{0.17}_{0.14}$ \\
$[2,3]$ & $1.78$ & $^+_-$ & $^{0.02}_{0.02}$&$^+_-$ & $^{0.15}_{0.17}$ & $1.94$ & $^+_-$ & $^{0.01}_{0.01}$&$^+_-$ & $^{0.14}_{0.10}$ & $2.20$ & $^+_-$ & $^{0.02}_{0.02}$&$^+_-$ & $^{0.13}_{0.11}$ & $2.35$ & $^+_-$ & $^{0.02}_{0.02}$&$^+_-$ & $^{0.15}_{0.12}$ & $2.70$ & $^+_-$ & $^{0.05}_{0.05}$&$^+_-$ & $^{0.17}_{0.17}$ \\
$[3,4]$ & $2.03$ & $^+_-$ & $^{0.02}_{0.02}$&$^+_-$ & $^{0.14}_{0.16}$ & $2.20$ & $^+_-$ & $^{0.02}_{0.02}$&$^+_-$ & $^{0.14}_{0.11}$ & $2.35$ & $^+_-$ & $^{0.02}_{0.02}$&$^+_-$ & $^{0.14}_{0.12}$ & $2.69$ & $^+_-$ & $^{0.03}_{0.03}$&$^+_-$ & $^{0.19}_{0.16}$ & $3.33$ & $^+_-$ & $^{0.09}_{0.09}$&$^+_-$ & $^{0.32}_{0.31}$ \\
$[4,5]$ & $2.19$ & $^+_-$ & $^{0.03}_{0.03}$&$^+_-$ & $^{0.17}_{0.13}$ & $2.48$ & $^+_-$ & $^{0.03}_{0.02}$&$^+_-$ & $^{0.16}_{0.12}$ & $2.65$ & $^+_-$ & $^{0.03}_{0.03}$&$^+_-$ & $^{0.16}_{0.14}$ & $3.07$ & $^+_-$ & $^{0.05}_{0.05}$&$^+_-$ & $^{0.17}_{0.19}$ & $3.04$ & $^+_-$ & $^{0.16}_{0.15}$&$^+_-$ & $^{0.63}_{0.55}$ \\
$[5,6]$ & $2.57$ & $^+_-$ & $^{0.05}_{0.04}$&$^+_-$ & $^{0.21}_{0.15}$ & $2.66$ & $^+_-$ & $^{0.04}_{0.04}$&$^+_-$ & $^{0.17}_{0.13}$ & $2.90$ & $^+_-$ & $^{0.05}_{0.04}$&$^+_-$ & $^{0.16}_{0.18}$ & $3.15$ & $^+_-$ & $^{0.08}_{0.08}$&$^+_-$ & $^{0.22}_{0.21}$ & $5.6$ & $^+_-$ & $^{1.0}_{0.7}$&$^+_-$ & $^{2.2}_{1.6}$ \\
$[6,7]$ & $2.55$ & $^+_-$ & $^{0.06}_{0.05}$&$^+_-$ & $^{0.19}_{0.16}$ & $2.95$ & $^+_-$ & $^{0.06}_{0.06}$&$^+_-$ & $^{0.19}_{0.16}$ & $3.27$ & $^+_-$ & $^{0.07}_{0.07}$&$^+_-$ & $^{0.18}_{0.21}$ & $3.85$ & $^+_-$ & $^{0.16}_{0.15}$&$^+_-$ & $^{0.44}_{0.44}$ & \multicolumn{5}{c}{ } \\
$[7,8]$ & $2.67$ & $^+_-$ & $^{0.08}_{0.08}$&$^+_-$ & $^{0.22}_{0.17}$ & $3.00$ & $^+_-$ & $^{0.08}_{0.07}$&$^+_-$ & $^{0.20}_{0.18}$ & $3.44$ & $^+_-$ & $^{0.11}_{0.10}$&$^+_-$ & $^{0.21}_{0.26}$ & $3.93$ & $^+_-$ & $^{0.32}_{0.28}$&$^+_-$ & $^{0.91}_{0.73}$ & \multicolumn{5}{c}{ } \\
\bottomrule\end{tabular}

}{The ratios of differential production cross-sections, $R_{13/5}$, for 
prompt $\Dz + \Dzbar$ mesons in bins of \pTy.
The first uncertainty is statistical, and the second is the total 
systematic.}{table:differential_ratio:D0}

\vspace*{\fill}

\newpage
\vspace*{\fill}
\hvFloat[
  nonFloat=true,
  objectAngle=90,
  capPos=l,
  capAngle=90,
  capWidth=h]{table}{
    \centering
    \renewcommand{\arraystretch}{1.3}
\begin{tabular}{c|r@{\hskip+0.2em}c@{\hskip+0.2em}r@{\hskip+0.2em}c@{\hskip+0.2em}rr@{\hskip+0.2em}c@{\hskip+0.2em}r@{\hskip+0.2em}c@{\hskip+0.2em}rr@{\hskip+0.2em}c@{\hskip+0.2em}r@{\hskip+0.2em}c@{\hskip+0.2em}rr@{\hskip+0.2em}c@{\hskip+0.2em}r@{\hskip+0.2em}c@{\hskip+0.2em}rr@{\hskip+0.2em}c@{\hskip+0.2em}r@{\hskip+0.2em}c@{\hskip+0.2em}r}
\toprule&\multicolumn{25}{c}{$\text{$y$}$}\\
$\text{$p_{\text{T}}$ [\gevc]}$ & \multicolumn{5}{c}{$[2.0,2.5]$} & \multicolumn{5}{c}{$[2.5,3.0]$} & \multicolumn{5}{c}{$[3.0,3.5]$} & \multicolumn{5}{c}{$[3.5,4.0]$} & \multicolumn{5}{c}{$[4.0,4.5]$} \\
\midrule$[0,1]$ & \multicolumn{5}{c}{ } & $2.10$ & $^+_-$ & $^{0.23}_{0.19}$&$^+_-$ & $^{0.69}_{0.44}$ & $1.83$ & $^+_-$ & $^{0.12}_{0.11}$&$^+_-$ & $^{0.32}_{0.28}$ & $1.66$ & $^+_-$ & $^{0.13}_{0.12}$&$^+_-$ & $^{0.28}_{0.25}$ & \multicolumn{5}{c}{ } \\
$[1,2]$ & $1.80$ & $^+_-$ & $^{0.07}_{0.07}$&$^+_-$ & $^{0.44}_{0.38}$ & $1.69$ & $^+_-$ & $^{0.02}_{0.02}$&$^+_-$ & $^{0.31}_{0.20}$ & $1.88$ & $^+_-$ & $^{0.02}_{0.02}$&$^+_-$ & $^{0.18}_{0.19}$ & $2.08$ & $^+_-$ & $^{0.03}_{0.03}$&$^+_-$ & $^{0.17}_{0.17}$ & $2.25$ & $^+_-$ & $^{0.06}_{0.06}$&$^+_-$ & $^{0.24}_{0.22}$ \\
$[2,3]$ & $1.72$ & $^+_-$ & $^{0.03}_{0.03}$&$^+_-$ & $^{0.30}_{0.27}$ & $1.97$ & $^+_-$ & $^{0.01}_{0.01}$&$^+_-$ & $^{0.25}_{0.18}$ & $2.09$ & $^+_-$ & $^{0.01}_{0.01}$&$^+_-$ & $^{0.15}_{0.17}$ & $2.25$ & $^+_-$ & $^{0.02}_{0.02}$&$^+_-$ & $^{0.15}_{0.16}$ & $2.64$ & $^+_-$ & $^{0.05}_{0.04}$&$^+_-$ & $^{0.22}_{0.19}$ \\
$[3,4]$ & $1.87$ & $^+_-$ & $^{0.03}_{0.03}$&$^+_-$ & $^{0.28}_{0.26}$ & $2.21$ & $^+_-$ & $^{0.02}_{0.02}$&$^+_-$ & $^{0.22}_{0.17}$ & $2.41$ & $^+_-$ & $^{0.02}_{0.02}$&$^+_-$ & $^{0.18}_{0.15}$ & $2.60$ & $^+_-$ & $^{0.03}_{0.03}$&$^+_-$ & $^{0.18}_{0.15}$ & $2.81$ & $^+_-$ & $^{0.06}_{0.05}$&$^+_-$ & $^{0.21}_{0.21}$ \\
$[4,5]$ & $2.12$ & $^+_-$ & $^{0.03}_{0.03}$&$^+_-$ & $^{0.27}_{0.26}$ & $2.48$ & $^+_-$ & $^{0.02}_{0.02}$&$^+_-$ & $^{0.22}_{0.17}$ & $2.55$ & $^+_-$ & $^{0.03}_{0.03}$&$^+_-$ & $^{0.19}_{0.14}$ & $2.96$ & $^+_-$ & $^{0.04}_{0.04}$&$^+_-$ & $^{0.21}_{0.17}$ & $3.59$ & $^+_-$ & $^{0.11}_{0.10}$&$^+_-$ & $^{0.23}_{0.29}$ \\
$[5,6]$ & $2.40$ & $^+_-$ & $^{0.04}_{0.04}$&$^+_-$ & $^{0.25}_{0.27}$ & $2.68$ & $^+_-$ & $^{0.03}_{0.03}$&$^+_-$ & $^{0.22}_{0.17}$ & $2.95$ & $^+_-$ & $^{0.04}_{0.04}$&$^+_-$ & $^{0.21}_{0.16}$ & $3.25$ & $^+_-$ & $^{0.06}_{0.06}$&$^+_-$ & $^{0.24}_{0.20}$ & $4.51$ & $^+_-$ & $^{0.24}_{0.22}$&$^+_-$ & $^{0.47}_{0.48}$ \\
$[6,7]$ & $2.54$ & $^+_-$ & $^{0.06}_{0.06}$&$^+_-$ & $^{0.25}_{0.25}$ & $2.80$ & $^+_-$ & $^{0.05}_{0.05}$&$^+_-$ & $^{0.22}_{0.17}$ & $3.01$ & $^+_-$ & $^{0.06}_{0.06}$&$^+_-$ & $^{0.22}_{0.17}$ & $3.92$ & $^+_-$ & $^{0.11}_{0.11}$&$^+_-$ & $^{0.25}_{0.29}$ & $4.27$ & $^+_-$ & $^{0.48}_{0.39}$&$^+_-$ & $^{0.61}_{0.50}$ \\
$[7,8]$ & $2.72$ & $^+_-$ & $^{0.08}_{0.07}$&$^+_-$ & $^{0.28}_{0.24}$ & $3.10$ & $^+_-$ & $^{0.07}_{0.07}$&$^+_-$ & $^{0.25}_{0.19}$ & $3.39$ & $^+_-$ & $^{0.09}_{0.08}$&$^+_-$ & $^{0.27}_{0.20}$ & $3.70$ & $^+_-$ & $^{0.16}_{0.14}$&$^+_-$ & $^{0.29}_{0.34}$ & \multicolumn{5}{c}{ } \\
\bottomrule\end{tabular}

}{The ratios of differential production cross-sections, $R_{13/5}$, for 
prompt $\Dp + \Dm$ mesons in bins of \pTy.
The first uncertainty is statistical, and the second is the total 
systematic.}{table:differential_ratio:Dp}

\vspace*{\fill}

\newpage
\vspace*{\fill}
\hvFloat[
  nonFloat=true,
  objectAngle=90,
  capPos=l,
  capAngle=90,
  capWidth=h]{table}{
    \centering
    \renewcommand{\arraystretch}{1.3}
\begin{tabular}{c|r@{\hskip+0.2em}c@{\hskip+0.2em}r@{\hskip+0.2em}c@{\hskip+0.2em}rr@{\hskip+0.2em}c@{\hskip+0.2em}r@{\hskip+0.2em}c@{\hskip+0.2em}rr@{\hskip+0.2em}c@{\hskip+0.2em}r@{\hskip+0.2em}c@{\hskip+0.2em}rr@{\hskip+0.2em}c@{\hskip+0.2em}r@{\hskip+0.2em}c@{\hskip+0.2em}rr@{\hskip+0.2em}c@{\hskip+0.2em}r@{\hskip+0.2em}c@{\hskip+0.2em}r}
\toprule&\multicolumn{25}{c}{$\text{$y$}$}\\
$\text{$p_{\text{T}}$ [\gevc]}$ & \multicolumn{5}{c}{$[2.0,2.5]$} & \multicolumn{5}{c}{$[2.5,3.0]$} & \multicolumn{5}{c}{$[3.0,3.5]$} & \multicolumn{5}{c}{$[3.5,4.0]$} & \multicolumn{5}{c}{$[4.0,4.5]$} \\
\midrule$[1,2]$ & $1.30$ & $^+_-$ & $^{0.28}_{0.21}$&$^+_-$ & $^{0.54}_{0.37}$ & $1.48$ & $^+_-$ & $^{0.12}_{0.10}$&$^+_-$ & $^{0.42}_{0.24}$ & $1.83$ & $^+_-$ & $^{0.14}_{0.12}$&$^+_-$ & $^{0.26}_{0.24}$ & $2.30$ & $^+_-$ & $^{0.26}_{0.22}$&$^+_-$ & $^{0.41}_{0.31}$ & \multicolumn{5}{c}{ } \\
$[2,3]$ & $1.58$ & $^+_-$ & $^{0.10}_{0.09}$&$^+_-$ & $^{0.41}_{0.32}$ & $1.76$ & $^+_-$ & $^{0.05}_{0.05}$&$^+_-$ & $^{0.24}_{0.17}$ & $2.10$ & $^+_-$ & $^{0.07}_{0.06}$&$^+_-$ & $^{0.19}_{0.17}$ & $2.77$ & $^+_-$ & $^{0.13}_{0.13}$&$^+_-$ & $^{0.22}_{0.22}$ & $2.71$ & $^+_-$ & $^{0.36}_{0.28}$&$^+_-$ & $^{0.45}_{0.38}$ \\
$[3,4]$ & $1.94$ & $^+_-$ & $^{0.10}_{0.09}$&$^+_-$ & $^{0.35}_{0.30}$ & $2.06$ & $^+_-$ & $^{0.06}_{0.06}$&$^+_-$ & $^{0.21}_{0.19}$ & $2.44$ & $^+_-$ & $^{0.08}_{0.08}$&$^+_-$ & $^{0.20}_{0.16}$ & $2.58$ & $^+_-$ & $^{0.12}_{0.11}$&$^+_-$ & $^{0.20}_{0.18}$ & $2.75$ & $^+_-$ & $^{0.30}_{0.25}$&$^+_-$ & $^{0.32}_{0.27}$ \\
$[4,5]$ & $2.09$ & $^+_-$ & $^{0.12}_{0.11}$&$^+_-$ & $^{0.33}_{0.26}$ & $2.35$ & $^+_-$ & $^{0.09}_{0.08}$&$^+_-$ & $^{0.23}_{0.17}$ & $2.38$ & $^+_-$ & $^{0.10}_{0.09}$&$^+_-$ & $^{0.18}_{0.15}$ & $3.42$ & $^+_-$ & $^{0.21}_{0.19}$&$^+_-$ & $^{0.25}_{0.26}$ & $3.28$ & $^+_-$ & $^{0.48}_{0.38}$&$^+_-$ & $^{0.41}_{0.33}$ \\
$[5,6]$ & $2.82$ & $^+_-$ & $^{0.20}_{0.18}$&$^+_-$ & $^{0.35}_{0.33}$ & $3.26$ & $^+_-$ & $^{0.18}_{0.16}$&$^+_-$ & $^{0.33}_{0.20}$ & $2.56$ & $^+_-$ & $^{0.14}_{0.13}$&$^+_-$ & $^{0.20}_{0.18}$ & $3.39$ & $^+_-$ & $^{0.29}_{0.25}$&$^+_-$ & $^{0.38}_{0.26}$ & $4.0$ & $^+_-$ & $^{1.0}_{0.7}$&$^+_-$ & $^{0.7}_{0.6}$ \\
$[6,7]$ & $2.60$ & $^+_-$ & $^{0.23}_{0.19}$&$^+_-$ & $^{0.32}_{0.28}$ & $2.94$ & $^+_-$ & $^{0.20}_{0.18}$&$^+_-$ & $^{0.28}_{0.22}$ & $3.59$ & $^+_-$ & $^{0.30}_{0.26}$&$^+_-$ & $^{0.29}_{0.28}$ & $3.58$ & $^+_-$ & $^{0.47}_{0.37}$&$^+_-$ & $^{0.49}_{0.34}$ & \multicolumn{5}{c}{ } \\
$[7,8]$ & $2.50$ & $^+_-$ & $^{0.27}_{0.22}$&$^+_-$ & $^{0.37}_{0.22}$ & $2.79$ & $^+_-$ & $^{0.26}_{0.22}$&$^+_-$ & $^{0.24}_{0.25}$ & $3.47$ & $^+_-$ & $^{0.38}_{0.32}$&$^+_-$ & $^{0.37}_{0.29}$ & $4.12$ & $^+_-$ & $^{0.72}_{0.55}$&$^+_-$ & $^{0.76}_{0.56}$ & \multicolumn{5}{c}{ } \\
\bottomrule\end{tabular}

}{The ratios of differential production cross-sections, $R_{13/5}$, for 
prompt $\Dsp + \Dsm$ mesons in bins of \pTy.
The first uncertainty is statistical, and the second is the total 
systematic.}{table:differential_ratio:Dsp}

\vspace*{\fill}

\newpage
\vspace*{\fill}
\hvFloat[
  nonFloat=true,
  objectAngle=90,
  capPos=l,
  capAngle=90,
  capWidth=h]{table}{
    \centering
    \renewcommand{\arraystretch}{1.3}
\begin{tabular}{c|r@{\hskip+0.2em}c@{\hskip+0.2em}r@{\hskip+0.2em}c@{\hskip+0.2em}rr@{\hskip+0.2em}c@{\hskip+0.2em}r@{\hskip+0.2em}c@{\hskip+0.2em}rr@{\hskip+0.2em}c@{\hskip+0.2em}r@{\hskip+0.2em}c@{\hskip+0.2em}rr@{\hskip+0.2em}c@{\hskip+0.2em}r@{\hskip+0.2em}c@{\hskip+0.2em}rr@{\hskip+0.2em}c@{\hskip+0.2em}r@{\hskip+0.2em}c@{\hskip+0.2em}r}
\toprule&\multicolumn{25}{c}{$\text{$y$}$}\\
$\text{$p_{\text{T}}$ [\gevc]}$ & \multicolumn{5}{c}{$[2.0,2.5]$} & \multicolumn{5}{c}{$[2.5,3.0]$} & \multicolumn{5}{c}{$[3.0,3.5]$} & \multicolumn{5}{c}{$[3.5,4.0]$} & \multicolumn{5}{c}{$[4.0,4.5]$} \\
\midrule$[1,2]$ & \multicolumn{5}{c}{ } & $1.33$ & $^+_-$ & $^{0.07}_{0.06}$&$^+_-$ & $^{0.15}_{0.15}$ & $1.70$ & $^+_-$ & $^{0.04}_{0.04}$&$^+_-$ & $^{0.19}_{0.17}$ & $1.81$ & $^+_-$ & $^{0.06}_{0.05}$&$^+_-$ & $^{0.21}_{0.17}$ & $2.52$ & $^+_-$ & $^{0.17}_{0.15}$&$^+_-$ & $^{0.40}_{0.25}$ \\
$[2,3]$ & $1.51$ & $^+_-$ & $^{0.18}_{0.16}$&$^+_-$ & $^{0.32}_{0.30}$ & $1.70$ & $^+_-$ & $^{0.04}_{0.04}$&$^+_-$ & $^{0.20}_{0.14}$ & $2.01$ & $^+_-$ & $^{0.04}_{0.03}$&$^+_-$ & $^{0.21}_{0.18}$ & $2.29$ & $^+_-$ & $^{0.06}_{0.05}$&$^+_-$ & $^{0.27}_{0.19}$ & $2.47$ & $^+_-$ & $^{0.13}_{0.12}$&$^+_-$ & $^{0.33}_{0.29}$ \\
$[3,4]$ & $1.60$ & $^+_-$ & $^{0.09}_{0.08}$&$^+_-$ & $^{0.20}_{0.21}$ & $1.99$ & $^+_-$ & $^{0.04}_{0.04}$&$^+_-$ & $^{0.18}_{0.19}$ & $2.08$ & $^+_-$ & $^{0.04}_{0.04}$&$^+_-$ & $^{0.21}_{0.17}$ & $2.49$ & $^+_-$ & $^{0.07}_{0.07}$&$^+_-$ & $^{0.28}_{0.21}$ & $3.00$ & $^+_-$ & $^{0.20}_{0.18}$&$^+_-$ & $^{0.35}_{0.26}$ \\
$[4,5]$ & $1.82$ & $^+_-$ & $^{0.09}_{0.08}$&$^+_-$ & $^{0.21}_{0.21}$ & $2.18$ & $^+_-$ & $^{0.05}_{0.05}$&$^+_-$ & $^{0.20}_{0.20}$ & $2.25$ & $^+_-$ & $^{0.05}_{0.05}$&$^+_-$ & $^{0.20}_{0.18}$ & $2.54$ & $^+_-$ & $^{0.09}_{0.08}$&$^+_-$ & $^{0.30}_{0.24}$ & $3.24$ & $^+_-$ & $^{0.42}_{0.33}$&$^+_-$ & $^{0.77}_{0.62}$ \\
$[5,6]$ & $1.96$ & $^+_-$ & $^{0.10}_{0.09}$&$^+_-$ & $^{0.21}_{0.22}$ & $2.43$ & $^+_-$ & $^{0.08}_{0.07}$&$^+_-$ & $^{0.22}_{0.22}$ & $2.72$ & $^+_-$ & $^{0.09}_{0.09}$&$^+_-$ & $^{0.25}_{0.22}$ & $3.22$ & $^+_-$ & $^{0.17}_{0.15}$&$^+_-$ & $^{0.33}_{0.25}$ & $2.61$ & $^+_-$ & $^{0.16}_{0.16}$&$^+_-$ & $^{0.65}_{0.50}$ \\
$[6,7]$ & $2.07$ & $^+_-$ & $^{0.13}_{0.11}$&$^+_-$ & $^{0.22}_{0.23}$ & $2.72$ & $^+_-$ & $^{0.11}_{0.10}$&$^+_-$ & $^{0.25}_{0.24}$ & $2.95$ & $^+_-$ & $^{0.13}_{0.12}$&$^+_-$ & $^{0.26}_{0.25}$ & $3.10$ & $^+_-$ & $^{0.25}_{0.22}$&$^+_-$ & $^{0.32}_{0.26}$ & \multicolumn{5}{c}{ } \\
$[7,8]$ & $2.67$ & $^+_-$ & $^{0.20}_{0.18}$&$^+_-$ & $^{0.30}_{0.29}$ & $2.71$ & $^+_-$ & $^{0.14}_{0.13}$&$^+_-$ & $^{0.27}_{0.24}$ & $3.23$ & $^+_-$ & $^{0.20}_{0.18}$&$^+_-$ & $^{0.31}_{0.27}$ & $3.89$ & $^+_-$ & $^{0.53}_{0.42}$&$^+_-$ & $^{0.61}_{0.47}$ & \multicolumn{5}{c}{ } \\
\bottomrule\end{tabular}

}{The ratios of differential production cross-sections, $R_{13/5}$, for 
prompt $\Dstp + \Dstm$ mesons in bins of \pTy.
The first uncertainty is statistical, and the second is the total 
systematic.}{table:differential_ratio:Dstp}

\vspace*{\fill}

\clearpage

\clearpage
\section{Cross-section ratios for different mesons}
\label{app:MesonRatio}

The numerical values of the cross-section ratios between mesons, described in 
Sec.~\ref{sec:ratios}, are given in \pTy bins in
Tables~\ref{table:meson_ratio:DpD0}--\ref{table:meson_ratio:DsDst}.

\vspace*{\fill}
\hvFloat[
  nonFloat=true,
  objectAngle=90,
  capPos=l,
  capAngle=90,
  capWidth=h]{table}{
    \centering
    \renewcommand{\arraystretch}{1.3}
\begin{tabular}{c|r@{\hskip+0.2em}c@{\hskip+0.2em}r@{\hskip+0.2em}c@{\hskip+0.2em}rr@{\hskip+0.2em}c@{\hskip+0.2em}r@{\hskip+0.2em}c@{\hskip+0.2em}rr@{\hskip+0.2em}c@{\hskip+0.2em}r@{\hskip+0.2em}c@{\hskip+0.2em}rr@{\hskip+0.2em}c@{\hskip+0.2em}r@{\hskip+0.2em}c@{\hskip+0.2em}rr@{\hskip+0.2em}c@{\hskip+0.2em}r@{\hskip+0.2em}c@{\hskip+0.2em}r}
\toprule&\multicolumn{25}{c}{$\text{$y$}$}\\
$\text{$p_{\text{T}}$ [\gevc]}$ & \multicolumn{5}{c}{$[2.0,2.5]$} & \multicolumn{5}{c}{$[2.5,3.0]$} & \multicolumn{5}{c}{$[3.0,3.5]$} & \multicolumn{5}{c}{$[3.5,4.0]$} & \multicolumn{5}{c}{$[4.0,4.5]$} \\
\midrule$[0,1]$ & \multicolumn{5}{c}{ } & $101$ & $^+_-$ & $^{9}_{9}$&$^+_-$ & $^{12}_{14}$ & $83.9$ & $^+_-$ & $^{5.0}_{5.0}$&$^+_-$ & $^{8.4}_{6.6}$ & $101.0$ & $^+_-$ & $^{7.2}_{7.1}$&$^+_-$ & $^{9.9}_{8.9}$ & \multicolumn{5}{c}{ } \\
$[1,2]$ & $84$ & $^+_-$ & $^{3}_{3}$&$^+_-$ & $^{11}_{10}$ & $88.2$ & $^+_-$ & $^{1.1}_{1.1}$&$^+_-$ & $^{3.6}_{4.1}$ & $94.9$ & $^+_-$ & $^{1.0}_{1.0}$&$^+_-$ & $^{5.4}_{3.2}$ & $89.4$ & $^+_-$ & $^{1.2}_{1.2}$&$^+_-$ & $^{4.2}_{3.0}$ & $98.1$ & $^+_-$ & $^{2.9}_{2.8}$&$^+_-$ & $^{7.2}_{6.3}$ \\
$[2,3]$ & $91.9$ & $^+_-$ & $^{1.7}_{1.6}$&$^+_-$ & $^{7.3}_{6.7}$ & $94.0$ & $^+_-$ & $^{0.8}_{0.8}$&$^+_-$ & $^{2.8}_{3.7}$ & $98.7$ & $^+_-$ & $^{0.9}_{0.9}$&$^+_-$ & $^{3.9}_{2.8}$ & $95.9$ & $^+_-$ & $^{1.1}_{1.1}$&$^+_-$ & $^{4.0}_{3.1}$ & $99.4$ & $^+_-$ & $^{2.4}_{2.3}$&$^+_-$ & $^{5.2}_{5.1}$ \\
$[3,4]$ & $98.8$ & $^+_-$ & $^{1.7}_{1.7}$&$^+_-$ & $^{4.8}_{5.8}$ & $99.0$ & $^+_-$ & $^{1.0}_{1.0}$&$^+_-$ & $^{2.0}_{3.9}$ & $101.0$ & $^+_-$ & $^{1.1}_{1.1}$&$^+_-$ & $^{2.9}_{3.5}$ & $99.0$ & $^+_-$ & $^{1.4}_{1.4}$&$^+_-$ & $^{3.4}_{3.9}$ & $111.8$ & $^+_-$ & $^{3.6}_{3.6}$&$^+_-$ & $^{8.1}_{8.1}$ \\
$[4,5]$ & $106.2$ & $^+_-$ & $^{2.0}_{2.0}$&$^+_-$ & $^{4.4}_{5.2}$ & $103.1$ & $^+_-$ & $^{1.4}_{1.4}$&$^+_-$ & $^{2.5}_{3.7}$ & $104.9$ & $^+_-$ & $^{1.5}_{1.5}$&$^+_-$ & $^{3.3}_{3.8}$ & $102.6$ & $^+_-$ & $^{2.1}_{2.0}$&$^+_-$ & $^{3.4}_{3.8}$ & $106$ & $^+_-$ & $^{6}_{6}$&$^+_-$ & $^{16}_{15}$ \\
$[5,6]$ & $107.2$ & $^+_-$ & $^{2.6}_{2.5}$&$^+_-$ & $^{4.4}_{4.7}$ & $107.5$ & $^+_-$ & $^{1.9}_{1.9}$&$^+_-$ & $^{3.5}_{3.9}$ & $104.1$ & $^+_-$ & $^{2.1}_{2.1}$&$^+_-$ & $^{3.3}_{4.2}$ & $103.2$ & $^+_-$ & $^{3.1}_{3.0}$&$^+_-$ & $^{5.1}_{5.7}$ & $184$ & $^+_-$ & $^{32}_{24}$&$^+_-$ & $^{58}_{44}$ \\
$[6,7]$ & $106.6$ & $^+_-$ & $^{3.2}_{3.1}$&$^+_-$ & $^{4.5}_{5.1}$ & $106.7$ & $^+_-$ & $^{2.6}_{2.5}$&$^+_-$ & $^{3.8}_{4.3}$ & $111.8$ & $^+_-$ & $^{3.1}_{3.0}$&$^+_-$ & $^{4.3}_{5.0}$ & $110.7$ & $^+_-$ & $^{5.4}_{5.0}$&$^+_-$ & $^{9.7}_{9.8}$ & \multicolumn{5}{c}{ } \\
$[7,8]$ & $113.6$ & $^+_-$ & $^{4.4}_{4.2}$&$^+_-$ & $^{5.5}_{6.4}$ & $99.4$ & $^+_-$ & $^{3.3}_{3.0}$&$^+_-$ & $^{3.4}_{5.1}$ & $106.4$ & $^+_-$ & $^{4.1}_{3.9}$&$^+_-$ & $^{5.2}_{6.1}$ & $130$ & $^+_-$ & $^{11}_{10}$&$^+_-$ & $^{24}_{21}$ & \multicolumn{5}{c}{ } \\
$[8,9]$ & $108.4$ & $^+_-$ & $^{5.4}_{5.1}$&$^+_-$ & $^{6.4}_{6.9}$ & $106.4$ & $^+_-$ & $^{4.7}_{4.3}$&$^+_-$ & $^{4.9}_{6.8}$ & $102.7$ & $^+_-$ & $^{5.7}_{5.5}$&$^+_-$ & $^{8.2}_{7.9}$ & $202$ & $^+_-$ & $^{35}_{28}$&$^+_-$ & $^{53}_{37}$ & \multicolumn{5}{c}{ } \\
$[9,10]$ & $91.3$ & $^+_-$ & $^{5.6}_{5.4}$&$^+_-$ & $^{7.0}_{6.6}$ & $101.9$ & $^+_-$ & $^{5.8}_{5.5}$&$^+_-$ & $^{6.3}_{7.7}$ & $125$ & $^+_-$ & $^{10}_{\phantom{0}9}$&$^+_-$ & $^{17}_{15}$ & \multicolumn{5}{c}{ } & \multicolumn{5}{c}{ } \\
\bottomrule\end{tabular}

}{%
  The ratios of the measurements of cross-section times branching fraction 
  for prompt \Dp and \Dz mesons in bins of \pTy.
  The first uncertainty is statistical, and the second is the total 
  systematic. All values are given in percent.
}{table:meson_ratio:DpD0}
\vspace*{\fill}

\newpage
\vspace*{\fill}
\hvFloat[
  nonFloat=true,
  objectAngle=90,
  capPos=l,
  capAngle=90,
  capWidth=h]{table}{
    \centering
    \renewcommand{\arraystretch}{1.3}
\begin{tabular}{c|r@{\hskip+0.2em}c@{\hskip+0.2em}r@{\hskip+0.2em}c@{\hskip+0.2em}rr@{\hskip+0.2em}c@{\hskip+0.2em}r@{\hskip+0.2em}c@{\hskip+0.2em}rr@{\hskip+0.2em}c@{\hskip+0.2em}r@{\hskip+0.2em}c@{\hskip+0.2em}rr@{\hskip+0.2em}c@{\hskip+0.2em}r@{\hskip+0.2em}c@{\hskip+0.2em}rr@{\hskip+0.2em}c@{\hskip+0.2em}r@{\hskip+0.2em}c@{\hskip+0.2em}r}
\toprule&\multicolumn{25}{c}{$\text{$y$}$}\\
$\text{$p_{\text{T}}$ [\gevc]}$ & \multicolumn{5}{c}{$[2.0,2.5]$} & \multicolumn{5}{c}{$[2.5,3.0]$} & \multicolumn{5}{c}{$[3.0,3.5]$} & \multicolumn{5}{c}{$[3.5,4.0]$} & \multicolumn{5}{c}{$[4.0,4.5]$} \\
\midrule$[1,2]$ & $6.8$ & $^+_-$ & $^{1.0}_{1.0}$&$^+_-$ & $^{1.5}_{1.1}$ & $9.56$ & $^+_-$ & $^{0.57}_{0.58}$&$^+_-$ & $^{0.54}_{0.60}$ & $9.37$ & $^+_-$ & $^{0.58}_{0.59}$&$^+_-$ & $^{0.72}_{0.52}$ & $7.91$ & $^+_-$ & $^{0.72}_{0.73}$&$^+_-$ & $^{0.66}_{0.49}$ & \multicolumn{5}{c}{ } \\
$[2,3]$ & $10.7$ & $^+_-$ & $^{0.6}_{0.6}$&$^+_-$ & $^{1.7}_{1.3}$ & $11.31$ & $^+_-$ & $^{0.32}_{0.32}$&$^+_-$ & $^{0.38}_{0.47}$ & $11.39$ & $^+_-$ & $^{0.34}_{0.35}$&$^+_-$ & $^{0.52}_{0.40}$ & $9.17$ & $^+_-$ & $^{0.41}_{0.41}$&$^+_-$ & $^{0.48}_{0.35}$ & $9.1$ & $^+_-$ & $^{1.0}_{1.0}$&$^+_-$ & $^{1.2}_{0.9}$ \\
$[3,4]$ & $10.80$ & $^+_-$ & $^{0.52}_{0.53}$&$^+_-$ & $^{0.98}_{0.79}$ & $11.76$ & $^+_-$ & $^{0.35}_{0.34}$&$^+_-$ & $^{0.30}_{0.45}$ & $11.10$ & $^+_-$ & $^{0.36}_{0.36}$&$^+_-$ & $^{0.43}_{0.33}$ & $11.53$ & $^+_-$ & $^{0.50}_{0.52}$&$^+_-$ & $^{0.52}_{0.45}$ & $12.0$ & $^+_-$ & $^{1.2}_{1.2}$&$^+_-$ & $^{1.0}_{1.0}$ \\
$[4,5]$ & $11.30$ & $^+_-$ & $^{0.59}_{0.58}$&$^+_-$ & $^{0.67}_{0.77}$ & $12.67$ & $^+_-$ & $^{0.45}_{0.46}$&$^+_-$ & $^{0.32}_{0.51}$ & $12.70$ & $^+_-$ & $^{0.52}_{0.52}$&$^+_-$ & $^{0.49}_{0.45}$ & $10.84$ & $^+_-$ & $^{0.62}_{0.62}$&$^+_-$ & $^{0.51}_{0.40}$ & $13.0$ & $^+_-$ & $^{1.7}_{1.7}$&$^+_-$ & $^{2.3}_{2.0}$ \\
$[5,6]$ & $11.41$ & $^+_-$ & $^{0.75}_{0.75}$&$^+_-$ & $^{0.63}_{0.65}$ & $10.62$ & $^+_-$ & $^{0.55}_{0.53}$&$^+_-$ & $^{0.39}_{0.41}$ & $14.23$ & $^+_-$ & $^{0.75}_{0.72}$&$^+_-$ & $^{0.61}_{0.62}$ & $11.88$ & $^+_-$ & $^{0.96}_{0.94}$&$^+_-$ & $^{0.72}_{0.76}$ & $23.1$ & $^+_-$ & $^{6.3}_{5.2}$&$^+_-$ & $^{7.6}_{5.7}$ \\
$[6,7]$ & $11.75$ & $^+_-$ & $^{0.92}_{0.94}$&$^+_-$ & $^{0.70}_{0.75}$ & $12.98$ & $^+_-$ & $^{0.86}_{0.82}$&$^+_-$ & $^{0.71}_{0.54}$ & $11.05$ & $^+_-$ & $^{0.85}_{0.85}$&$^+_-$ & $^{0.59}_{0.52}$ & $12.8$ & $^+_-$ & $^{1.5}_{1.5}$&$^+_-$ & $^{1.5}_{1.3}$ & \multicolumn{5}{c}{ } \\
$[7,8]$ & $14.7$ & $^+_-$ & $^{1.4}_{1.4}$&$^+_-$ & $^{0.9}_{1.2}$ & $11.47$ & $^+_-$ & $^{0.97}_{0.97}$&$^+_-$ & $^{0.78}_{0.56}$ & $11.6$ & $^+_-$ & $^{1.2}_{1.2}$&$^+_-$ & $^{0.8}_{0.8}$ & $14.8$ & $^+_-$ & $^{2.5}_{2.3}$&$^+_-$ & $^{3.5}_{2.6}$ & \multicolumn{5}{c}{ } \\
$[8,9]$ & $9.8$ & $^+_-$ & $^{1.4}_{1.4}$&$^+_-$ & $^{0.8}_{0.7}$ & $10.5$ & $^+_-$ & $^{1.2}_{1.2}$&$^+_-$ & $^{0.8}_{0.7}$ & $11.6$ & $^+_-$ & $^{1.6}_{1.5}$&$^+_-$ & $^{1.3}_{1.1}$ & \multicolumn{5}{c}{ } & \multicolumn{5}{c}{ } \\
$[9,10]$ & $11.4$ & $^+_-$ & $^{1.8}_{1.8}$&$^+_-$ & $^{1.3}_{1.2}$ & $12.5$ & $^+_-$ & $^{1.8}_{1.7}$&$^+_-$ & $^{1.3}_{1.0}$ & $15.2$ & $^+_-$ & $^{2.9}_{2.7}$&$^+_-$ & $^{2.7}_{2.1}$ & \multicolumn{5}{c}{ } & \multicolumn{5}{c}{ } \\
\bottomrule\end{tabular}

}{%
  The ratios of the measurements of cross-section times branching fraction 
  for prompt \Dsp and \Dz mesons in bins of \pTy.
  The first uncertainty is statistical, and the second is the total 
  systematic. All values are given in percent.
}{table:meson_ratio:DsD0}
\vspace*{\fill}

\newpage
\vspace*{\fill}
\hvFloat[
  nonFloat=true,
  objectAngle=90,
  capPos=l,
  capAngle=90,
  capWidth=h]{table}{
    \centering
    \renewcommand{\arraystretch}{1.3}
\begin{tabular}{c|r@{\hskip+0.2em}c@{\hskip+0.2em}r@{\hskip+0.2em}c@{\hskip+0.2em}rr@{\hskip+0.2em}c@{\hskip+0.2em}r@{\hskip+0.2em}c@{\hskip+0.2em}rr@{\hskip+0.2em}c@{\hskip+0.2em}r@{\hskip+0.2em}c@{\hskip+0.2em}rr@{\hskip+0.2em}c@{\hskip+0.2em}r@{\hskip+0.2em}c@{\hskip+0.2em}rr@{\hskip+0.2em}c@{\hskip+0.2em}r@{\hskip+0.2em}c@{\hskip+0.2em}r}
\toprule&\multicolumn{25}{c}{$\text{$y$}$}\\
$\text{$p_{\text{T}}$ [\gevc]}$ & \multicolumn{5}{c}{$[2.0,2.5]$} & \multicolumn{5}{c}{$[2.5,3.0]$} & \multicolumn{5}{c}{$[3.0,3.5]$} & \multicolumn{5}{c}{$[3.5,4.0]$} & \multicolumn{5}{c}{$[4.0,4.5]$} \\
\midrule$[1,2]$ & \multicolumn{5}{c}{ } & $28.6$ & $^+_-$ & $^{1.3}_{1.3}$&$^+_-$ & $^{2.1}_{1.2}$ & $26.9$ & $^+_-$ & $^{0.6}_{0.6}$&$^+_-$ & $^{1.5}_{1.7}$ & $25.2$ & $^+_-$ & $^{0.7}_{0.7}$&$^+_-$ & $^{1.5}_{1.7}$ & $23.4$ & $^+_-$ & $^{1.4}_{1.4}$&$^+_-$ & $^{1.6}_{1.9}$ \\
$[2,3]$ & $31.7$ & $^+_-$ & $^{3.2}_{3.1}$&$^+_-$ & $^{3.0}_{3.2}$ & $30.5$ & $^+_-$ & $^{0.7}_{0.7}$&$^+_-$ & $^{1.1}_{1.2}$ & $29.8$ & $^+_-$ & $^{0.5}_{0.5}$&$^+_-$ & $^{1.5}_{1.9}$ & $26.5$ & $^+_-$ & $^{0.7}_{0.7}$&$^+_-$ & $^{1.5}_{1.8}$ & $28.3$ & $^+_-$ & $^{1.5}_{1.4}$&$^+_-$ & $^{2.9}_{2.7}$ \\
$[3,4]$ & $35.4$ & $^+_-$ & $^{1.8}_{1.8}$&$^+_-$ & $^{1.5}_{2.1}$ & $31.9$ & $^+_-$ & $^{0.7}_{0.7}$&$^+_-$ & $^{1.3}_{1.1}$ & $32.4$ & $^+_-$ & $^{0.7}_{0.6}$&$^+_-$ & $^{1.9}_{1.7}$ & $28.2$ & $^+_-$ & $^{0.8}_{0.8}$&$^+_-$ & $^{1.6}_{1.9}$ & $29.7$ & $^+_-$ & $^{2.0}_{2.0}$&$^+_-$ & $^{2.1}_{2.5}$ \\
$[4,5]$ & $34.5$ & $^+_-$ & $^{1.6}_{1.6}$&$^+_-$ & $^{1.6}_{1.7}$ & $33.2$ & $^+_-$ & $^{0.8}_{0.8}$&$^+_-$ & $^{1.2}_{1.4}$ & $33.6$ & $^+_-$ & $^{0.9}_{0.8}$&$^+_-$ & $^{2.0}_{1.6}$ & $34.1$ & $^+_-$ & $^{1.2}_{1.2}$&$^+_-$ & $^{2.4}_{2.6}$ & $28.4$ & $^+_-$ & $^{3.5}_{3.4}$&$^+_-$ & $^{7.5}_{5.5}$ \\
$[5,6]$ & $35.8$ & $^+_-$ & $^{1.8}_{1.8}$&$^+_-$ & $^{1.8}_{1.9}$ & $32.2$ & $^+_-$ & $^{1.0}_{1.0}$&$^+_-$ & $^{1.4}_{1.4}$ & $31.9$ & $^+_-$ & $^{1.1}_{1.1}$&$^+_-$ & $^{1.8}_{1.7}$ & $31.4$ & $^+_-$ & $^{1.7}_{1.6}$&$^+_-$ & $^{2.3}_{2.0}$ & $67$ & $^+_-$ & $^{11}_{\phantom{0}9}$&$^+_-$ & $^{23}_{17}$ \\
$[6,7]$ & $38.1$ & $^+_-$ & $^{2.2}_{2.2}$&$^+_-$ & $^{2.0}_{2.2}$ & $33.4$ & $^+_-$ & $^{1.4}_{1.4}$&$^+_-$ & $^{1.5}_{1.6}$ & $32.9$ & $^+_-$ & $^{1.6}_{1.6}$&$^+_-$ & $^{1.8}_{1.8}$ & $36.6$ & $^+_-$ & $^{3.0}_{3.0}$&$^+_-$ & $^{3.5}_{3.7}$ & \multicolumn{5}{c}{ } \\
$[7,8]$ & $33.2$ & $^+_-$ & $^{2.4}_{2.4}$&$^+_-$ & $^{2.0}_{2.2}$ & $32.9$ & $^+_-$ & $^{1.8}_{1.8}$&$^+_-$ & $^{1.5}_{2.0}$ & $33.9$ & $^+_-$ & $^{2.2}_{2.1}$&$^+_-$ & $^{2.3}_{2.2}$ & $37.2$ & $^+_-$ & $^{5.2}_{5.0}$&$^+_-$ & $^{7.1}_{6.3}$ & \multicolumn{5}{c}{ } \\
$[8,9]$ & $29.9$ & $^+_-$ & $^{2.8}_{2.8}$&$^+_-$ & $^{2.2}_{2.4}$ & $34.2$ & $^+_-$ & $^{2.5}_{2.3}$&$^+_-$ & $^{1.9}_{2.4}$ & $33.2$ & $^+_-$ & $^{3.0}_{2.8}$&$^+_-$ & $^{3.2}_{2.8}$ & $59$ & $^+_-$ & $^{14}_{12}$&$^+_-$ & $^{24}_{13}$ & \multicolumn{5}{c}{ } \\
$[9,10]$ & $35.8$ & $^+_-$ & $^{3.9}_{3.8}$&$^+_-$ & $^{3.1}_{3.0}$ & $33.8$ & $^+_-$ & $^{3.1}_{3.0}$&$^+_-$ & $^{2.3}_{2.9}$ & $57.2$ & $^+_-$ & $^{6.5}_{6.3}$&$^+_-$ & $^{9.6}_{8.1}$ & \multicolumn{5}{c}{ } & \multicolumn{5}{c}{ } \\
\bottomrule\end{tabular}

}{%
  The ratios of the measurements of cross-section times branching fraction 
  for prompt \Dstarp and \Dz mesons in bins of \pTy.
  The first uncertainty is statistical, and the second is the total 
  systematic. All values are given in percent.
}{table:meson_ratio:DstD0}
\vspace*{\fill}

\newpage
\vspace*{\fill}
\hvFloat[
  nonFloat=true,
  objectAngle=90,
  capPos=l,
  capAngle=90,
  capWidth=h]{table}{
    \centering
    \renewcommand{\arraystretch}{1.3}
\begin{tabular}{c|r@{\hskip+0.2em}c@{\hskip+0.2em}r@{\hskip+0.2em}c@{\hskip+0.2em}rr@{\hskip+0.2em}c@{\hskip+0.2em}r@{\hskip+0.2em}c@{\hskip+0.2em}rr@{\hskip+0.2em}c@{\hskip+0.2em}r@{\hskip+0.2em}c@{\hskip+0.2em}rr@{\hskip+0.2em}c@{\hskip+0.2em}r@{\hskip+0.2em}c@{\hskip+0.2em}rr@{\hskip+0.2em}c@{\hskip+0.2em}r@{\hskip+0.2em}c@{\hskip+0.2em}r}
\toprule&\multicolumn{25}{c}{$\text{$y$}$}\\
$\text{$p_{\text{T}}$ [\gevc]}$ & \multicolumn{5}{c}{$[2.0,2.5]$} & \multicolumn{5}{c}{$[2.5,3.0]$} & \multicolumn{5}{c}{$[3.0,3.5]$} & \multicolumn{5}{c}{$[3.5,4.0]$} & \multicolumn{5}{c}{$[4.0,4.5]$} \\
\midrule$[1,2]$ & $8.0$ & $^+_-$ & $^{1.3}_{1.2}$&$^+_-$ & $^{2.1}_{1.5}$ & $10.84$ & $^+_-$ & $^{0.67}_{0.66}$&$^+_-$ & $^{0.55}_{0.54}$ & $9.87$ & $^+_-$ & $^{0.62}_{0.63}$&$^+_-$ & $^{0.47}_{0.55}$ & $8.84$ & $^+_-$ & $^{0.82}_{0.82}$&$^+_-$ & $^{0.60}_{0.58}$ & \multicolumn{5}{c}{ } \\
$[2,3]$ & $11.7$ & $^+_-$ & $^{0.7}_{0.6}$&$^+_-$ & $^{1.9}_{1.5}$ & $12.03$ & $^+_-$ & $^{0.33}_{0.34}$&$^+_-$ & $^{0.34}_{0.28}$ & $11.53$ & $^+_-$ & $^{0.35}_{0.34}$&$^+_-$ & $^{0.33}_{0.35}$ & $9.56$ & $^+_-$ & $^{0.42}_{0.42}$&$^+_-$ & $^{0.38}_{0.38}$ & $9.2$ & $^+_-$ & $^{1.0}_{1.0}$&$^+_-$ & $^{1.3}_{0.9}$ \\
$[3,4]$ & $10.9$ & $^+_-$ & $^{0.5}_{0.5}$&$^+_-$ & $^{1.1}_{0.8}$ & $11.87$ & $^+_-$ & $^{0.35}_{0.34}$&$^+_-$ & $^{0.36}_{0.17}$ & $10.99$ & $^+_-$ & $^{0.36}_{0.36}$&$^+_-$ & $^{0.40}_{0.21}$ & $11.63$ & $^+_-$ & $^{0.51}_{0.51}$&$^+_-$ & $^{0.54}_{0.33}$ & $10.7$ & $^+_-$ & $^{1.0}_{1.0}$&$^+_-$ & $^{0.9}_{0.8}$ \\
$[4,5]$ & $10.64$ & $^+_-$ & $^{0.56}_{0.56}$&$^+_-$ & $^{0.68}_{0.64}$ & $12.29$ & $^+_-$ & $^{0.45}_{0.44}$&$^+_-$ & $^{0.34}_{0.33}$ & $12.11$ & $^+_-$ & $^{0.49}_{0.50}$&$^+_-$ & $^{0.47}_{0.35}$ & $10.57$ & $^+_-$ & $^{0.61}_{0.60}$&$^+_-$ & $^{0.52}_{0.36}$ & $12.3$ & $^+_-$ & $^{1.6}_{1.6}$&$^+_-$ & $^{1.1}_{0.9}$ \\
$[5,6]$ & $10.64$ & $^+_-$ & $^{0.72}_{0.71}$&$^+_-$ & $^{0.54}_{0.53}$ & $9.88$ & $^+_-$ & $^{0.51}_{0.50}$&$^+_-$ & $^{0.38}_{0.28}$ & $13.68$ & $^+_-$ & $^{0.70}_{0.69}$&$^+_-$ & $^{0.61}_{0.45}$ & $11.52$ & $^+_-$ & $^{0.91}_{0.90}$&$^+_-$ & $^{0.71}_{0.56}$ & $12.6$ & $^+_-$ & $^{2.6}_{2.6}$&$^+_-$ & $^{1.7}_{1.3}$ \\
$[6,7]$ & $11.01$ & $^+_-$ & $^{0.89}_{0.87}$&$^+_-$ & $^{0.64}_{0.57}$ & $12.17$ & $^+_-$ & $^{0.79}_{0.77}$&$^+_-$ & $^{0.64}_{0.37}$ & $9.87$ & $^+_-$ & $^{0.75}_{0.74}$&$^+_-$ & $^{0.63}_{0.40}$ & $11.5$ & $^+_-$ & $^{1.3}_{1.3}$&$^+_-$ & $^{1.1}_{0.8}$ & \multicolumn{5}{c}{ } \\
$[7,8]$ & $13.0$ & $^+_-$ & $^{1.2}_{1.2}$&$^+_-$ & $^{0.9}_{0.9}$ & $11.53$ & $^+_-$ & $^{0.97}_{0.97}$&$^+_-$ & $^{0.88}_{0.39}$ & $10.9$ & $^+_-$ & $^{1.1}_{1.1}$&$^+_-$ & $^{0.8}_{0.5}$ & $11.4$ & $^+_-$ & $^{1.7}_{1.7}$&$^+_-$ & $^{1.8}_{1.3}$ & \multicolumn{5}{c}{ } \\
$[8,9]$ & $9.1$ & $^+_-$ & $^{1.3}_{1.3}$&$^+_-$ & $^{0.8}_{0.6}$ & $9.9$ & $^+_-$ & $^{1.2}_{1.1}$&$^+_-$ & $^{0.9}_{0.4}$ & $11.4$ & $^+_-$ & $^{1.5}_{1.5}$&$^+_-$ & $^{1.1}_{0.9}$ & \multicolumn{5}{c}{ } & \multicolumn{5}{c}{ } \\
$[9,10]$ & $12.5$ & $^+_-$ & $^{1.9}_{1.9}$&$^+_-$ & $^{1.4}_{1.1}$ & $12.3$ & $^+_-$ & $^{1.7}_{1.7}$&$^+_-$ & $^{1.4}_{0.7}$ & $12.2$ & $^+_-$ & $^{2.2}_{2.1}$&$^+_-$ & $^{1.5}_{1.3}$ & \multicolumn{5}{c}{ } & \multicolumn{5}{c}{ } \\
\bottomrule\end{tabular}

}{%
  The ratios of the measurements of cross-section times branching fraction 
  for prompt \Dsp and \Dp mesons in bins of \pTy.
  The first uncertainty is statistical, and the second is the total 
  systematic. All values are given in percent.
}{table:meson_ratio:DsDp}
\vspace*{\fill}

\newpage
\vspace*{\fill}
\hvFloat[
  nonFloat=true,
  objectAngle=90,
  capPos=l,
  capAngle=90,
  capWidth=h]{table}{
    \centering
    \renewcommand{\arraystretch}{1.3}
\begin{tabular}{c|r@{\hskip+0.2em}c@{\hskip+0.2em}r@{\hskip+0.2em}c@{\hskip+0.2em}rr@{\hskip+0.2em}c@{\hskip+0.2em}r@{\hskip+0.2em}c@{\hskip+0.2em}rr@{\hskip+0.2em}c@{\hskip+0.2em}r@{\hskip+0.2em}c@{\hskip+0.2em}rr@{\hskip+0.2em}c@{\hskip+0.2em}r@{\hskip+0.2em}c@{\hskip+0.2em}rr@{\hskip+0.2em}c@{\hskip+0.2em}r@{\hskip+0.2em}c@{\hskip+0.2em}r}
\toprule&\multicolumn{25}{c}{$\text{$y$}$}\\
$\text{$p_{\text{T}}$ [\gevc]}$ & \multicolumn{5}{c}{$[2.0,2.5]$} & \multicolumn{5}{c}{$[2.5,3.0]$} & \multicolumn{5}{c}{$[3.0,3.5]$} & \multicolumn{5}{c}{$[3.5,4.0]$} & \multicolumn{5}{c}{$[4.0,4.5]$} \\
\midrule$[1,2]$ & \multicolumn{5}{c}{ } & $32.5$ & $^+_-$ & $^{1.5}_{1.5}$&$^+_-$ & $^{2.2}_{0.8}$ & $28.4$ & $^+_-$ & $^{0.7}_{0.7}$&$^+_-$ & $^{0.5}_{1.6}$ & $28.2$ & $^+_-$ & $^{0.9}_{0.9}$&$^+_-$ & $^{0.9}_{1.6}$ & $23.9$ & $^+_-$ & $^{1.6}_{1.5}$&$^+_-$ & $^{1.6}_{2.0}$ \\
$[2,3]$ & $34.5$ & $^+_-$ & $^{3.4}_{3.5}$&$^+_-$ & $^{3.8}_{3.9}$ & $32.4$ & $^+_-$ & $^{0.8}_{0.8}$&$^+_-$ & $^{1.1}_{0.9}$ & $30.2$ & $^+_-$ & $^{0.5}_{0.5}$&$^+_-$ & $^{0.8}_{1.5}$ & $27.6$ & $^+_-$ & $^{0.7}_{0.7}$&$^+_-$ & $^{0.9}_{1.5}$ & $28.5$ & $^+_-$ & $^{1.5}_{1.4}$&$^+_-$ & $^{2.8}_{2.5}$ \\
$[3,4]$ & $35.8$ & $^+_-$ & $^{1.8}_{1.8}$&$^+_-$ & $^{2.2}_{2.2}$ & $32.2$ & $^+_-$ & $^{0.7}_{0.7}$&$^+_-$ & $^{1.4}_{0.4}$ & $32.1$ & $^+_-$ & $^{0.6}_{0.6}$&$^+_-$ & $^{1.6}_{1.2}$ & $28.5$ & $^+_-$ & $^{0.8}_{0.8}$&$^+_-$ & $^{1.4}_{1.5}$ & $26.5$ & $^+_-$ & $^{1.7}_{1.7}$&$^+_-$ & $^{1.4}_{1.7}$ \\
$[4,5]$ & $32.5$ & $^+_-$ & $^{1.5}_{1.5}$&$^+_-$ & $^{1.7}_{1.4}$ & $32.2$ & $^+_-$ & $^{0.8}_{0.8}$&$^+_-$ & $^{1.1}_{0.9}$ & $32.0$ & $^+_-$ & $^{0.8}_{0.8}$&$^+_-$ & $^{1.7}_{1.1}$ & $33.2$ & $^+_-$ & $^{1.1}_{1.1}$&$^+_-$ & $^{2.2}_{2.2}$ & $26.9$ & $^+_-$ & $^{3.1}_{3.1}$&$^+_-$ & $^{5.8}_{4.0}$ \\
$[5,6]$ & $33.4$ & $^+_-$ & $^{1.7}_{1.7}$&$^+_-$ & $^{1.5}_{1.5}$ & $29.9$ & $^+_-$ & $^{1.0}_{0.9}$&$^+_-$ & $^{1.1}_{1.0}$ & $30.6$ & $^+_-$ & $^{1.1}_{1.1}$&$^+_-$ & $^{1.6}_{1.2}$ & $30.5$ & $^+_-$ & $^{1.6}_{1.6}$&$^+_-$ & $^{1.9}_{1.3}$ & $36.6$ & $^+_-$ & $^{2.2}_{2.0}$&$^+_-$ & $^{5.8}_{4.7}$ \\
$[6,7]$ & $35.8$ & $^+_-$ & $^{2.1}_{2.1}$&$^+_-$ & $^{1.8}_{1.7}$ & $31.3$ & $^+_-$ & $^{1.3}_{1.2}$&$^+_-$ & $^{1.2}_{1.2}$ & $29.4$ & $^+_-$ & $^{1.4}_{1.4}$&$^+_-$ & $^{1.6}_{1.3}$ & $33.1$ & $^+_-$ & $^{2.6}_{2.5}$&$^+_-$ & $^{2.1}_{2.1}$ & \multicolumn{5}{c}{ } \\
$[7,8]$ & $29.2$ & $^+_-$ & $^{2.1}_{2.1}$&$^+_-$ & $^{1.8}_{1.6}$ & $33.1$ & $^+_-$ & $^{1.7}_{1.7}$&$^+_-$ & $^{1.6}_{1.4}$ & $31.8$ & $^+_-$ & $^{2.0}_{1.9}$&$^+_-$ & $^{2.0}_{1.5}$ & $28.6$ & $^+_-$ & $^{3.6}_{3.5}$&$^+_-$ & $^{3.1}_{3.0}$ & \multicolumn{5}{c}{ } \\
$[8,9]$ & $27.6$ & $^+_-$ & $^{2.5}_{2.5}$&$^+_-$ & $^{1.9}_{1.8}$ & $32.2$ & $^+_-$ & $^{2.2}_{2.2}$&$^+_-$ & $^{1.9}_{1.6}$ & $32.3$ & $^+_-$ & $^{2.7}_{2.6}$&$^+_-$ & $^{2.5}_{2.1}$ & $29.0$ & $^+_-$ & $^{5.4}_{5.3}$&$^+_-$ & $^{8.8}_{5.6}$ & \multicolumn{5}{c}{ } \\
$[9,10]$ & $39.3$ & $^+_-$ & $^{4.2}_{4.1}$&$^+_-$ & $^{3.2}_{2.9}$ & $33.2$ & $^+_-$ & $^{3.0}_{2.9}$&$^+_-$ & $^{2.4}_{2.2}$ & $45.8$ & $^+_-$ & $^{4.8}_{4.6}$&$^+_-$ & $^{5.8}_{5.1}$ & \multicolumn{5}{c}{ } & \multicolumn{5}{c}{ } \\
\bottomrule\end{tabular}

}{%
  The ratios of the measurements of cross-section times branching fraction 
  for prompt \Dstarp and \Dp mesons in bins of \pTy.
  The first uncertainty is statistical, and the second is the total 
  systematic. All values are given in percent.
}{table:meson_ratio:DstDp}
\vspace*{\fill}

\newpage
\vspace*{\fill}
\hvFloat[
  nonFloat=true,
  objectAngle=90,
  capPos=l,
  capAngle=90,
  capWidth=h]{table}{
    \centering
    \renewcommand{\arraystretch}{1.3}
\begin{tabular}{c|r@{\hskip+0.2em}c@{\hskip+0.2em}r@{\hskip+0.2em}c@{\hskip+0.2em}rr@{\hskip+0.2em}c@{\hskip+0.2em}r@{\hskip+0.2em}c@{\hskip+0.2em}rr@{\hskip+0.2em}c@{\hskip+0.2em}r@{\hskip+0.2em}c@{\hskip+0.2em}rr@{\hskip+0.2em}c@{\hskip+0.2em}r@{\hskip+0.2em}c@{\hskip+0.2em}rr@{\hskip+0.2em}c@{\hskip+0.2em}r@{\hskip+0.2em}c@{\hskip+0.2em}r}
\toprule&\multicolumn{25}{c}{$\text{$y$}$}\\
$\text{$p_{\text{T}}$ [\gevc]}$ & \multicolumn{5}{c}{$[2.0,2.5]$} & \multicolumn{5}{c}{$[2.5,3.0]$} & \multicolumn{5}{c}{$[3.0,3.5]$} & \multicolumn{5}{c}{$[3.5,4.0]$} & \multicolumn{5}{c}{$[4.0,4.5]$} \\
\midrule$[1,2]$ & \multicolumn{5}{c}{ } & $33.4$ & $^+_-$ & $^{2.6}_{2.4}$&$^+_-$ & $^{1.3}_{2.6}$ & $34.8$ & $^+_-$ & $^{2.3}_{2.3}$&$^+_-$ & $^{2.7}_{1.2}$ & $31.4$ & $^+_-$ & $^{3.0}_{3.0}$&$^+_-$ & $^{3.3}_{1.9}$ & \multicolumn{5}{c}{ } \\
$[2,3]$ & $33.8$ & $^+_-$ & $^{4.2}_{3.5}$&$^+_-$ & $^{7.2}_{4.6}$ & $37.0$ & $^+_-$ & $^{1.4}_{1.3}$&$^+_-$ & $^{1.3}_{1.3}$ & $38.2$ & $^+_-$ & $^{1.3}_{1.3}$&$^+_-$ & $^{2.3}_{1.0}$ & $34.6$ & $^+_-$ & $^{1.7}_{1.7}$&$^+_-$ & $^{2.7}_{1.6}$ & $32.1$ & $^+_-$ & $^{3.9}_{3.8}$&$^+_-$ & $^{5.5}_{4.0}$ \\
$[3,4]$ & $30.5$ & $^+_-$ & $^{2.2}_{2.1}$&$^+_-$ & $^{3.3}_{2.1}$ & $36.9$ & $^+_-$ & $^{1.3}_{1.3}$&$^+_-$ & $^{0.8}_{1.4}$ & $34.3$ & $^+_-$ & $^{1.3}_{1.2}$&$^+_-$ & $^{1.7}_{1.6}$ & $40.8$ & $^+_-$ & $^{2.1}_{2.0}$&$^+_-$ & $^{3.2}_{2.1}$ & $40.4$ & $^+_-$ & $^{4.6}_{4.5}$&$^+_-$ & $^{4.2}_{2.6}$ \\
$[4,5]$ & $32.8$ & $^+_-$ & $^{2.3}_{2.1}$&$^+_-$ & $^{2.1}_{2.1}$ & $38.2$ & $^+_-$ & $^{1.6}_{1.6}$&$^+_-$ & $^{1.3}_{1.4}$ & $37.9$ & $^+_-$ & $^{1.7}_{1.7}$&$^+_-$ & $^{1.8}_{2.2}$ & $31.9$ & $^+_-$ & $^{2.0}_{2.0}$&$^+_-$ & $^{2.9}_{2.1}$ & $45.7$ & $^+_-$ & $^{8.3}_{7.1}$&$^+_-$ & $^{9.5}_{7.7}$ \\
$[5,6]$ & $31.9$ & $^+_-$ & $^{2.7}_{2.5}$&$^+_-$ & $^{1.9}_{1.7}$ & $33.0$ & $^+_-$ & $^{1.9}_{1.8}$&$^+_-$ & $^{1.4}_{1.2}$ & $44.6$ & $^+_-$ & $^{2.7}_{2.5}$&$^+_-$ & $^{2.6}_{2.4}$ & $37.8$ & $^+_-$ & $^{3.5}_{3.3}$&$^+_-$ & $^{2.6}_{2.7}$ & $34.3$ & $^+_-$ & $^{6.9}_{6.8}$&$^+_-$ & $^{6.3}_{4.9}$ \\
$[6,7]$ & $30.8$ & $^+_-$ & $^{3.0}_{2.8}$&$^+_-$ & $^{2.0}_{1.8}$ & $38.9$ & $^+_-$ & $^{2.8}_{2.8}$&$^+_-$ & $^{2.5}_{1.5}$ & $33.6$ & $^+_-$ & $^{2.9}_{2.8}$&$^+_-$ & $^{2.4}_{1.9}$ & $34.8$ & $^+_-$ & $^{4.8}_{4.4}$&$^+_-$ & $^{3.9}_{2.6}$ & \multicolumn{5}{c}{ } \\
$[7,8]$ & $44.5$ & $^+_-$ & $^{5.2}_{5.0}$&$^+_-$ & $^{3.5}_{3.0}$ & $34.9$ & $^+_-$ & $^{3.3}_{3.3}$&$^+_-$ & $^{3.0}_{1.4}$ & $34.3$ & $^+_-$ & $^{4.0}_{3.7}$&$^+_-$ & $^{2.9}_{2.3}$ & $40.0$ & $^+_-$ & $^{8.1}_{7.1}$&$^+_-$ & $^{7.6}_{4.8}$ & \multicolumn{5}{c}{ } \\
$[8,9]$ & $32.9$ & $^+_-$ & $^{5.5}_{5.2}$&$^+_-$ & $^{3.3}_{2.7}$ & $30.6$ & $^+_-$ & $^{4.1}_{3.9}$&$^+_-$ & $^{3.1}_{1.5}$ & $35.1$ & $^+_-$ & $^{5.4}_{5.2}$&$^+_-$ & $^{3.8}_{3.0}$ & \multicolumn{5}{c}{ } & \multicolumn{5}{c}{ } \\
$[9,10]$ & $31.9$ & $^+_-$ & $^{6.0}_{5.4}$&$^+_-$ & $^{4.3}_{3.3}$ & $37.0$ & $^+_-$ & $^{6.0}_{5.5}$&$^+_-$ & $^{4.8}_{2.6}$ & $26.6$ & $^+_-$ & $^{5.5}_{4.9}$&$^+_-$ & $^{4.3}_{3.2}$ & \multicolumn{5}{c}{ } & \multicolumn{5}{c}{ } \\
\bottomrule\end{tabular}

}{%
  The ratios of the measurements of cross-section times branching fraction 
  for prompt \Dsp and \Dstarp mesons in bins of \pTy.
  The first uncertainty is statistical, and the second is the total 
  systematic. All values are given in percent.
}{table:meson_ratio:DsDst}
\vspace*{\fill}

\clearpage





\addcontentsline{toc}{section}{References}
\setboolean{inbibliography}{true}
\bibliographystyle{LHCb}
\bibliography{main,LHCb-PAPER,LHCb-CONF,LHCb-DP,LHCb-TDR,theory}

\newpage

\newpage
\centerline{\large\bf LHCb collaboration}
\begin{flushleft}
\small
R.~Aaij$^{40}$,
B.~Adeva$^{39}$,
M.~Adinolfi$^{48}$,
Z.~Ajaltouni$^{5}$,
S.~Akar$^{6}$,
J.~Albrecht$^{10}$,
F.~Alessio$^{40}$,
M.~Alexander$^{53}$,
S.~Ali$^{43}$,
G.~Alkhazov$^{31}$
% ...

{\footnotesize \it
% ...
$ ^{5}$Clermont Universit{\'e}, Universit{\'e} Blaise Pascal, CNRS/IN2P3, LPC, Clermont-Ferrand, France\\
$ ^{6}$CPPM, Aix-Marseille Universit{\'e}, CNRS/IN2P3, Marseille, France\\
% ...
$ ^{10}$Fakult{\"a}t Physik, Technische Universit{\"a}t Dortmund, Dortmund, Germany\\
% ...
$ ^{31}$Petersburg Nuclear Physics Institute (PNPI), Gatchina, Russia\\
% ...
$ ^{39}$Universidad de Santiago de Compostela, Santiago de Compostela, Spain\\
$ ^{40}$European Organization for Nuclear Research (CERN), Geneva, Switzerland\\
% ...
$ ^{43}$Nikhef National Institute for Subatomic Physics, Amsterdam, The Netherlands\\
% ...
$ ^{48}$H.H. Wills Physics Laboratory, University of Bristol, Bristol, United Kingdom\\
% ...
$ ^{53}$School of Physics and Astronomy, University of Glasgow, Glasgow, United Kingdom\\
% ...
}
\end{flushleft}

\end{document}